\documentclass{aa}
\usepackage{graphicx}
\usepackage{amsmath}
\usepackage{txfonts}
\usepackage{lscape}
\usepackage{longtable}
\usepackage{rotating}
\usepackage{appendix}
\begin{document}
\def\HII{H\,{\sc{ii}}}
\def\HI{H\,{\sc{i}}}

\def\Ks{$K_{\rm{s}}$}
\def\sun{\hbox{$\odot$}}
\def\degr{\hbox{$^\circ$}}
\def\h{\hbox{$^{\reset@font\r@mn{h}}$}}
\def\m{\hbox{$^{\reset@font\r@mn{m}}$}}
\def\s{\hbox{$^{\reset@font\r@mn{s}}$}}
\def\msol{\hbox{\kern 0.20em $M_\odot$}}

\def\kms{\hbox{\kern 0.20em km\kern 0.20em s$^{-1}$}}
\def\pc{\hbox{\kern 0.20em pc$^{2}$}}
\def\halpha{\hbox{H$\alpha$ }}

\newcommand\micron{\mbox{$\mu$m}}%
   \title{A gallery of bubbles}
   
   \subtitle{The nature of the bubbles observed by Spitzer and \\
   what ATLASGAL tells us about the surrounding neutral material}
\author{L. Deharveng\inst{1}
	\and F. Schuller\inst{2}
	\and L.D. Anderson\inst{1}
	\and A. Zavagno\inst{1}
	\and F. Wyrowski\inst{2}	
	\and K.M. Menten\inst{2}
	\and L. Bronfman\inst{3}
	\and L. Testi\inst{4}
	\and C.M. Walmsley\inst{5}
	\and M. Wienen\inst{2}
	}
	

\institute{Laboratoire d'Astrophysique de Marseille (UMR 6110 CNRS \& 
Universit\'e de Provence), 38 rue F. Joliot-Curie, 13388 Marseille 
Cedex 13, France
	\and Max-Planck Institut f\"{u}r Radioastronomie, 
	Auf dem H\"{u}gel 69, D-53121 Bonn, Germany
	\and Departamento de Astronomia, Universidad de Chile, 
	Casilla 36-D, Santiago, Chile
	\and ESO, Karl Schwarzschild-Strasse 2, D-85748 Garching 
	bei M\"{u}nchen, germany
	\and Osservatorio Astrofisico di Arcetri, Largo E. Fermi, 5, 
	I-50125 Firence, Italy
	}

   \date{Received ; Accepted }


  \abstract
   {This study deals with infrared bubbles, the \HII\ regions they enclose, and triggered 
   massive-star formation on their borders.}
   {We attempt to determine the nature of the bubbles 
   observed by Spitzer in the Galactic plane, mainly to establish if possible  
   their association with massive stars. We take advantage of 
   the very simple morphology of these objects to search for star formation 
   triggered by \HII\ regions, and to estimate the importance of this mode 
   of star formation.}   
   {We consider a sample of 102 bubbles detected by Spitzer-GLIMPSE, 
   and 
   catalogued by Churchwell et al.~(2006; hereafter CH06). We 
   use mid-infrared and radio-continuum public data (respectively the 
   {\it Spitzer}-GLIMPSE and -MIPSGAL surveys and the MAGPIS and 
   VGPS surveys) to discuss their nature. We use the ATLASGAL 
   survey at 870~$\mu$m  to search for dense neutral material collected 
   on their borders. The 870~$\mu$m data traces the distribution of cold dust, 
   thus of the dense neutral material where stars may form.}   
   {We find that 86\% of the bubbles contain ionized gas detected by means  
   of its radio-continuum emission at 20-cm. Thus, most of the bubbles 
   observed at 8.0~$\mu$m enclose \HII\ regions ionized by O-B2 stars. 
   This finding differs from the earlier CH06 results 
   ($\sim$25\% of the bubbles enclosing \HII\ regions). 
   Ninety-eight percent of the bubbles exhibit 24~$\mu$m emission 
   in their central regions. The ionized regions at the center of 
   the 8.0~$\mu$m bubbles seem to be devoid of PAHs but contain hot 
   dust. PAH emission at 8.0~$\mu$m is observed in the 
   direction of the photodissociation regions surrounding the 
   ionized gas. 
      
   Among the 65 regions for which the  
   angular resolution of the observations is high enough to resolve the 
   spatial distribution of cold dust 
   at 870~$\mu$m, we find that 40\% are surrounded by cold dust, 
   and that another 28\% contain interacting condensations. 
   The former are good candidates for the collect and collapse process, 
   as they 
   display an accumulation of dense material at their borders. The latter 
   are good candidates for the compression of pre-existing 
   condensations by the ionized gas. 
   Thirteen bubbles exhibit associated ultracompact \HII\ regions 
   in the direction of dust condensations adjacent to their ionization 
   fronts. Another five show methanol masers in similar condensations.}   
   {Our results suggest that 
  more than a quarter of the bubbles may have triggered the formation 
  of massive objects.  Therefore, star 
  formation triggered by \HII\ regions may be an important process, 
  especially for massive-star formation.}

\keywords{Stars: formation -- Stars: early-type -- ISM: \HII\
regions }

\maketitle

 \titlerunning{A gallery of bubbles}
 \authorrunning{L.~Deharveng et al.}

\maketitle


\section{Introduction}

The {\it Spitzer}-GLIMPSE images at 8.0~$\mu$m (Benjamin et
al. \cite{ben03}) have revealed a ``bubbling Galactic disk''
(Churchwell et al.~\cite{chu06}; hereafter CH06). About 600 ring
structures have been detected and catalogued between longitudes
$-60\degr$ and $+60\degr$ by CH06 and Churchwell et
al.~(\cite{chu07}). Most of these bubbles have angular diameters smaller 
than $10\arcmin$; according to Churchwell et al.~(\cite{chu07}),
many more bubbles are present, with smaller or larger diameters.
Thus, understanding the origin of these bubbles is important. After
comparison with catalogues of known radio \HII\ regions and catalogues
of star clusters, CH06 conclude that ``{\it about 25\% of the bubbles
  coincide with known radio \HII\ regions, and about 13\% enclose
  known star clusters. It appears that B4-B9 stars probably produce
  about three-quarters of the bubbles in our sample.}''  This
conclusion, which we disagree with, was based mainly on a comparison
with the Paladini et al.~(\cite{pal03}) catalog of diffuse and compact
\HII\ regions, which have a typical angular extent of a few
arcmins. This catalogue is based on previously published lists of
\HII\ regions obtained from single-dish medium resolution
observations, typically with beamwidths of a few arcmin. Thus, many
ultracompact (UC) or compact \HII\ regions, which could possibly be
associated with {\it Spitzer} bubbles, are absent from the Paladini et
al.~(\cite{pal03}) catalogue.  To improve this situation, we
have used the Multi-Array Galactic Plane Imaging Survey (MAGPIS), a
radio-continuum survey conducted with the NRAO Very Large Array (VLA)
at 20-cm.  The angular resolution of MAGPIS (~$5\arcsec$), allows us
to detect compact and ultracompact (UC) \HII\ regions with small
angular sizes.  We have also utilized data from the {\it
  Spitzer}-MIPSGAL survey at 24~$\mu$m (Carey et al.~\cite{car09}).
These data show the extended emission of hot dust and also possibly
the emission from associated young stellar objects (YSOs). This study
is presented in Section~4, after a general introduction about the
formation and evolution of an \HII\ region in Section~2, and the
description of the surveys in Section~3. We have studied one hundred
and two bubbles, allowing us to derive statistical conclusions.

In Section~5, we take advantage of the very simple morphology of the
bubbles to search for dense neutral shells and condensations
surrounding the \HII\ regions. These condensations are potential sites
of star formation.  This work is based on the APEX\footnote{This publication 
is based on data acquired with the Atacama Pathfinder EXperiment (APEX). 
APEX is a collaboration between the Max-Planck-Institut f\"ur 
Radioastronomie, the European Southern Observatory, and the Onsala 
Space Observatory} Telescope Large
Area Survey of the Galactic plane at 870~$\mu$m (ATLASGAL; Schuller et
al.~\cite{sch09}). Because it is sensitive to the emission from cold
dust, the ATLASGAL survey detects the dense cores inside which stars
form.  The distribution of cold dust (and thus the distribution
of dense molecular material) also reflects the interaction between the
bubbles and their surroundings.

In Section~6, once again, we take advantage of the simple morphology of
these regions to discuss the influence of the environment on the shape
of the bubbles, and the distribution of various types of dust
grains. We present several possible cases of massive-star formation
triggered by the expansion of \HII\ regions. Detailed studies of the
most interesting regions in terms of triggered star formation will be
given in a forthcoming paper.

\section{Morphology and evolution of an \HII\ region}

\begin{figure}[tb]

 \includegraphics[angle=0,width=90mm]{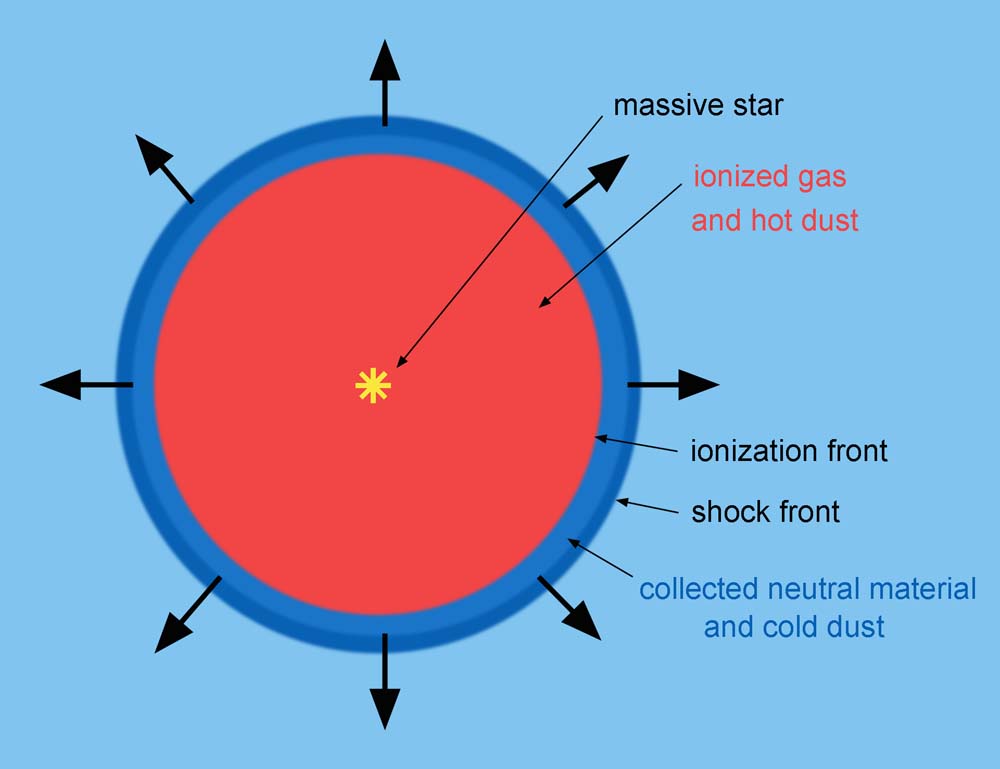}
  \caption{Model of a spherical \HII\ region expanding into a 
  homogeneous medium. The ionized region is surrounded by a shell 
  of dense neutral material collected during the expansion phase.}
  \label{model1}
\end{figure}

We consider a simple model of \HII\ region evolution in which an
\HII\ region forms and evolves in a uniform medium of density 
$n_0$ atoms~cm$^{-3}$ (Fig.~\ref{model1}). The ionizing flux of the
central exciting star is $N_*$~photons~s$^{-1}$. According to Dyson \&
Williams (\cite{dys97}), two main phases can be distinguished during the
life of the \HII\ region:

$\bullet$ First, there is a short formation phase during which 
the central star
rapidly ionizes the neutral medium. Because of the speed at which the
neutral medium is ionized, the neutral and ionized gases are at rest
with respect to each other and at the same density ($n_e=n_0$). For
$n_0=10^4$~cm$^{-3}$ and $N_*=10^{49}$~photons~s$^{-1}$ (the ionizing
photon flux of an O6V star), this phase lasts about 50~years. At the
end of this phase, for these values of $n_e$ and $N_*$, the radius of
the Str{\"o}mgren sphere is $R_0 \sim 0.15$~pc.

$\bullet$ The formation phase is followed by a long expansion
phase. Because of the higher pressure in the warm ionized gas
($T_e\sim10^4$~K) than in the cold neutral
surrounding medium ($T$ in the range 10--100~K), the \HII\ region
expands. This expansion is supersonic; the ionization front (IF) is
preceded by a shock front (SF) on the neutral side. With time, the
size and mass of the \HII\ region increase, whereas the electron
density and the expansion velocity decrease. The expansion ceases when
the ionized gas reaches pressure equilibrium with the surrounding
neutral gas. Except in a high density medium, this equilibrium is not
reached before the death of the central exciting star. For example,
for expansion in a uniform medium of density $n_0=100$~cm$^{-3}$, and
for $N_*=10^{49}$~photons~s$^{-1}$, equilibrium is reached after
$80 \times 10^6$~yr when the radius of the \HII\ region is $R\sim 34
R_0$, and the electron density $n_e \sim 0.5$~cm$^{-3}$. The exciting
star has left the main-sequence at this time, as the lifetime of an O6
star is of the order of $8 \times 10^6$~yr (Schaller et
al.~\cite{sch92}).

During the expansion of the \HII\ region, neutral material accumulates
between the IF and the SF. A layer of dense neutral material builds up
around the ionized region. With time, this layer becomes very massive,
perhaps several thousand solar masses (Hosokawa \& Inutsuka
\cite{hos06b}). A new generation of stars may form in the collected
layer (see Sect.~5). The dynamical expansion of ionization and
dissociation fronts around a massive star has been simulated by
Hosokawa \& Inutsuka~(\cite{hos05}, \cite{hos06a}, \cite{hos06b}),
taking into account the transfer of UV and FUV radiation, the thermal
equilibrium of the gas, and the dissociation of the H$_2$ and CO
molecules. These simulations show that the density in the shell is 10
to 100 times higher than the ambient gas density and that the highest
densities are found in the outer parts of the shell.  They show that
the shell becomes gravitationally unstable, first in its outer parts,
and that the gas is mainly molecular at the time of gravitational
collapse. \\

This model is very simple and neglects several well-known facets of
\HII\ region evolution:

$\bullet$ Inhomogeneus neutral material. It is difficult to believe
that an \HII\ region of a several parsec diameter can evolve in a
perfectly uniform medium. The surrounding medium is most likely turbulent
(Falgarone \& Phillips, \cite{fal96}), and thus highly
inhomogeneous. \HII\ regions forming and evolving in a turbulent
medium were simulated by Mellema et al.~(\cite{mel06}), who illustrated 
the highly irregular shape of the \HII\ region's IF. Some other
simulations of a shock propagating into a turbulent medium (Elmegreen
et al. \cite{elm95}) show that the clumps formed by turbulence are
squeezed and collected into the compressed layer, and merge into a few
massive cores. These cores protrude slightly into the ionized gas,
forming bright rims.  Massive stars form in dense cores that 
contain density gradients. Icke (\cite{ick79}), for example, considers
the initial formation of \HII\ regions (their first phase) in a medium
presenting an exponential density distribution. The \HII\ regions are
egg-shaped (elongated in the directions of lower density); they may
even be open in these directions if the ionizing photon flux is high 
or if the ambient density is low. The formation and expansion of
\HII\ regions in clouds with a power-law density stratification,
$r^{-w}$, and in isothermal self-gravitating disks was described
by Franco et al.~(\cite{fra90}). For power-law profiles, there is a
critical value of $w$ above which the cloud is entirely
ionized. During the expansion phase, $w_{crit}=3/2$. For clouds with
$w<3/2$, the \HII\ region expands and neutral material is collected in
a shell around the ionized region. For clouds with $w\geq3/2$, a
``champagne phase'' occurs, as described in Tenorio-Tagle
(\cite{ten79}): the dense ionized part of the cloud expands
supersonically into the low-density surrounding ionized medium. After
the champagne phase, there is no additional accumulation of neutral
material around the \HII\ region. Bisbas et al.~(\cite{bis09})
simulated another situation of the expansion of an off-center
\HII\ region formed in a homogeneous cloud.  Here again, the
\HII\ region opens on the outside when the IF reaches the edge of the
cloud.  All these departures from evolution in a homogeneous medium
act to destroy the spherical symmetry of the \HII\ region, and 
reduce the efficiency of collecting material during its expansion
phase.

$\bullet$ Stellar winds emitted by the exciting stars. Capriotti \&
Kozminski~(\cite{cap01}) and Freyer et al.~(\cite{fre03},
\cite{fre06}) developed analytical models and hydrodynamical
simulations of \HII\ regions, including stellar winds. At the
beginning of its evolution, the young \HII\ region should not be significantly 
affected by the wind of its exciting star(s). With time, the wind
strengthens (especially at the end of the star's lifetime, in the
Wolf-Rayet stage). A stellar-wind bubble filled of very hot
(T$\sim$10$^6$~K), low density, shocked gas forms inside the \HII\
region. The X-ray emission of this gas has been observed for M17 and
the Rosette Nebula (Townsley et al.~\cite{tow03}), and the Orion Nebula
(G\"{u}del et al.~\cite{gud08}). It is often claimed that the many
bubbles observed in the Galactic plane are the result of the action of
powerful winds (CH06), but this remains to be established. The
presence of a central hole in the distribution of the ionized gas may
be the indirect signature of the action of stellar winds, as discussed
by Watson et al.~(\cite{wat08}, \cite{wat09}).

$\bullet$ A magnetic field. Krumholz, Stone \& Gardiner~(\cite{kru07})
present a simulation of an \HII\ region expanding into a magnetised
gas.  This simulation shows that magnetic fields suppress the sweeping
up of gas perpendicular to magnetic field lines.  This leads to a
non-spherical \HII\ region, elongated in the direction of the magnetic 
field, bounded by a dense shell of swept-up gas in
the direction along the field, but not perpendicular to it. The
authors conclude that this may reduce the efficiency of star formation
triggered by \HII\ regions.

$\bullet$ The radiation-pressure of the exciting stars. Krumholz \&
Matzner~(\cite{kru09}) evaluated the role of radiation pressure
in the dynamics of \HII\ regions. They demonstrated that it is
unimportant for \HII\ regions driven by a handful of
massive stars (which is the case for the \HII\ regions discussed 
in this paper).

We describe the morphology of the 102 bubbles discussed here in Sect.~6.1.

\section{Surveys used in this study}

\subsection{The {\it Spitzer}-GLIMPSE and -MIPSGAL surveys}

The bubbles catalogued by CH06 have been detected at 8.0~$\mu$m in the
{\it Spitzer}-GLIMPSE survey of the Galactic plane (Benjamin et
al.~\cite{ben03}). This survey extends from $l = 295\degr$ to
$65\degr$, $|b| \le 1\degr$, with additional latitude coverage near
the Galactic center.  At 8.0~$\mu$m, the angular resolution of the
{\it Spitzer} IRAC instrument is $1\farcs9$ (Fazio et
al.~\cite{faz04}). We also use images at 24~$\mu$m from the {\it
  Spitzer}-MIPSGAL survey (Carey et al.~\cite{car09}), which has the
same Galactic coverage. The resolution of the {\it Spitzer} MIPS
instrument is $6\arcsec$ at 24~$\mu$m (Rieke et al.~\cite{rie04}).

The {\it Spitzer}-IRAC filter centered on 8.0~$\mu$m contains the
7.7~$\mu$m and 8.6~$\mu$m emission bands generally attributed to
polycyclic aromatic hydrocarbon (PAH) molecules or clusters of
molecules. Infrared emission from PAHs is observed in the direction of
photo-dissociation regions (PDRs), where they are excited by the
absorption of far-UV photons leaking from \HII\ regions. Thus, PAH
emission is a good tracer of ionization fronts. The PAH emission bands
are superimposed on continuum emission generally attributed to the
PAHs and/or to very small grains (VSGs) that are out of thermal
equilibrium after absorption of high energy photons (L\'eger \& Puget,
\cite{leg84}; Sellgren, \cite{sel84}; Desert, Boulanger \& Puget,
\cite{des90}; Draine \& Li, \cite{dra07}; Tielens, \cite{tie08} and
references therein).  The distribution of PAHs is discussed in
Sect.~6.2.

The Spitzer-MIPS filter centered on 24~$\mu$m is dominated by
continuum emission, which may be due to very small grains (VSGs) being out of
thermal equilibrium, or to big grains (BGs) being in thermal equilibrium.
(The maximum of the Planck function B$_{\nu}$ for dust grains in
equilibrium at 120~K is at $\lambda = 24~\mu$m). This point is 
discussed in Sect.6.3.

\subsection{The MAGPIS survey}

The Multi-Array Galactic Plane Imaging Survey (MAGPIS; Helfand et
al.~\cite{hel06}) is a 20-cm VLA continuum survey of the Galactic
plane.  This survey spans $l = 5\degr$ to $48\fdg$ and $|b| <
0.8\degr$.  It has a resolution of $\sim5\arcsec$, a detection
threshold in the range 1 -- 2 mJy, and a high dynamic range
$\sim$1000:1. The VLA images were combined with images from the
Effelsberg 100-m telescope to account for the flux from large-scale
structures. The high resolution of the MAGPIS survey allows the
detection of the free-free emission from compact and ultracompact
\HII\ regions that are small in angular size. It is especially useful
when looking for (second-generation) UC \HII\ regions at the borders
of large (first-generation) \HII\ regions (Sect.~6.5).

We used the MAGPIS
website\footnote{http://third.ucllnl.org/gps/}, and
especially the image ``cutout'' facility to extract, for each bubble,
MAGPIS 20-cm, {\it Spitzer}-GLIMPSE 8.0~$\mu$m, 
and {\it Spitzer}-MIPSGAL 24~$\mu$m images. These images have 
user-specified center and size. They are regridded to have the 
same number of pixels (with a pixel size of 2$\arcsec$), which 
makes the comparison of the same object at different wavelengths 
easier. The sizes of the extracted fields are multiples of
5$\arcmin$ (up to 30$\arcmin$), depending on the size of the
bubbles. The sizes of these fields, and thus of the colour images
presented in Sect.~4, are given in Table~A.1.

\subsection{The VGPS survey}

For MAGPIS fields that are too noisy or not sensitive enough, we 
used the continuum data from the VLA Galactic Plane Survey at 21-cm
(VGPS, Stil et al.~\cite{sti06}). This survey extends from $l =
18\degr$ to $67\degr$, with $|b|$ varying from $1.3\degr$ to
$2.6\degr$ over the longitude range. The \HI\ line and continuum
emission observations at 21-cm have a resolution (FWHM) of
$1\arcmin$. In our analysis, we have oversampled the VGPS maps from
their native $18\arcsec$ pixels to $2\arcsec$ pixels to allow
comparison with the 8.0~$\mu$m and 24~$\mu$m images obtained with the
MAGPIS cut-out facility.

\subsection{The ATLASGAL survey at 870~$\mu$m}

ATLASGAL is the APEX Telescope Large Area Survey of the Galaxy at
870~$\mu$m (Schuller et al.~\cite{sch09}). This survey covers the
inner Galactic plane, $l = 300\degr$ to
$60\degr$, $|b| \le 1.5\degr$, with an rms noise in the range
0.05-0.07~Jy/beam.  The calibration uncertainty in the final maps is
of the order of 15\%.

LABOCA, the Large Apex BOlometer CAmera used for these observations,
is a 295-pixel bolometer array developed by the Max-Planck-Institut
f\"ur Radioastronomie (Siringo et al.~\cite{sir09}). The beam size at
870~$\mu$m is 19\farcs2. The observations are reduced using the
Bolometer array data Analysis package (BoA; Schuller et al.\ in
preparation). During the reduction, and as a result of the correction
for correlated noise, uniformly extended emission (on scales larger
than $2.5\arcmin$) that mimics variations in the sky emission
(skynoise) is filtered out. This can be a problem for our analysis, as
we search for extended structures.

The continuum emission at 870~$\mu$m is dominated by the
thermal emission from cold dust.  Low temperature dust is contained in
dense material: dense molecular cores or filaments. These structures
are the places where stars form.

\section{The nature of the bubbles observed by {\it Spitzer}}

We study one hundred and two bubbles from $10\degr$ to $48\degr$, with
$|b| \le -0\fdg8$.  This longitude range represents the overlap
between the CH06 catalogue and the MAGPIS survey. A few bubbles from
the list of CH06 are missing in our sample. Bubbles N5 and N19 are too
high in latitude to be mapped by MAGPIS. Bubbles N17, N30, and N38 are
entirely surrounded by the larger N16, N29, and N39 bubbles. Bubbles
N63 and N88, are supernova remnants (SNRs) and we do not include them
in the present study (see CH06 and references therein). We added
11 new bubbles found in the observed fields close to the CH06 bubbles
(called Nxxbis, if in the field of Nxx; otherwise, named by their
Galactic longitude). Three of the bubbles of our sample are included in 
the study of Watson et al. (\cite{wat08}); they are N10, N21, and N49.

\begin{figure*}[tb]
 \includegraphics[angle=0,width=180mm]{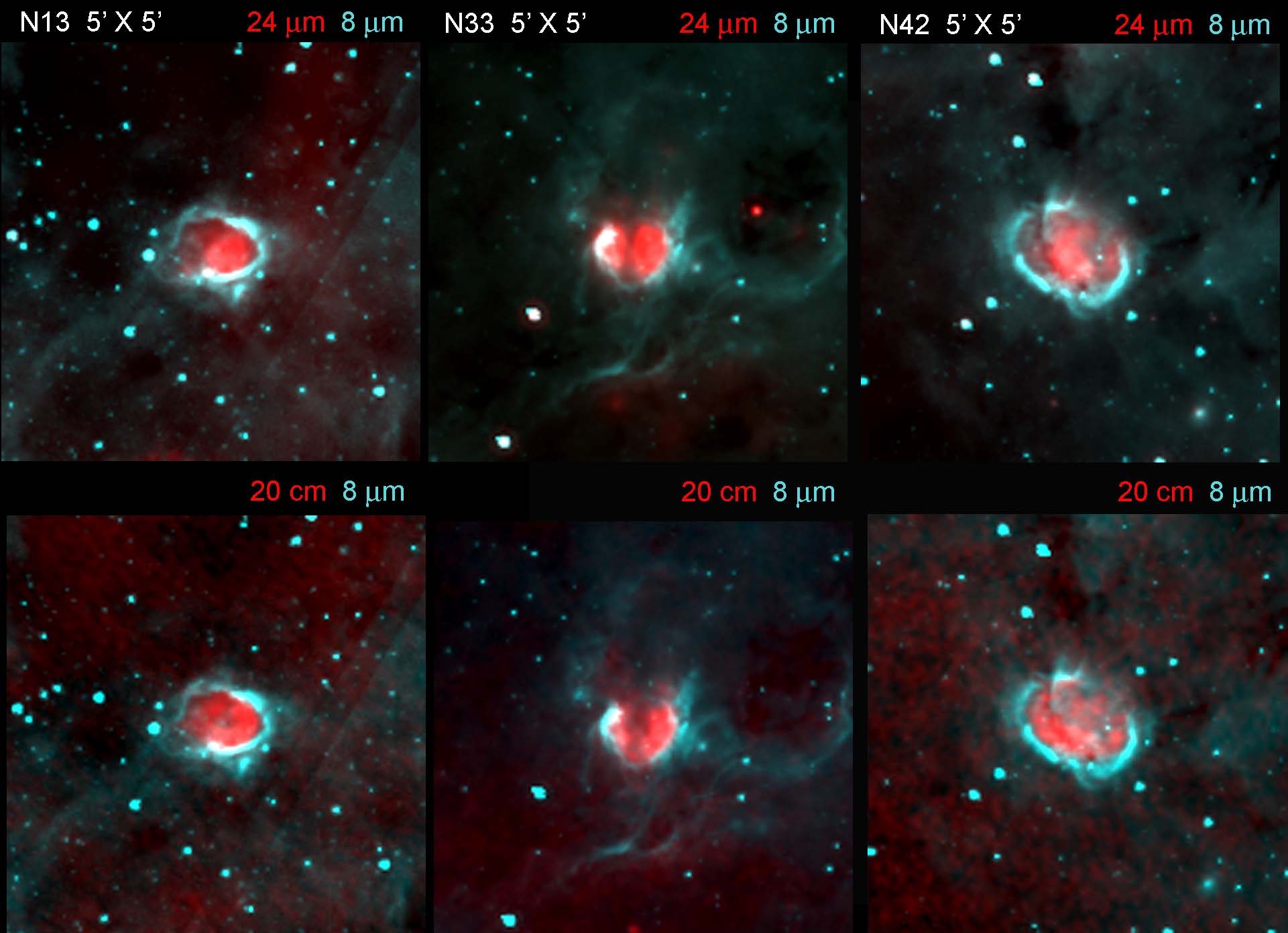}
  \caption{Examples of three bubbles that are small in angular size. All the fields are 
  5\arcmin $\times$ 5\arcmin. Turquoise is the 
  {\it Spitzer}-GLIMPSE images at 8.0~$\mu$m, which defines the bubbles. 
  Red is the {\it Spitzer}-MIPSGAL emission at 
  24~$\mu$m from hot dust (top row) or MAGPIS 20-cm radio
  continuum emission from the ionized gas (bottom row). Directions 
  in which both the 8~$\mu$m and 24~$\mu$m emission (or both the 
  8~$\mu$m and radio emission) are present appear in white, 
  independent of the intensity of the emission. Equivalent 
  images for the entire sample are available online. }
  \label{bulles1}
\end{figure*}

\begin{figure*}[tb]
 \includegraphics[angle=0,width=180mm]{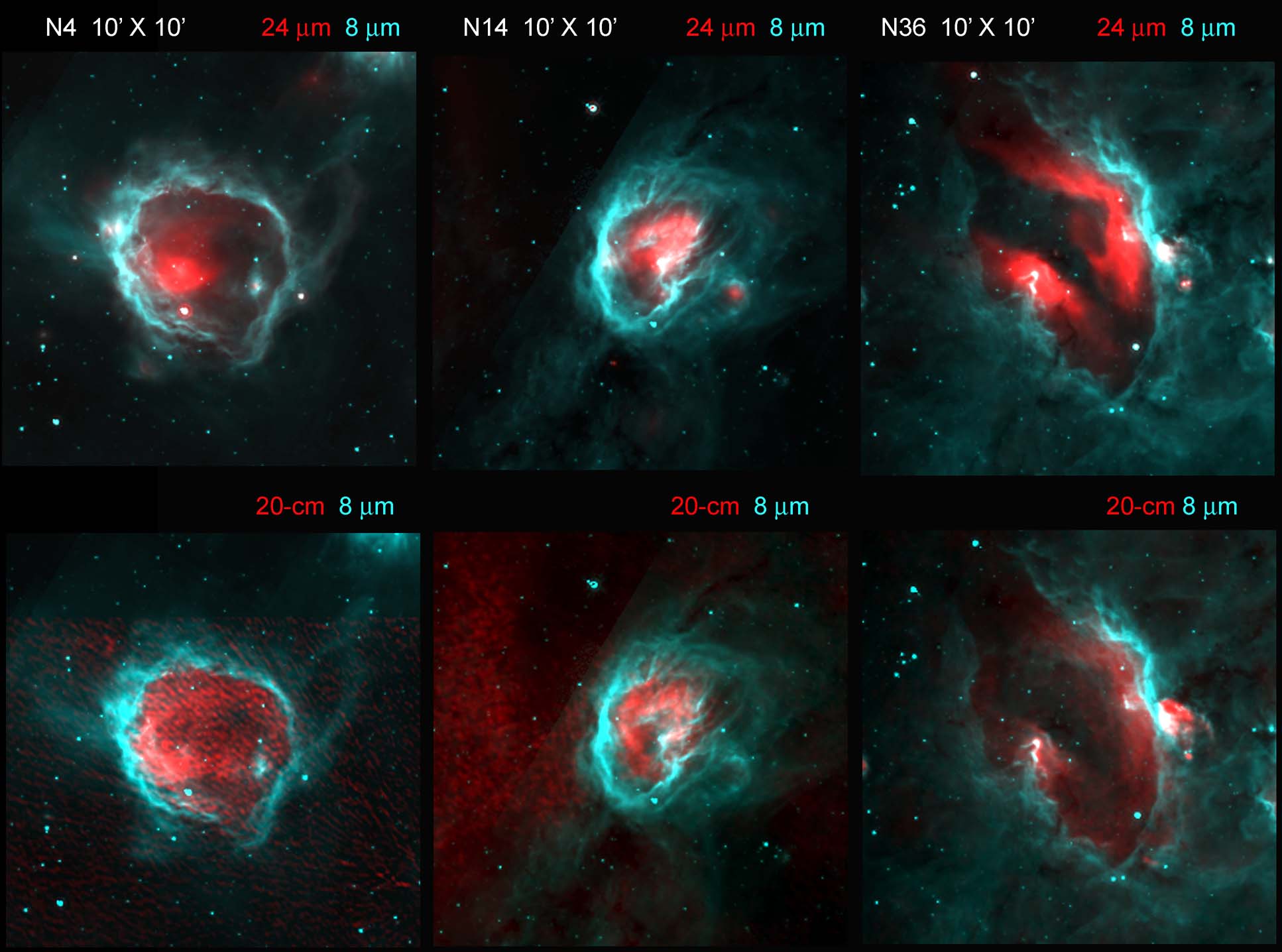}
  \caption{Examples of three bubbles that are large in angular
  size. All the fields are $10\arcmin \times 10\arcmin$. The colours
  are the same as in Fig.~\ref{bulles1}.}
  \label{bulles2}
\end{figure*}

For each of these bubbles, we compiled two-color composite images
with PHOTOSHOP to compare the emission of the ionized gas to
that of the dust at 8.0~$\mu$m and 24~$\mu$m.  We show twelve of these
images in Fig.~\ref{bulles1} (N13, N33, and N42) and
Fig.~\ref{bulles2} (N4, N14, and N36). In these 
figures\footnote{These images are RGB color images; the Red channel 
contains the 24~$\mu$m or radio-continuum image; the Green and 
Blue channels contain the same 8$\mu$m image, thus the turquoise color 
for the 8~$\mu$m emitting regions. The RGB color model is an additive 
color model; thus a region with both 24~$\mu$m and 8~$\mu$m emission 
(or both radio and 8~$\mu$m emission) will appear in white, independently 
of the intensity of these emissions.}, red shows the
hot dust emission at 24~$\mu$m (top rows) or the ionized gas free-free
emission at 20-cm (bottom rows). Turquoise shows the 8.0~$\mu$m
emission tracing the bubbles.

Images for all the bubbles can be found 
online\footnote{http://lamwws.oamp.fr/bubbles}. The name of these
images begins with the name of the bubble, followed by ``\_\,8+24'' or
``\_\,8+MAGPIS'' or ``\_\,8+VGPS'', indicating which frames have been combined
(8 and 24 are for 8.0~$\mu$m and 24~$\mu$m). For example, N1\_\,8+24.eps
and N1\_\,8+MAGPIS.eps are the names of the two color images relating to
the N1 bubble.  When several sources are present in the same field
they have been identified; otherwise the bubble is always in the
center of the field.\\

Based on the MAGPIS or VGPS data, we determined whether the
bubbles are surrounding \HII\ regions. Our conclusions are presented
in Table~A.1 (Appendix A), available online. Column 1 is their identification
name.  Their Galactic coordinates from CH06 are given in Cols.~2 and
3. Column 4 gives the angular size of the fields analyzed (these are
also the size of the colour images).  Columns 5 and 6 comment on 
the presence or absence of the radio-continuum and the 24~$\mu$m
extended emissions. Column 7 provides our conclusions about the nature of
the bubbles (whether they are surrounding \HII\ regions).\\

Of the 102 bubbles studied here, 88 (86\%) enclose \HII\ regions, as
they contain ionized gas detected via their radio-continuum emission. 
This strongly differs from the conclusions of CH06 who stated that
only one-quarter of the bubbles enclose \HII\ regions (the other
three-quarter being associated with non-ionizing B4-B9 stars).

Furthermore, the percentage of bubbles enclosing \HII\ regions is
probably still underestimated here, and may increase to 93\%.  The
VGPS survey shows radio emission in the direction of a number of small
bubbles not detected by MAGPIS: N48, N71, N84, N85, and N87.  However,
because radio emission is also observed outside these bubbles, and the
resolution of the VGPS ($1\arcmin$) is comparable to the size of these
bubbles, their nature as \HII\ regions is questionable; we are unsure
wether these bubbles enclose \HII\ regions (thus the ``?'' in Table~A.1,
Col.~7, for their nature). In addition, N64bis is observed in the direction
of a very bright 20-cm radio source (the SNR W44 or G034.6-00.5),
preventing the detection of any faint emission associated with the
bubble. The nature of N6, a complex bubble, is also unclear.  This
bubble is composed of two different structures: an open and faint
bubble in the northeast, and a filamentary ionization front in the
southwest. Both structures appear as filaments at 24~$\mu$m, but they
are possibly not linked.\\

Extended 24~$\mu$m emission is observed in the direction of the PDRs
and inside the bubbles. Ninety-eight percent of the bubbles exhibit 
extended 24~$\mu$m emission enclosed within the bubble. The only
exceptions are N7 and N15. The bubble N7 lies on the border of a
bright molecular condensation enclosing the most active parts of the
active and complex star-forming region W33 (Beck, Kelly \&
Lacy~\cite{bec98}).  The bubble N15 lies on the border of M17, an
extremely luminous massive star-formation region. Both N7 and N15 do
not seem to be individual bubbles but appear to be filaments at the
periphery of active \HII\ regions; they are possibly caused by the
expansion of very hot gas.\\

The origin and distribution of the 8.0~$\mu$m and 24~$\mu$m emissions are 
discussed in Secs.~6.2 and 6.3. 

\section{The neutral environment of the bubbles as seen by ATLASGAL}

\begin{figure}[tb]
 \includegraphics[angle=0,width=90mm]{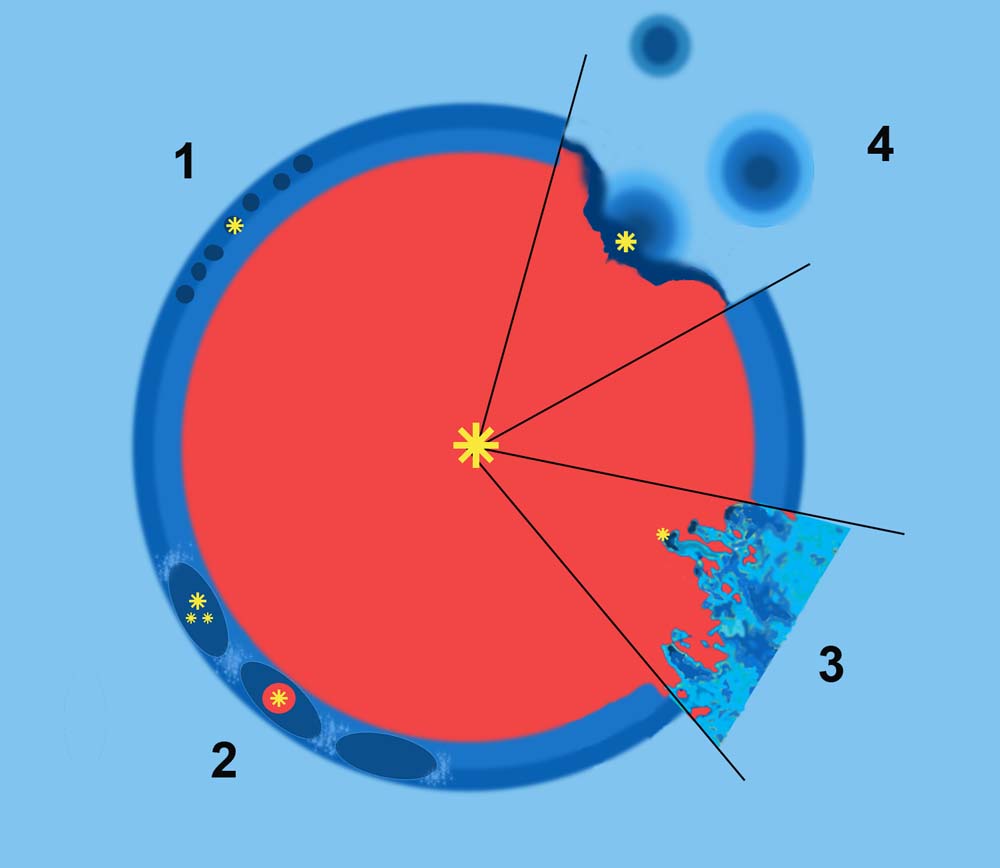}
  \caption{Schematic view of a spherical expanding \HII\ region and 
  of its neutral environment. Different processes of triggered star 
  formation are depicted: 1 - small-scale gravitational instabilities; 
  2 - large-scale gravitational instabilities leading to the formation 
  of high-mass fragments; 3 - ionizing radiation acting on a turbulent 
  medium; 4 - radiation-driven compression of pre-existing dense clumps.}
  \label{model2}
\end{figure}

We now  study the distribution of the dense
neutral material traced using the ATLASGAL survey at 870~$\mu$m that
is associated with the bubbles and their enclosed \HII\ regions. We 
take advantage of their simple morphology to detect the
material collected during their expansion phase. \\

During the expansion of an \HII\ region, neutral material accumulates
between the ionization and shock fronts, forming a shell of dense
material (10 to 100 times the density of the neutral surrounding material) 
surrounding the ionized gas. With time the mass of this shell
increases, reaching thousands of solar masses (see Sect.~2).  The shell
is dense and is mainly molecular. Thus, it should contain cold dust,
which radiates in the (sub-)millimeter range. It is the thermal emission 
from this cold dust that is observed with the APEX-LABOCA camera at
870~$\mu$m.

This shell is an important component of the photodissociation region
(PDR). It may become gravitationally unstable, leading to star
formation. In theory, stars of all masses can form in this shell
around \HII\ regions (cf Fig.~\ref{model2}). Small-scale gravitational 
instabilities (for example Jeans instabilities) can lead to the 
formation of low-mass stars. Large-scale gravitational instabilities 
along the collected shell 
can lead to the formation of massive fragments, potential sites of
massive-star formation; this is the ``collect \& collapse'' process
(Elmegreen \& Lada \cite{elm77}). The ionizing radiation can also act 
on a turbulent medium, forming peculiar structures such as fingers or
pillars (Gritschneder et al.~\cite{gri09}), and pre-existing
condensations can be compressed by the pressure of the ionized gas.
(This mechanism is often called ``radiation-driven implosion'' of a
condensation, but ``radiation-driven compression'' would be better as
no observations indicate that star formation occurs during the
implosion). Good illustrations of star formation triggered by the
expansion of \HII\ regions are given by the RCW~79 (Zavagno et
al.~\cite{zav06}) and RCW~120 (Deharveng et al.~\cite{deh09}) bubbles.

In the following, we use the ATLASGAL survey to search for the
following signatures:

$\bullet$ A shell of cold dust emission
material, surrounding the ionized \HII\ region and adjacent to the
8.0~$\mu$m bubble.

$\bullet$ Massive fragments, observed elongated along the IF. These 
fragments are expected if the shell is already fragmented by means of the
collect \& collapse process; they should have masses of several tens
or hundreds of solar masses. These fragments should not distort the IF
because they are formed from collected material that moves at the 
same velocity as the ionization and shock fronts.

$\bullet$ Massive condensations protruding inside the ionized region. 
This is the signature of pre-existing massive condensations; they  
distort the shape of the ionization front because their velocity
differs from that of the IF. They are dense, thus   
bordered by bright rims at 8.0~$\mu$m and by bright radio emission on the 
ionized side.

$\bullet$ Thin structures, such as fingers or pillars, protuding inside
the ionized gas.  These are signatures of a turbulent medium. These
structures are expected to be of low mass, and should therefore be
more difficult to detect for distant bubbles. These dense pillars 
should also be bordered by bright rims at 8.0~$\mu$m.

Young stellar objects (YSOs) should be present in the PDRs surrounding
the \HII\ regions. They can be identified based on their infrared (IR)
colours (see the discussion in Robitaille et al.~\cite{rob08}). Since
the stars share the velocity of the material inside which they form,
the second-generation stars, formed in the collected layer by
gravitational instabilities, should be observed later on in the
direction of the collected material (or slightly in front, on the
neutral side, if they have formed early in the evolution of the
\HII\ region, when its expansion velocity was large). The stars formed
in compressed pre-existing globules should be seen in the direction of
the \HII\ region.  In this paper, we try to identify only the
most massive second-generation stars, which are those able to ionize
second-generation \HII\ regions.

\subsection{The ATLASGAL images}

We first used the interactive software sky atlas
Aladin\footnote{http://aladin.u-strasbg.fr/} to superimpose the
ATLASGAL images (and isocontours) on the {\it Spitzer}-GLIMPSE
8.0~$\mu$m and 24~$\mu$m images.  In a second step, we used the Kang
IDL software\footnote{http://www.bu.edu/iar/kang/.}  to superimpose
ATLASGAL isocontours on GLIMPSE 8.0~$\mu$m, MIPSGAL 24~$\mu$m, and
MAGPIS 20-cm data. For the GLIMPSE and MIPSGAL data, we used a
logarithmic scaling to create the background images in the
figures. For the MAGPIS data, however, a logarithmic scaling
accentuated the noise in the images and we instead used a linear
scaling. We calculated the standard deviation over each fields
($\sigma$ value) using the IDL routine
``mmm''\footnote{http://idlastro.gsfc.nasa.gov/ftp/pro/idlphot/mmm.pro}
found in the IDL Astronomy User's Library.  The lowest contour level
in our figures is the 1-$\sigma$ level and the contour levels increase
in multiples of the 1-$\sigma$ level.

All of these images are available 
online\footnote{http://lamwws.oamp.fr/bubbles}.  The name of each image 
begins with the name of the bubble, followed by ``\_\,870+8'', or ``\_\,870+24'',
or ``\_\,870+MAGPIS'', or ``\_\,870+VGPS'' to indicate the frame on which the
870~$\mu$m contours have been superimposed. For example, N1\_\,870+8.eps,
N1\_\,870+24.eps, and N1\_\,870+MAGPIS.eps are the images relating to the
N1 bubble. The 870~$\mu$m 1-$\sigma$ level is given for each field.

\subsection{The distribution of dense neutral material}

\begin{figure*}[tb]
 \includegraphics[angle=0,width=180mm]{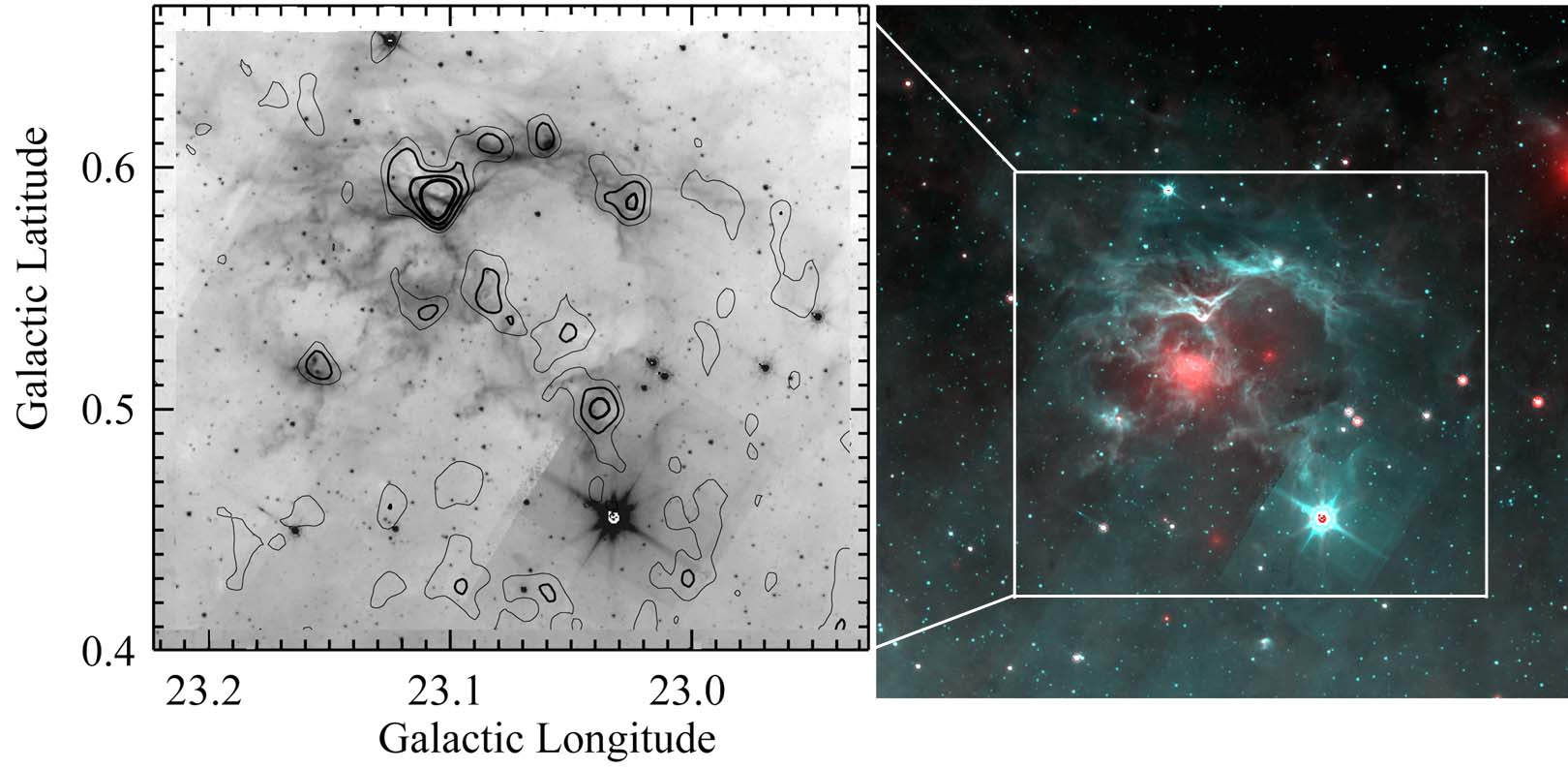}
  \caption{N29: {\it Left:} Contours of the 870~$\mu$m emission
    superimposed on the 8.0~$\mu$m bubble ($\sigma$=0.05~Jy/beam;
    contour levels 1, 2, 3, 5$\sigma$); {\it Right:} {\it
      Spitzer}-GLIMPSE 8.0~$\mu$m emission in turquoise and {\it
      Spitzer}-MIPSGAL 24~$\mu$m emission in red. Several faint
    870~$\mu$m condensations are present in the direction of the
    PDR. The brightest one (at 23.104+00.584) is bordered by a bright
    rim at 8.0~$\mu$m. Thus, it is associated with the bubble and its
    central \HII\ region. This condensation probably pre-existed
    as it distorts the IF; it is presently compressed by the ionized
    gas. Some other faint condensations, near $b \sim 0.6\degr$, 
    may have formed from collected material.}
  \label{N29}
\end{figure*}

We must be very cautious about associating the cold dust
condensations with the bubbles. A cold dust condensation observed in
the direction of a bubble may not be associated with it, but may be, 
by chance, situated along the same line of sight. This problem can be mediated
with velocity information. The velocity of the ionized gas is measured
using hydrogen recombination lines. The velocities of the central
\HII\ regions, when known, are given in Table~A.2 (Appendix~A).  The
velocities of some bright dust condensations have been measured using
the NH$_3$ (1,1) inversion line, at Effelsberg 
(M. Wienen, Wyrowski, Menten et al. in preparation). About 80\%  
of these bright dust condensations adjacent to bubbles have 
a NH$_3$ velocity that differs by less than 
5 km~s$^{-1}$ from that of the enclosed \HII\ region.   
In addition to the velocity information, we
regard the association of a dust condensation with a bubble as highly
probable if we have signs of interaction between the ionized gas and
the condensation (for example, the presence of a bright rim, observed
at 8.0~$\mu$m, bordering the condensation and part of the 
ionization front).  For example, the
brightest condensation in N29 (Fig.~\ref{N29}) is clearly associated
with the bubble because it is bordered by a bright rim at 8.0~$\mu$m.
Finally, if a (partial) shell of dust or if numerous condensations
surround the bubble we consider the association probable. \\

\begin{figure*}[tb]
 \includegraphics[angle=0,width=180mm]{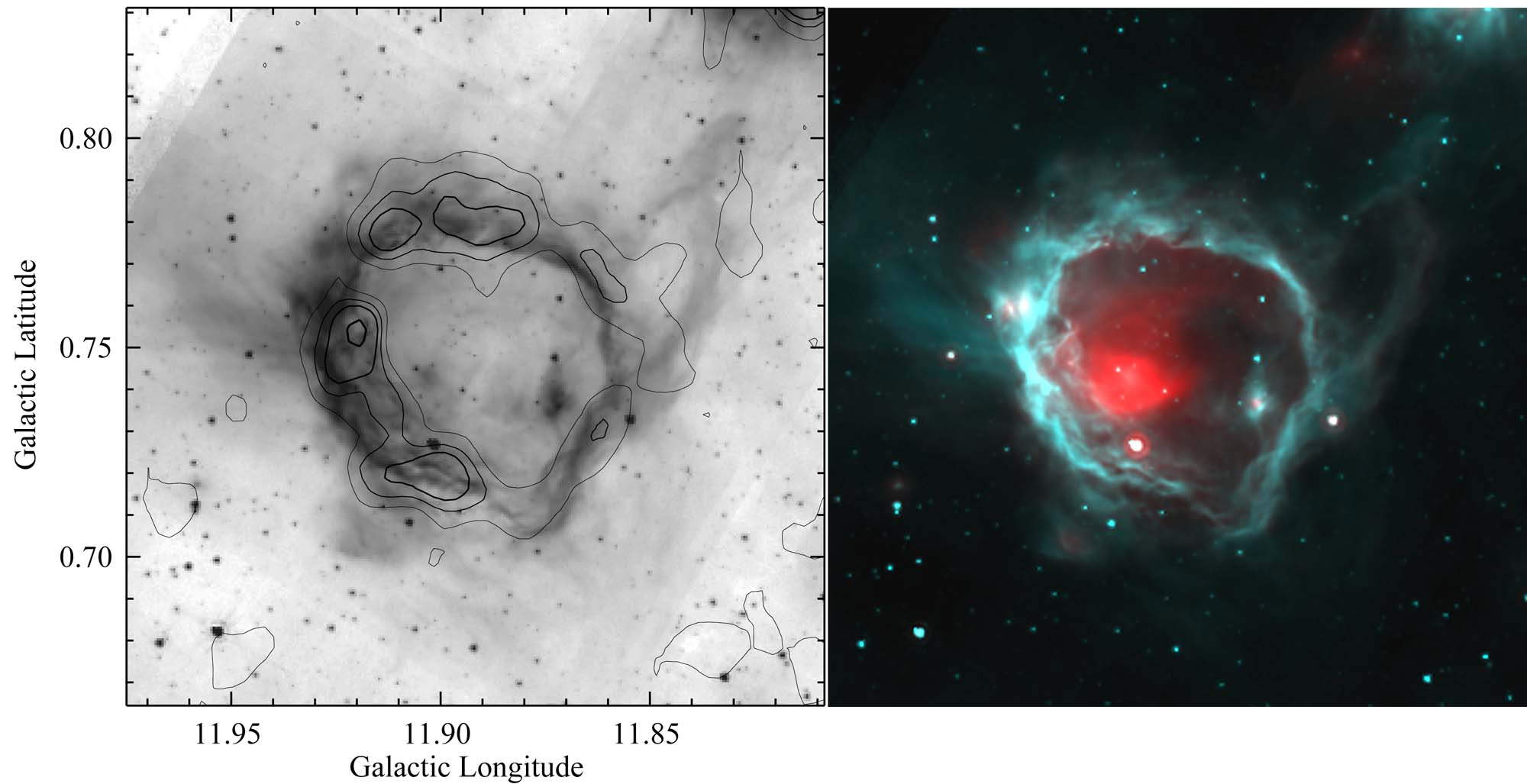}
  \caption{N4: This bubble is surrounded by a shell of collected material. 
  Its mass is $\sim1150$~M\sun\ (for D=3.14~kpc). 
  {\it Left:} Contours of ATLASGAL 870~$\mu$m emission superimposed on the 
  {\it Spitzer}-GLIMPSE 8.0~$\mu$m bubble ($\sigma$=0.04~Jy/beam; contour 
  levels 1, 3, 5, 10, 15~$\sigma$); {\it Right:} 
  {\it Spitzer}-GLIMPSE 8.0~$\mu$m emission in turquoise and 
  {\it Spitzer}-MIPSGAL 24~$\mu$m emission in red.}
  \label{N4}
\end{figure*}

We present in Tables~\ref{atlasgal1}, \ref{atlasgal2}, and
\ref{atlasgal3} our conclusions derived from the comparison between 
the cold dust emission, as seen by ATLASGAL, and the morphology and 
location of the bubbles enclosing \HII\ regions. 
We consider several cases:

$\bullet$ The presence of a partial shell of collected material, or 
of numerous condensations surrounding the ionized region. 
The corresponding regions are listed in Table~\ref{atlasgal1}. 
Twenty-six regions (Table~\ref{atlasgal1}) show a shell of cold dust
surrounding the bubble, or numerous condensations observed in the
direction of the PDR surrounding the ionized region. These regions are
good candidates for the collect process (accumulation of dense
material around the ionized region during its expansion). N4
(Fig.~\ref{N4}), N14 (Fig.~\ref{N14}),and N49 (Sect.~6.5 and Fig.~\ref{N49}) 
are good examples of such regions.

$\bullet$ The presence of a few dust condensations in the
direction of the PDR surrounding the \HII\ region, some interacting  
with the ionized gas (as evidenced by bright rims seen at 8~$\mu$m). 
The corresponding 18 regions are listed in Table~\ref{atlasgal2}. 
The condensations presently interacting with the ionized gas may be
pre-existing condensations reached by the ionization front and
compressed by the ionized gas. In this case, they should protude 
inside the ionized region. The bright condensation in N29 
(Fig.~\ref{N29}) was probably already present in the neutral gas
before being reached by the ionization front. Some of the 18 
bubbles present in Table~\ref{atlasgal2} may also show condensations 
formed from collected material; and both types of condensation may 
be present around the same region. For example, the faint condensations 
present around N29, near $b \sim 0.6\degr$, probably formed  from 
collected material. For most of these regions and condensations, 
it is difficult to determine which of the two processes is at
work: collect and collapse or radiation-driven compression of a
pre-existing condensation. Both processes can induce star
formation.

$\bullet$ Presence of a cold dust condensation, often bright, 
in the direction of the \HII\ region or adjacent to it. The 
association is most often uncertain: velocities are missing and/or we 
lack the angular resolution to see the signatures of collected material 
or of interactions with the ionized gas. The corresponding regions are 
listed in Table~\ref{atlasgal3}.

\begin{figure*}[tb]
 \includegraphics[angle=0,width=180mm]{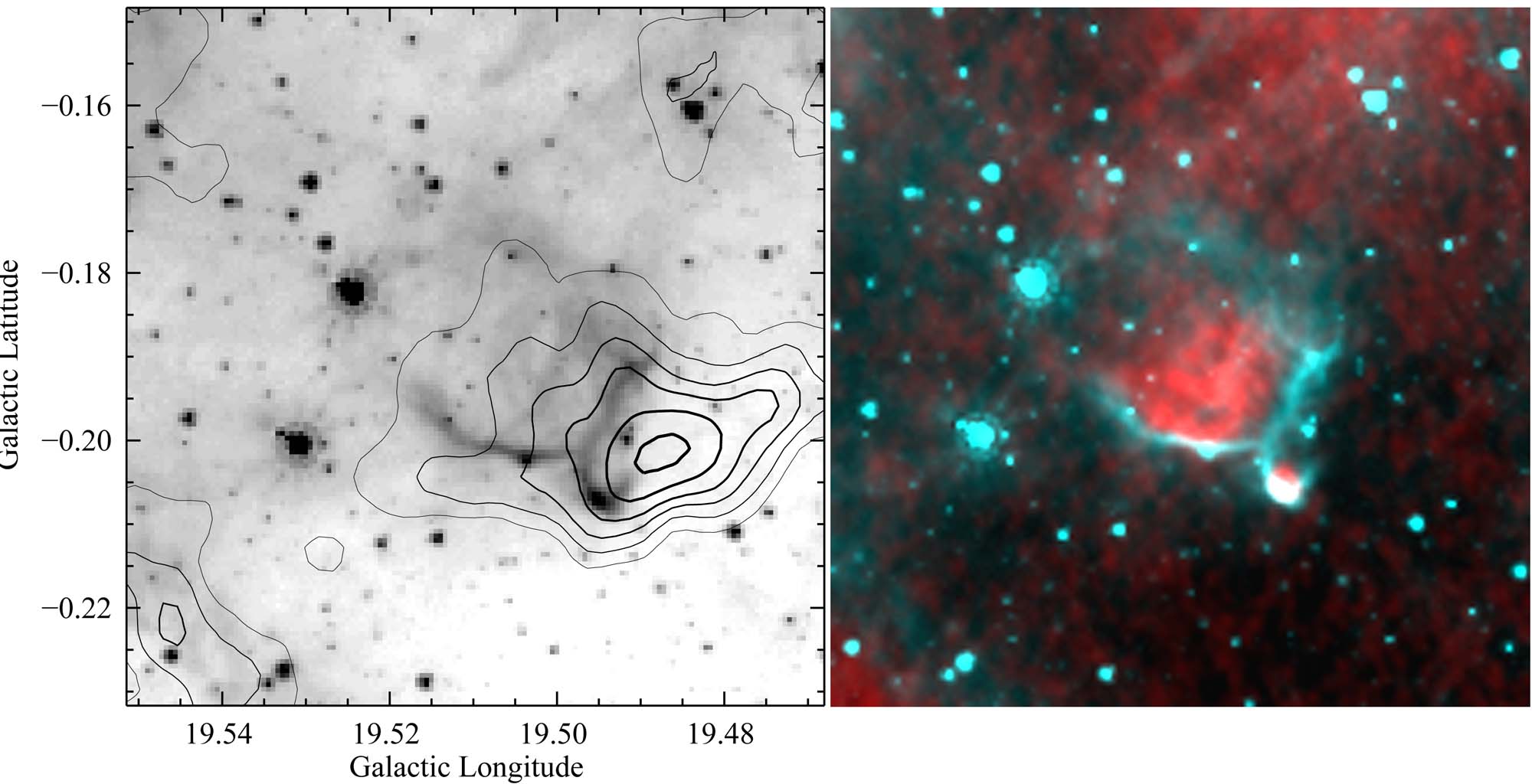}
  \caption{N25. {\it Left:} Contours of the 870~$\mu$m emission
    superimposed to the 8.0~$\mu$m bubble ($\sigma$=0.06~Jy/beam;
    contour levels 1, 3, 5, 7, 10, 14~$\sigma$); {\it Right:} {\it
      Spitzer}-GLIMPSE 8.0~$\mu$m emission in turquoise and MAGPIS
    20-cm emission in red. N25 is adjacent to a bright condensation.
    The bubble is bright on the side of the dense material, and opens
    on the low-density side. But we lack angular resolution to get 
    details about the distribution of neutral material wih respect 
    to the bubble.}
  \label{N25}
\end{figure*}

Forty-seven bubbles are listed in Table~\ref{atlasgal3}; most of them
are of small apparent size (and thus are either young or distant).  In
the absence of velocity measurements, the association between the
bubbles and the dust condensations is uncertain; this is the case for
N8 and N25 (Fig.~\ref{N25}).  For angularly small bubbles, we also 
lack the angular resolution required to assess whether the bubbles and
the condensations are interacting (this applies to 70\% of these 48
regions).  These dust condensations may be massive, and they often
contain several massive young objects (UC \HII\ regions, 
6.7~GHz methanol masers). The bright condensations 
adjacent to N25
(Fig.~\ref{N25}), N32 (Sect.~6.1, Fig.~\ref{N32}), or N67bis
(Sect.~6.6, Fig.~\ref{N67bis}) are good illustrations of such a
configuration.

$\bullet$ A few bubbles have very faint, or lack entirely,
associated 870 \micron\ cold dust emission. They are N11, N65bis, and
N67. Furthermore, two large bubbles with very faint dust emission
present a clearly associated compact \HII\ region at their
periphery. They are N81 (with the compact \HII\ region enclosed by N82, 
Fig.~\ref{N81}), and N97 (with N96).

\begin{table*}
\caption{Bubbles as seen by ATLASGAL: Presence of collected material 
(partial shell or numerous condensations surrounding the ionized region)}
\begin{tabular}{ll}
\hline\hline
Identification & Comments \\
\hline
N2 & collected material along the open bubble (faint emission) + several 
bright condensations at one end, containing  several UC \\
   & and compact \HII\ regions and two 6.7~GHz methanol masers \\
N4 & shell of collected material \\
N12 & several adjacent condensations, some in the direction of ``fingers'' or of IRDCs \\
N14 & shell of collected material \\
G15.68 & square bubble; shell of collected material; most of the condensations are IRDCs \\
N18 & several faint condensations following the PDR, two UC \HII\ regions  
     (uncertain association) \\
N21 & collected material (faint emission); one bright adjacent interacting condensation \\
N22 & collected material (faint emission); two bright adjacent interacting 
    condensations \\
N24 & collected material (faint emission); several bright condensations,  
     one adjacent to the N24bis1 \HII\ region, \\
     & another one surrounding the N24bis2 \HII\ region \\
N24bis2 & shell of collected material; a 6.7~GHz methanol maser \\
N29 & several faint condensations along the PDR; one bright interacting condensation \\
N35 & several faint condensations along the PDR; one interacting condensation; 
     two nearby compact \HII\ regions\\
N36 & collected material + a very bright condensation containing an UC and 
     a compact \HII\ regions;  \\    
     & OH, H$_2$O and methanol masers associated with the UC \HII\ region \\
N45 & several bright condensations along the PDR; one interacting condensation \\ 
N49 & half a ring of collected material with massive fragments; one UC 
     \HII\ region on the border of the brightest fragment;  \\
    & two 6.7~GHz methanol masers \\
N50 & a partial shell of collected material \\
N52 & bright dust emission along the 8~$\mu$m filaments surrounding the bipolar 
     nebula; very bright condensations at its "waist", \\ 
    & containing an UC and a compact \HII\ regions; four 6.7~GHz methanol masers\\
G31.16 & faint emission following the PDR; one condensation in the direction of N53\\
N59 & numerous dust condensations along the PDR; three UC \HII\ regions in 
    the direction of the brightest condensation; \\
    & is the N58 bubble associated? is the bipolar nebula at 33.201$-$00.011 
     associated? \\
N65 & collected material (faint emission) on the border of the two bubbles; 
     one  bright condensation \\
     & with a hyper or UC \HII\ region and a 6.7~GHz methanol maser \\
N68 & similar to N36; several faint condensations adjacent to the PDR +  
     a bright condensation containing an UC and a compact \HII\ region\\
N69 & collected material (faint emission) along the bright part of the PDR \\
N77 & several condensations adjacent to the PDR; the brightest one  
     in interaction with the ionized gas \\
N80 & several condensations around the bubble \\
N90 & several condensations around the bubble \\
N94 & several faint condensations along the PDR; some are bordered by bright rims \\
\hline
\label{atlasgal1}
\end{tabular}\\
\end{table*}

\begin{table*}
\caption{Bubbles as seen by ATLASGAL: Presence of at least two dust condensations 
adjacent to the PDR, with one in interaction with the ionized region }
\begin{tabular}{ll}
\hline\hline
Identification & Comments \\
\hline
N1  & two adjacent interacting condensations between N1 and the nearby 
     W31 \HII\ region \\
N3 & two adjacent interacting condensations (faint emission) \\
N10 & two adjacent interacting condensations \\
N27 & open bubble; one condensation elongated along the bright part of the PDR \\
N39 & bipolar nebula; three interacting condensations at its ``waist''; another bright 
     condensation on the border, \\
     & in the direction of a compact \HII\ region (probably not associated, as shown by its velocity) \\
N47 & several faint interacting condensations \\
N54 & faint condensations adjacent to the PDR; the brightest one interacting with the ionized region \\
N60 & two adjacent interacting condensations \\
N61 & faint condensations adjacent to the PDR (collected material?); a bright adjacent condensation \\
     & containing one UC and two compact \HII\ regions (no triggering) \\  
N62 & faint emission in the direction of the PDR + a bright adjacent filament (IRDC) \\
N73 & two adjacent interacting condensations \\
N74 & several faint adjacent condensations; one interacting condensation \\
N76 & several faint interacting condensations \\
N82 & condensations elongated along the bright part of the PDR \\
N91 & several condensations, one elongated along the bright PDR, another very 
      bright (also an IRDC) \\
    & containing an UC \HII\ region and a 6.7~GHz methanol maser \\
N93 & two condensations adjacent to the PDR (one in interaction) \\
N95 & several condensations along the bright PDR \\
N98 & a bright interacting dust filament extended along the PDR \\
\hline
\label{atlasgal2}
\end{tabular}\\
\end{table*}

\begin{table*}
\caption{Bubbles as seen by ATLASGAL: Presence of at least one dust 
condensation in the direction of the bubble or adjacent to it. The 
association condensation -- bubble may be uncertain.}
\begin{tabular}{ll}
\hline\hline
Identification & Comments \\
\hline
N8 \& N8bis & one elongated bright condensation, a 24~$\mu$m source in the 
             center, N8 and N8bis on each side \\
N9  & one bright adjacent condensation \\
N13 & at the tip of a dust filament \\
N16 & one interacting adjacent condensation; also an IRDC \\
N20 & one bright adjacent condensation \\
N23 & between two condensations, along a bright fragmented filament \\
N24bis1 & one bright adjacent condensation, with an UC \HII\ region in its center\\
N25 \& N25bis & one adjacent bright condensation \\
N26 & two faint adjacent condensations \\
G19.82 & a bright adjacent condensation \\
N28 & an adjacent condensation (also a filamentary IRDC) \\
N31 & two faint adjacent dust condensations \\
N32 & one bright adjacent condensation containing an UC \HII\ region \\ 
N33 & an adjacent condensation \\
N34 & one filament crossing the bubble \\
N37 & one adjacent interacting condensation \\
N40 & one adjacent filament (also an IRDC), with several fragments \\
N41 & one adjacent dust condensation \\
N42 & one dust condensation in its direction \\
N43 & one bright adjacent condensation \\
N44 & two condensations, one in its direction and one adjacent \\
N46 & two faint adjacent condensations \\
N48 & one bright adjacent condensation (also a filamentary IRDC) \\
N51 & faint condensations \\
G30.25 & faint adjacent condensation \\
N53 & two faint adjacent condensations; N53 lies on the border of the large G31.165$-$0.127 bubble \\
N55 & a bright condensation along an adjacent filament (IRDC); three \HII\ regions within the condensation (no triggering) \\ 
N56 & one condensation in its direction \\
N57 & one adjacent condensation \\
N58 & one bright condensation in its direction \\
N64 & several faint nearby condensations (also IRDCs)\\
N64bis & one adjacent condensation \\
N66 & one condensation in its direction\\
N67bis & one bright condensation in its direction, containing two UC \HII\ regions \\
N70 & one adjacent bright condensation, containing an UC \HII\ region in its center  \\
N71 & several faint condensations \\     
N72 & a bright adjacent condensation (also a filamentary IRDC) \\ 
N75 & two adjacent condensations \\
N78 & one bright condensation in its direction \\
N79 & one adjacent condensation \\
N83 & one condensation in its direction \\
N84 & two adjacent condensations \\
N85 & one adjacent condensation containing an UC \HII\ region \\
N86 & one adjacent condensation \\
N87 & one adjacent condensation \\
N89 & a faint adjacent condensation \\ 
N92 & one filamentary condensation (also an IRDC) crossing the nebula \\
N96 & a condensation in its direction; N96 lies on the border of the large N97 bubble \\
\hline
\label{atlasgal3}
\end{tabular}\\
\end{table*}

To conclude, for the 65 \HII\ regions enclosed by bubbles and for
which the angular resolution of the observations is high enough to
study the spatial distribution of cold dust, 26 (40\%) are clearly
surrounded by collected material, and 18 more (28\%) show 
interacting dust condensations and also possibly condensations formed 
from collected material. For 16 regions (24\%), the association 
between the bubble and the condensations is uncertain, and
5 regions (8\%) have no detectable associated dust emission. \\

At least 16 bubbles contain a compact or ultracompact \HII\ region (or
several of them; detected via their radio and 24~$\mu$m central
emission) in the direction of their PDRs. These regions are good
candidates for triggered massive-star formation. They are identified
in Tables~\ref{atlasgal1} and \ref{atlasgal2}, and are discussed in
Sect.~6.5.

\section{Discussion}

We now address the morphology of the bubbles (whether it 
is influenced by their environment), their dust content 
(PAHs and hot dust), and their CO environment. We then discuss 
massive-star formation at their borders, possibly triggered by the 
enclosed \HII\ regions.

We require the distance of the bubbles to estimate parameters such 
as the size or the mass of the associated dust condensations. Their 
distances, when available, are given in Appendix~A, Table A.2. CH06 
``emphasize that the near distances are most likely the correct ones'' 
(the argument is that nearby bubbles are more easily detected because 
distant ones are masked by the foreground emission of the Galactic plane). 
However this is not true: the kinematic distance ambiguity has been 
resolved for 57 bubbles in our sample; 39 of them (68\%) lie at the far 
distance, whereas only 15 (26\%) of them are located at the close-by 
position (3 are at the tangent point).

In the following, we use the 870~$\mu$m cold dust emission to estimate the amount of 
neutral material associated with the bubbles. We 
give in Appendix B the relations used to derive the column 
density $N(\mathrm H_2)$ and the mass of a condensation from the
observed surface brightness $F_{\mathrm{870\,\mu m}}$.

\subsection{The morphology of the bubbles} 

We categorized a few different types of bubbles:

$\bullet$ Nearly complete bubbles such as N1, N4 (Fig.~\ref{bulles2}),
N8, N12, N13 (Fig.~\ref{bulles1}), N23, N26, N28, N42
(Fig.~\ref{bulles1}), N43, N44, N49, G30.250, N53, N57, N58, N66, N70,
N74, N75, N78, N80, N83, N86, N90, and N93. As mentioned in Sect.~2, 
the interstellar medium is not homogeneous on large scales. In an
inhomogeneous medium, there are density gradients and therefore
maintaining circular symmetry over large scales would seem unlikely.
Circular bubbles should therefore be of small size, and thus should be
either young or evolving in a very dense medium that restricts their
expansion.  Many of the complete bubbles are indeed small, such as N13
(diameter $\sim$0.85~pc), N58 (0.95~pc), and N83 (1.7~pc) -- all three
of which are seen in the direction of bright cold dust condensations
and thus are probably associated with dense material. However, N1
(diameter $\sim$ 7.8~pc), N4 (3.5~pc), N43 (2.6~pc), N49 (4.0~pc), N53
(4.5~pc), N70 (6.6~pc), and N93 (3.8~pc, if at the distance of N94) are
medium-sized \HII\ regions. Why are they so circular? The most 
obvious answer is that these regions evolve in a rather homogeneous 
medium (on the scale of a few parsecs), where the turbulence - at the origin  
of most inhomogeneities - is low.

\begin{figure*}[tb]
 \includegraphics[angle=0,width=180mm]{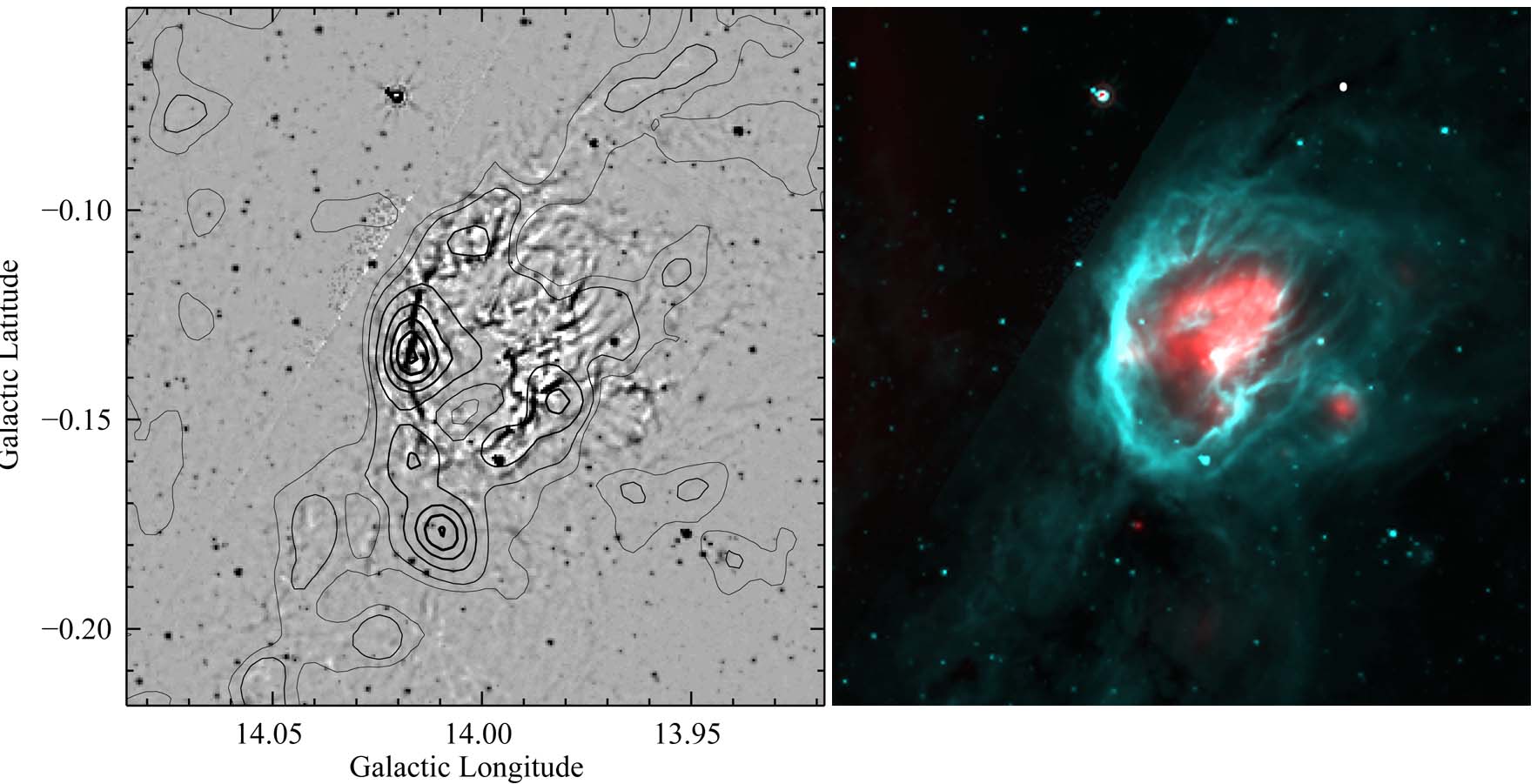}
  \caption{N14, a bubble possibly in the process of opening. The
    bubble is bright where it is adjacent to dense material, and opens
    in the direction of low density. {\it Left:} Contours of the
    870~$\mu$m emission superimposed on an unsharp-masked image at
    8.0~$\mu$m. We see the filaments shaped at the surface of the
    bubble by the ionized gas flowing out; (870~$\mu$m emission:
    $\sigma$=0.04~Jy/beam; the contour levels are 1, 2, 5, 10, 15, 20,
    and 25~$\sigma$) {\it Right:} {\it Spitzer}-GLIMPSE 8.0~$\mu$m
    emission in turquoise and {\it Spitzer}-MIPSGAL 24~$\mu$m emission
    in red.}
  \label{N14}
\end{figure*}

\begin{figure*}[tb]
 \includegraphics[angle=0,width=180mm]{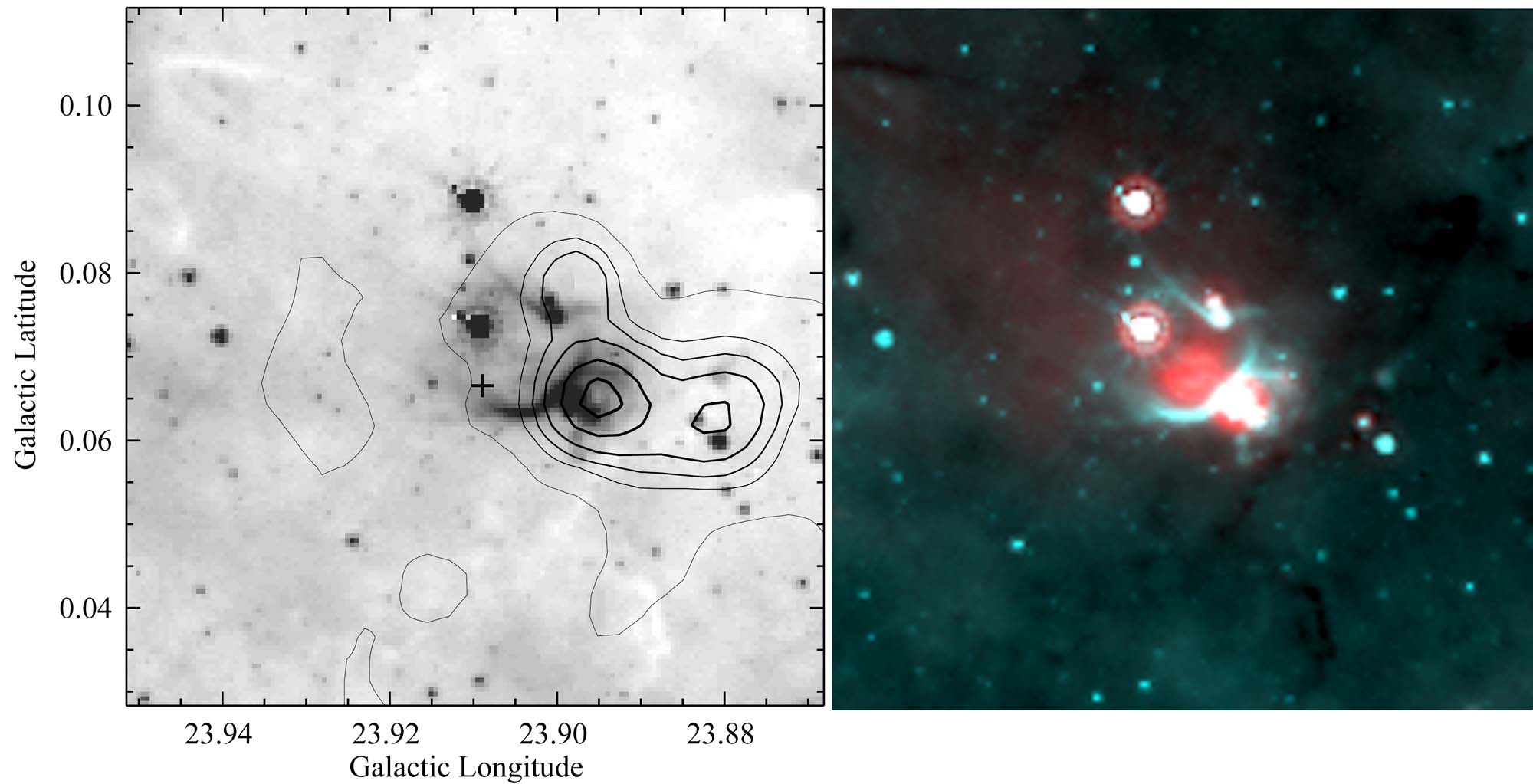}
  \caption{N32. {\it Left:} Contours of the 870~$\mu$m emission
    superimposed on the 8.0~$\mu$m image. The cross gives 
    the position of a methanol maser. {\it Right:} {\it
      Spitzer}-GLIMPSE 8.0~$\mu$m emission in turquoise and 
      {\it Spitzer}-MIPSGAL 24~$\mu$m emission in red. The 
      comparison of the velocity of the
    condensation with that of the \HII\ region suggests an
    association; the condensation at 23.8955+00.0638 has a velocity of
    39.25~km~s$^{-1}$ whereas the \HII\ region has a velocity of 
    32.8~km~s$^{-1}$ (Table~A.2).}
  \label{N32}
\end{figure*}

$\bullet$ Some bubbles are elongated, probably in the direction of
lower density, and are possibly in the process of opening (see, for
example, N10, N14 (Fig.~\ref{bulles2}), N29, N50, and N77). These
bubbles are, in general, well-defined on the side of high density
(where we see adjacent dust condensations). For example, N14
(Fig.~\ref{N14}) is surrounded by a partial shell of dense material,
and opens in the direction free of dense material. The bubbles
N24bis1, N25 (Fig.~\ref{N25}), G19.821, and N32 (Fig.~\ref{N32}) are
adjacent to one bright dust condensation.  They are well-defined on
the side of the condensation, and open on the opposite side,
presumably towards the low-density region.

$\bullet$ Some bubbles are open in one direction, such as N2
(Sect.~6.5, Fig.~\ref{N2}), N16, N21, N24 (wide opening,
Fig.~\ref{N24}), N27, N36 (Fig.~\ref{bulles2}), N37, N91, N92, N94,
and N97 (wide opening).  They are often of large size, such as N2
(38~pc $\times$ 19~pc, at a distance of 6.0~kpc), N16
(33~pc $\times$ 19~pc, at D=13.7~kpc), N24 ($\geq$30~pc, at 
D=4.6~kpc), N36 (16~pc $\times$ 8~pc, at D=6.4~kpc), N37
(16~pc $\times$ 9~pc, at D=12.6~kpc), N91 (31~pc $\times$ 19~pc, at 
D=8.1~kpc), and N97 (20~pc $\times$ 11~pc, at D=9.3~kpc). 
A large size may be indicative of either evolution in a low density medium  
or an old age. An \HII\ region of radius 10~pc, excited by an O6V star, 
may for example be 3.5~Myr old if evolving in a 
medium of $10^3$~cm$^{-3}$, or 1~Myr old if evolving in a medium 
of $10^2$~cm$^{-3}$.

\begin{figure*}[tb]
 \includegraphics[angle=0,width=180mm]{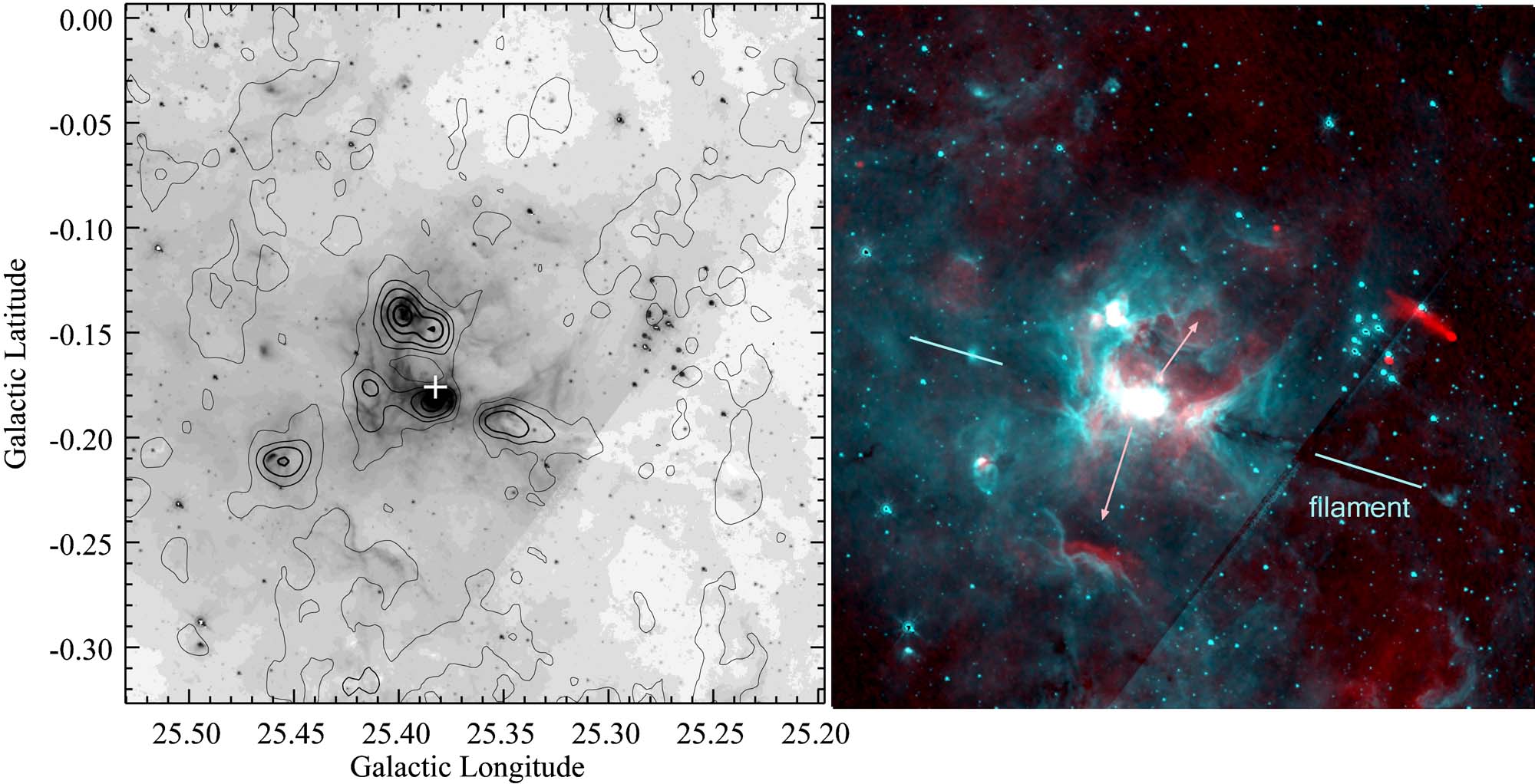}
  \caption{N39, a bipolar nebula: {\it Left:} Contours of the
    870~$\mu$m emission superimposed on the {\it Spitzer}-GLIMPSE 
    8~$\mu$m image. The white cross gives the position of a  
    methanol maser. {\it Right:} 
    {\it Spitzer}-GLIMPSE 8.0~$\mu$m emission in turquoise
    and MAGPIS 20-cm emission in red. The blue line shows the remains
    of the filament (or the sheet seen edge on) inside which the
    exciting star(s) of N39 possibly formed. The red arrows show the 
    probable preferential directions of expansion of the ionized gas.}
  \label{N39}
\end{figure*}

$\bullet$ Two bubbles (N39 and N52) are part of bipolar nebulae. N39
(Fig.~\ref{N39}) has three 870~$\mu$m condensations aligned along a
filament.  This filament is seen farther away from the bubble as an
infrared dark cloud (IRDC; line in Fig.~\ref{N39}). A
possible scenario is the following: the massive star or cluster
exciting the nebula formed inside this filament. Later, the expanding
\HII\ region broke through the edge of the cloud, simultaneously in
two directions, producing a bipolar nebula (arrows). This 
scenario was simulated by Bodenheimer et al. (\cite{bod79}, their
Fig.~4).  A bipolar nebula is also obtained when an \HII\ region forms
and expands near a dense filament, as simulated by Fukuda \& Hanawa
(\cite{fuk00}).  Their simulation shows that a gravitational
instability is induced to form dense cores along the filament, on each
side of the ionized region (their Fig.~1). This is another process by
which an expanding \HII\ region can trigger star formation. In the
case of N39, Fig.~\ref{N39} shows the presence of two condensations at
the waist of the bipolar nebula, along the filament, in the directions
25.410$-$00.176 and 25.351$-$00.191. A similar situation can explain
the N52 morphology (Sect.~6.5, Figs~\ref{N52}). In both nebulae,  
a 6.7~GHz methanol maser is present in a condensation situated at the 
waist of the  nebula, indicating that star formation is at work there. 
Two percent of the bubbles appear to be bipolar. If, as suggested by 
Beaumont \& Williams (\cite{bea10}; see Sect.~6.4), most molecular clouds 
were flat, more bipolar nebulae should be observed. We note however that 
an \HII\ region appears to be bipolar only when the line of sight is 
not too inclined with respect to the plane of the parental 
molecular cloud.\\

\begin{figure*}[tb]
 \includegraphics[angle=0,width=180mm]{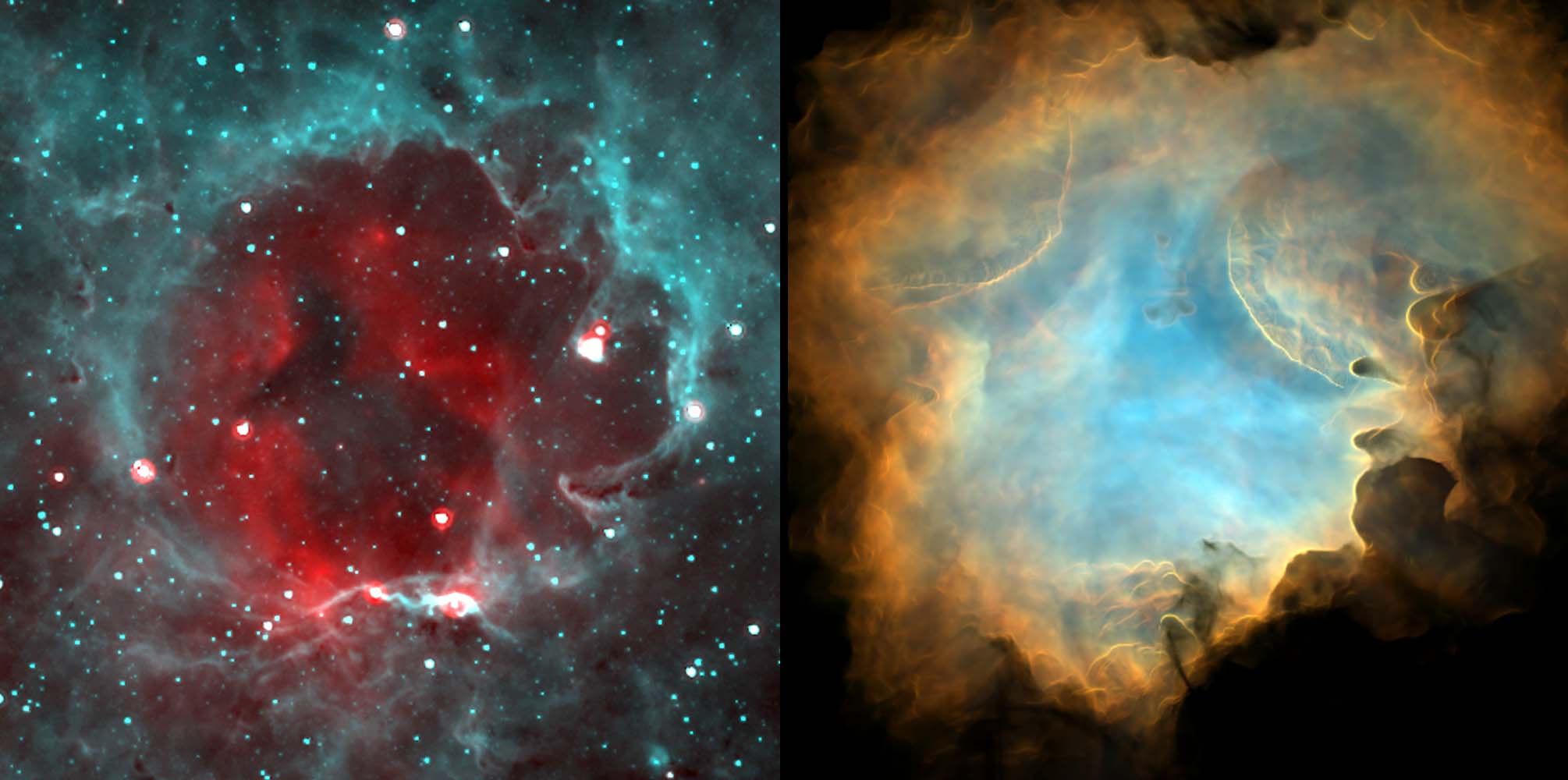}
  \caption{N12: {\it Left:} colour image with the {\it Spitzer}-GLIMPSE 
  image at 8~$\mu$m in turquoise and the {\it Spitzer}-MIPSGAL image at 
  24~$\mu$m in red. This bubble appears, at first glance, to be 
  rather circular; but 
  the ionization front that limits the central cavity shows many 
  structures. This is indicative of density inhomogeneities 
  in the medium  surrounding the bubble. {\it Right:} Synthetic optical 
  image of a simulated \HII\ region evolving in a turbulent molecular 
  cloud (Henney et al. \cite{hen10}). The [NII] 6584~\AA\ emission 
  appears in red, the H$\alpha$ 6563~\AA emission in green, and the 
  [OIII] 5007~\AA emission in blue.}
  \label{N12}
\end{figure*}

The inside boundary of many bubbles (in addition to the ionization front 
of the enclosed \HII\ regions) exhibits structures, such as 8~$\mu$m bright 
rims bordering condensations protruding inside the ionized regions, or 
pillars (also called fingers or elephant trunks). The N12 bubble, 
which is almost circular, displays such a  
bumpy IF (Fig.~\ref{N12}), as do the IFs of N35, N36 (Fig.~\ref{N36}), 
N68, N76, and N91 (Fig.~\ref{N91}). These structures are the signatures  
of small-scale density inhomogeneities in the medium surrounding the 
bubbles. These inhomogeneities are probably due to turbulence. 
The HD simulations of \HII\ regions expanding in a turbulent medium 
by Mellema et al. (\cite{mel06}), Gritschneder et al. (\cite{gri09}), 
and Arthur (\cite{art09}) reproduce rather well the appearance of 
these IFs.\\

One of the bubbles is not circular, but square (G15.68 -00.28).  It
is surrounded by a shell of collected material that follows the four
sides of the square and appears as IRDCs. The MHD simulations of 
Arthur (\cite{art10}) of an \HII\ region, excited by a B star, evolving  
in a magnetised medium show that the magnetic pressure is able to 
straighten the ionization front. Could a similar scenario explain the very 
peculiar shape of G15.68$-$00.28? \\

All these examples show that the birth and evolution of the bubbles,
and their present morphology, are strongly influenced by their
environment. Their environment plays a role at the time of their
birth, whether at the center or at the edge of a dense core, or in a
filament or a sheet; it also plays a role during their expansion phase,
in media with density gradients and/or turbulence.

\subsection{The distribution of PAHs as observed at 8.0~$\mu$m}

Several spectroscopic observations show that the PAH emission bands
are faint or absent in the direction of \HII\ regions, suggesting that
PAHs are destroyed inside the ionized region. Cesarsky et
al.~(\cite{ces96}) present ISOCAM spectra of the M17 \HII\ region and
find that the spectrum obtained in the direction of the adjacent
molecular cloud displays the PAH emission bands, whereas the spectrum
obtained in the direction of the ionized region is clearly dominated
by the continuum emission (the 7.7~$\mu$m and 8.6~$\mu$m PAH features
are almost absent). This observation has been confirmed by {\it
  Spitzer}-IRS spectra of M17 (Povich et
al.~\cite{pov07}). Lebouteiller et al.~(\cite{leb07}) present {\it
  Spitzer}-IRS spectra of the giant \HII\ region NGC3603. They compare
the intensity of the PAH emission bands with that of the underlying
continuum which is attributed to VSGs.  They find that the PAH/VSG ratio is
anticorrelated with the hardness of the radiation field, which implies 
that a destruction mechanism operates on the PAH molecules 
within the ionized gas. The
PAH emission band at 3.3~$\mu$m has been mapped in the Orion bar (a
nearby edge-on PDR) with a resolution of $1\arcsec$ (Giard et
al.~\cite{gia94}). The map shows a sudden drop of the 3.3~$\mu$m
emission feature at the ionization front, toward the \HII\ region.
The spatial extent of the transition zone is not resolved and is thus
thinner than 2.5 $\times 10^{-3}$~pc.

The destruction of PAHs in hard radiation fields, such as those
present in \HII\ regions, has been discussed by several authors (see
Tielens, \cite{tie08}, and references therein). Processes involving
the photodissociation of PAHs, for example by the loss of an
acetylenic group, show that small PAHs of fewer than about 50 carbon
atoms are destroyed by the radiation field in \HII\ regions, but that
larger PAHs may survive (Allain, Leach \& Sedlmayr~\cite{all96a},
\cite{all96b}). This process also involves photons with energies less
than 13.6~eV, and thus the PAH destruction does not occur exactly at
the limit of the ionized region.  PAHs may be doubly ionized in
\HII\ regions, and charge separation reactions (also called Coulomb
explosions or dissociations) do occur; this is another PAH destruction
process (Leach, \cite{lea89}). However, here again, photons with
energy near 13.6~eV or higher are not directly
involved. Chemisputtering by H$^+$ seems a promising PAH destruction
process in \HII\ regions, as proposed by Giard et al.~(\cite{gia94})
or Normand et al.~(\cite{nor95}), because it relies on the large
abundance of free protons found in \HII\ regions. (It was proposed by
Barlow \& Silk (\cite{bar77}) for the destruction of graphite grains,
and Omont (\cite{omo86}) proposed that it was also valid for
PAHs). However, the study by Micelotta et al. (\cite{mic10}) of the
interaction of PAHs with electrons or ions (H$^+$ and He$^+$) in a hot
or warm gas shows that PAH destruction is dominated by He$^+$
collisions at a temperature lower than 30000~K, thus is dominant in
\HII\ regions.\\

Many bubbles in our sample show that the carriers of the IR bands
responsible for the emission in the {\it Spitzer}-IRAC band at
8.0~$\mu$m are destroyed inside the ionized region.  For these
bubbles, the 8.0~$\mu$m emission stops where the 20-cm radio continuum
emission begins. This phenomenon is observed in \HII\ regions excited
by both massive, and also less massive stars. For example, it is
observed in: i) RCW~79, which is excited by a cluster of early O-stars
(Martins et al.~\cite{mar09}), and in N49, which is excited by an O5V
star (Watson et al.~\cite{wat08}); ii) RCW~120, which is excited by an
O8V star (Zavagno et al.~\cite{zav06}) and in N4, which is excited by
an O8V star (an estimation based on a
  radio flux of 2.02~Jy at 11-cm, Reich et al.~\cite{rei84}, and a
  distance of 3.14~kpc); iii) RCW~82, which is excited by two late
O-stars (Martins et al.~\cite{mar09}). It is also observed in bubbles 
enclosing regions of very
faint/or absent radio emission (as N8). Thus, the destruction
of PAHs does not require photons more energetic than necessary to
ionize hydrogen. This differs from the conclusions of
Lebouteiller et al.~(\cite{leb07}), who state that energies higher than
27.6~eV (the ionization potential of Ar$^+$) are required to destroy
PAHs containing more than 50 carbon atoms. Their argument, based on
observations of NGC3603, is that PAH emission is observed in regions
where Ar$^+$ is present. The morphology of NGC3603 is more complicated
than that of the bubbles studied here, and regions with very different
physical conditions can contribute to the emission along one line of
sight, this effect blurring the general picture.

In conclusion, PAH emission is observed in the direction of the PDRs,
outside the ionized region, independent of the hardness of the
radiation field of the exciting stars (at least from early to late O 
stars).  This  
indicates that the PAHs are destroyed in the \HII\ regions. The
destruction mechanism(s) remains uncertain.

\subsection{The extended dust emission at 24~$\mu$m}

Extended 24~$\mu$m emission associated with the bubbles is observed in
two locations.  First, it is seen in the direction of the center of
the bubbles. This emission is often bright. Spectroscopy of
\HII\ regions (Roelfsema et al. \cite{roe98}; Peeters et
al. \cite{pee02}) shows that no strong nebular emission lines are
present in the 24~$\mu$m band; the emission is mostly from the
continuum.  This emission can only be attributed to dust.  Therefore,
hot dust grains must be present in the ionized region, not far from
the exciting stars. Emission at 24~$\mu$m is also observed in the
direction of the dense PDR filaments that are also observed at
8.0~$\mu$m and delineate the bubble.  This emission is rather
faint. We lack the resolution required to determine whether the
24~$\mu$m and 8.0~$\mu$m emission observed in the direction of the
PDRs are cospatial.\\

There are several possible origins of the
24~$\mu$m emission. D\'esert et al.~(\cite{des90}) showed that the
cirrus continuum emission near 25~$\mu$m was due, in almost equal
proportions, to PAHs and VSGs. (VSGs have a size in the range 1~nm
to 10~nm, and are probably carbon dominated). These small particules
(PAHs and VSGs) are out of thermal equilibrium: they reach very high
temperatures (up to several hundreds of K) after absorbing high-energy
photons and radiate in the near- and mid-IR. In the direction of
the PDRs surrounding \HII\ regions, the radiation field hardens. It
has been shown, for example by Compi\`egne et al.~(\cite{com08}) for
the PDR tracing the Horsehead nebula, that the abundance ratio of PAHs
to VSGs varies.  Emission from VSGs dominates the continuum emission
near 24~$\mu$m. But what happens inside \HII\ regions where the
stellar radiation field is very intense and hard?  What kind of grains
can survive in this very harsh environment?

The 24~$\mu$m continuum emission in \HII\ regions may alternatively 
be caused by big dust grains (BGs) in thermal equilibrium (BGs have a
size of a few tens to a few hundreds of nm, and are mainly formed of
coated silicates). Cesarsky et al.~(\cite{ces00}) discussed mid-IR
imagery and spectroscopy obtained with the ISO satellite of a region
in the Orion Nebula. The images show extended emission from amorphous
silicate grains from the entire \HII\ region (ionized mainly by
$\theta$~$^1$ Ori, an O6 star), and from around $\theta$$^2$~OriA (an
O9.5V star). The spectra peak near 25~$\mu$m; Cesarsky et al. have
demonstrated that, at these wavelengths, the emission comes mostly from
amorphous silicate grains of temperatures 85~K to 145~K, with a small
contribution from amorphous carbon grains of temperatures 110~K to
200~K.\\

Thus, the mechanism behind the 24~$\mu$m emission in \HII\ regions 
is not clear. There are, however, several interesting properties of  
the 24~$\mu$m emission:

$\bullet$ The integrated 24~$\mu$m flux from within the ionized
regions represents half or more of the total 24~$\mu$m emission of the
bubbles.  We measured the fraction of 24~$\mu$m emission
coming from the central ionized region for selected well-defined
bubbles.  The ratios (flux from the ionized region)/(flux from the
ionized region + flux from the PDR) are 0.50 for N4, 0.51 for N13,
0.65 for N21, 0.49 for N25, 0.51 for N42, 0.66 for N49, and 0.60 for
N70.

\begin{figure}[tb]
 \includegraphics[angle=0,width=90mm]{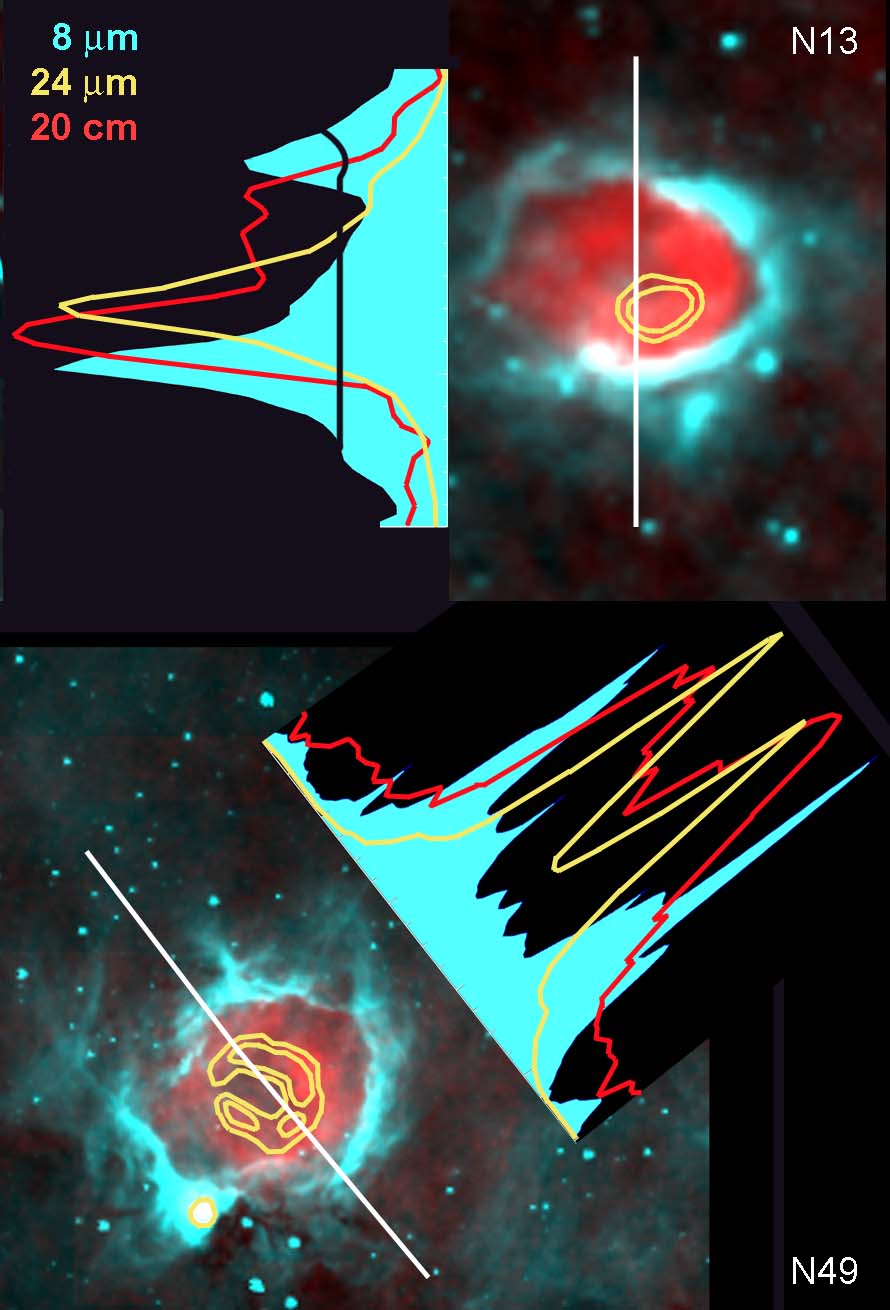}
  \caption{Variation in the 8.0~$\mu$m, 24~$\mu$m, and 20-cm
    brightness along the diameter of N13 and N49. The colour images
    are a composite of 8.0~$\mu$m (turquoise) and 20-cm emission
    (red). Yellow contours show the regions of maximum 24~$\mu$m
    emission. The 24~$\mu$m emission is more central than the 20-cm
    emission.}
  \label{dust}
\end{figure}

$\bullet$ In several bubbles, the observed 24~$\mu$m emission is more
central than the radio continuum emission; the ratio (surface
brightness at 24~$\mu$m)/(radio continuum brightness at 20-cm) peaks
at the very center of the bubbles. This is illustrated in 
Fig.~\ref{dust}, which compares the 24~$\mu$m and 20-cm brightness in
the N13 and N49 bubbles.  The ratio of the flux densities
$S_{24\mu}/S_{20-cm}$ averaged over the whole regions is respectively
30 and 97 for N10 and N49.  This ratio varies by a factor of more than
5 from the outside to the center of the \HII\ regions. The same
situation is observed in N21, N25, N42, and N70. The simplest explanation 
is that this effect is due to the dust temperature, which is higher near 
the radiation source. Inside the ionized
region, dust grains can be heated by Lyman $\alpha$ photons, and/or
directly by the Lyman continuum radiation of the central exciting
stars. That the 24~$\mu$m emission is centrally peaked favours
the explanation that the grains are mostly heated by absorption of
Lyman continuum photons, which are more numerous near the exciting star 
(Lyman $\alpha$ photons, more uniformly distributed inside 
the ionized region, would probably not produce this effect). However, 
using the 24~$\mu$m emission alone, we cannot determine wether it is 
caused by VSGs out of
equilibrium or by BGs in equilibrium. Additional observations at longer
wavelengths are required to answer this question.

\begin{figure*}[tb]
 \includegraphics[angle=0,width=180mm]{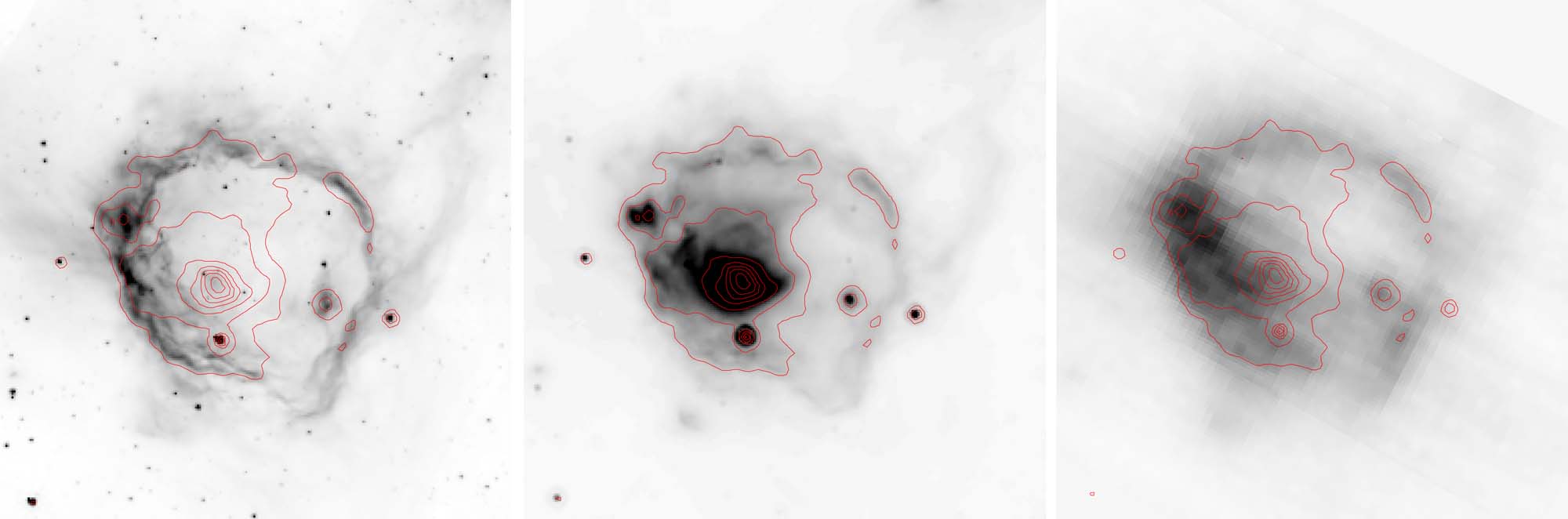}
  \caption{N4: comparison of the distribution of various 
  types of dust grains; {\it Left:} emission of the PAHs 
  at 8.0~$\mu$m from the PDR. {\it Middle:} emission at 
  24~$\mu$m of the hot grains from the ionized region (the red  
  contours correspond to this emission). 
  {\it Right:} emission at 70~$\mu$m from colder grains, distinct 
  from these emitting at 24~$\mu$m, and probably situated in the PDR.}
  \label{dustbis}
\end{figure*}

$\bullet$ Figure~\ref{dustbis} allows us to compare, in the well-defined
N4 bubble, the distribution of dust grains responsible for the {\it
  Spitzer}-MIPSGAL 24~$\mu$m and 70~$\mu$m emission (both bands trace
thermal emission from dust grains). It shows that the grains
responsible for this emission are not co-spatial. The 70~$\mu$m
emission is not central, but resembles the 8.0~$\mu$m emission; it
comes from the PDR rather than from the central ionized region. This
confirms that the dust grains located inside the ionized region are
very hot, as their emission is stronger at 24~$\mu$m than at 
70~$\mu$m.\\

Regardless of its origin (VSGs or BGs), the emission at 24~$\mu$m from
the center of bubbles is important for another reason: it allows us to
locate the central exciting stars (assuming that the 24~$\mu$m emission
comes from hot dust close to the exciting star).  The presence
of a small central hole, sometimes observed in this emission, may be
the signature of stellar winds emitted by the exciting star(s).  These
points are illustrated and discussed in Martins et
al.~(\cite{mar09}). Alternatively, the holes may be produced by the
radiation pressure of the central stars. In either case, the hole
indicates the location of the exciting stars. One example of a bubble
showing a central hole in both the radio-continuum emission and the
24~$\mu$m emission is N49. It has been shown by Watson et
al.~(\cite{wat08}) that the exciting star of N49 could be an O5V star
lying in the direction of the hole.\\
 
Three bubbles of small angular size, N8bis, N9, and N48, display central
24~$\mu$m emission but no central radio-continuum emission. What is
their nature? There are two possibilities.  First, they may be
\HII\ regions with faint radio continuum emission that is 
undetected by MAGPIS
or the VGPS.  This implies that they are either distant, of low
electron density, or excited by B1 or B2 stars. Alternatively, if they
are not \HII\ regions, they must contain a rather massive star because
they contain hot dust.  The massive star must be later than B2, or it
would have created an \HII\ region.

\subsection{The CO environment of the bubbles}

Beaumont \& Williams (\cite{bea10}; hereafter B\&W) mapped 38
bubbles from our sample in the $J= $3--2 line of $^{12}$CO, and 6 in
the $J= $4--3 line of HCO$^+$, with an angular resolution of
$16\arcsec$; these bubbles are identified in Table~A.1. Their study
provides information about the molecular environment, the three-dimensional
morphology and the column density and density of the 
material surrounding the bubbles. According to B\&W, the $^{12}$CO
(3-2) emission most likely originates in regions that are moderately
dense (n(H$_2$)$\sim$$10^3$--$10^4$~cm$^{-3}$) with temperatures in the
range 20~K--50~K. The HCO$^+$ (4-3) emission traces higher density
regions (n(H$_2$)$\geq10^5$~cm$^{-3}$) where star formation takes
place.

One feature showed by this study is the paucity of CO emission towards
the center of the bubbles. If the bubbles were two-dimensional
projections of spherical molecular expanding shells, we would expect
to see the front and back faces of the shells as relatively faint
blueshifted and redshifted emission regions, but this emission is not
observed. B\&W (\cite{bea10}) conclude that most bubbles are not
spherically symmetric, but rather rings formed in flattened molecular
clouds, with a thickness of a few parsecs.  We have shown in Sect.~6.1
that some of the bubbles are open towards the side of low
density. Thus, we agree that the front or back sides (or both) of some
bubbles may be missing, but is this a general feature? If most
molecular clouds had a thickness of a few parsecs, as predicted by
B\&W, numerous bipolar nebulae should be observed, which is not the
case. However, the bubbles in the GLIMPSE sample (CHU06) are quite
eccentric, with an eccentricity peaking at 0.65. This is indicative of 
anisotropy, and in turn of either density gradients, a magnetic field, 
or a flat (two-dimensional) bubble morphology.

Another result of the B\&W (\cite{bea10}) study are the low values
estimated for the column densities of the molecular material
surrounding the bubbles. The measured H$_2$ column densities are in
the range 5--10 $\times$ 10$^{20}$~cm$^{-2}$, averaged over the dense
regions with HCO$^+$ emission. These low column densities led the
authors to claim that the conditions for star formation are not met in
the immediate vicinity of the bubbles. \\

\begin{figure*}[tb]
 \includegraphics[angle=0,width=180mm]{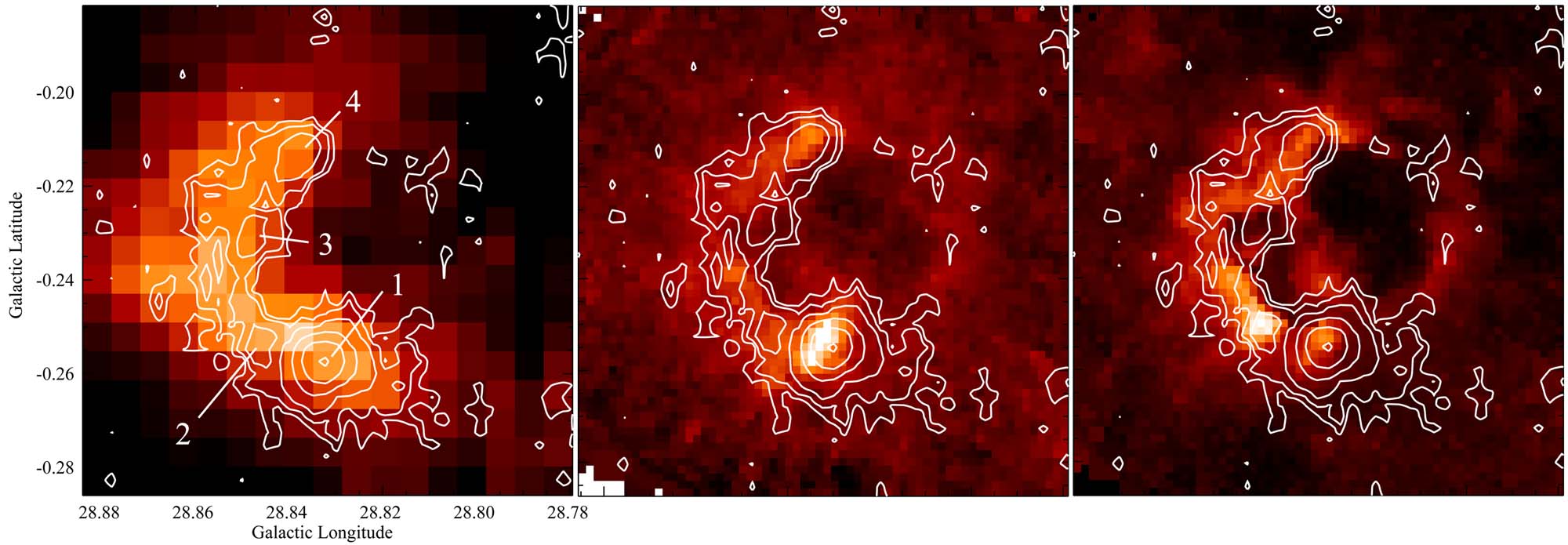}
  \caption{The CO environment of N49: The 870~$\mu$m contour
    levels (0.1, 0.2, 0.5, 1.0, 1.5, and 2.0~Jy/beam) are superimposed on the Galactic Ring Survey $^{13}$CO
    (1-0) emission (Jackson et al. \cite{jac06}) integrated over
    all velocities ({\it Left}), on the B\&W $^{12}$CO (3-2) emission
    integrated over the velocity ({\it Middle}), and on the B\&W
    $^{12}$CO (3-2) peak intensity map ({\it Right}). The dense dust
    condensation \#1 is almost absent from the CO peak intensity map,
    whereas condensation \#2, associated with the UC \HII\ region, is
    bright (presumably a temperature effect). The CO (1-0) and (3-2) emissions 
    integrated over velocity are more similar to the 870~$\mu$m 
    emission. However, the dust condensation \#1 is much brighter 
    at 870~$\mu$m than it is in CO.}
  \label{N49bis}
\end{figure*}

The CO (3-2) maps presented by B\&W in their Fig.~1 show the CO peak
intensity and appear to be very different from the 870~$\mu$m maps. The peak
intensity maps show the regions where the CO kinetic temperature is
high, at the borders of the \HII\ regions, hence clearly define the
bubble boundaries. Maps of CO emission integrated over all velocities
are more comparable to the 870~$\mu$m maps, which show all material
along the line of sight.  B\&W provided us with the data cubes of four
regions, and indeed, the maps of CO emission integrated over the
velocity are similar to the 870~$\mu$m maps. This is illustrated for
the N49 region in Fig.~\ref{N49bis}. However, the dust condensations,
which are dense and cold, are expected to be less apparent on the CO
maps than on the 870~$\mu$m maps for several reasons: i) the CO (3-2)
emission is insensitive to high densities; ii) the condensations are
probably optically thick in $^{12}$CO; and iii) the CO molecule is probably
depleted in the gaseous phase because it condenses onto the dust
grains in cold dense cores (Caselli et al.~\cite{cas99}; Bacmann
et al. \cite{bac02}).

The column densities that we estimate for the dust condensations
associated with the bubbles using the 870~$\mu$m emission, are much
higher than these estimated by B\&W from their observations. If the
column densities were of the order of those estimated by B\&W, almost no
emission would be detected by ATLASGAL.  At
the 870~$\mu$m rms noise of $\sim 0.07$~Jy/beam, the derived column
density of N(H$_2$) $\sim 1.8 \times 10^{21}$~cm$^{-2}$ is indeed 
higher than
the values estimated by B\&W for the HCO$^+$ emitting regions (see
their fig.~7).  The column densities that we measure for dust
condensations associated with massive young objects, for 
example the condensations containing class II methanol masers, 
are in the range $5 \times 10^{22}$ to $1.3 \times 10^{24}$~cm$^{-2}$ 
(see Sect.6.6). These values  
are compatible with the theoretical predictions of Whitworth et al. 
(\cite{whi94}), who show that, for a wide range of input parameters, 
the gravitational fragmentation of a shocked layer occurs when the column 
density of this layer reaches a value $\sim 6 \times 10^{21}$~cm$^{-2}$. 
Furthermore, the mean densities that we
estimate for the dust condensations are lower than these obtained by
B\&W, even for the shell surrounding the bubbles (their fig.~7). These
high densities obtained by B\&W probably result from a strong
underestimate of the CO column densities, possibly due to depletion or
from the assumption of small optical depth used in the LVG
formalism.

For example, for N65
and the associated dust condensation containing an UC \HII\ region and
a methanol maser (Fig.~\ref{N65}), we measure a peak intensity of
5.15~Jy/beam.  This corresponds to a column density of N(H$_2$)$=1.33
\times 10^{23}$~cm$^{-2}$.  The size at half intensity of this condensation 
is 0.3~pc (for D=3.6~kpc, Table~A.2) and its integrated
flux density is 14.5~Jy, which imply  a mass of 1065~\msol\  and 
a mean density of $7 \times 10^5$~cm$^{-3}$. These numbers
are very different from the values estimated by B\&W for this region
(respectively $\sim 10^{21}$~cm$^{-2}$ and $\geq10^7$~cm$^{-3}$ for the
column density and density in the HCO$^+$ region).

\subsection{Massive-star formation triggered by the expanding bubbles}

Sixteen bubbles display compact or UC \HII\ regions (detected in
radio-continuum and 24~$\mu$m emission) at their periphery, often in
the direction of bright adjacent dust condensations. The column
densities, visual extinctions, and masses given in this section are
based on the formulae and assumptions given in Appendix~B, and on the
distances given in Table~A.2. Detailed studies of the most interesting
regions, in terms of triggered star formation, will be given in a
forthcoming paper.

\subsubsection{N2}

\begin{figure*}[tb]
 \includegraphics[angle=0,width=180mm]{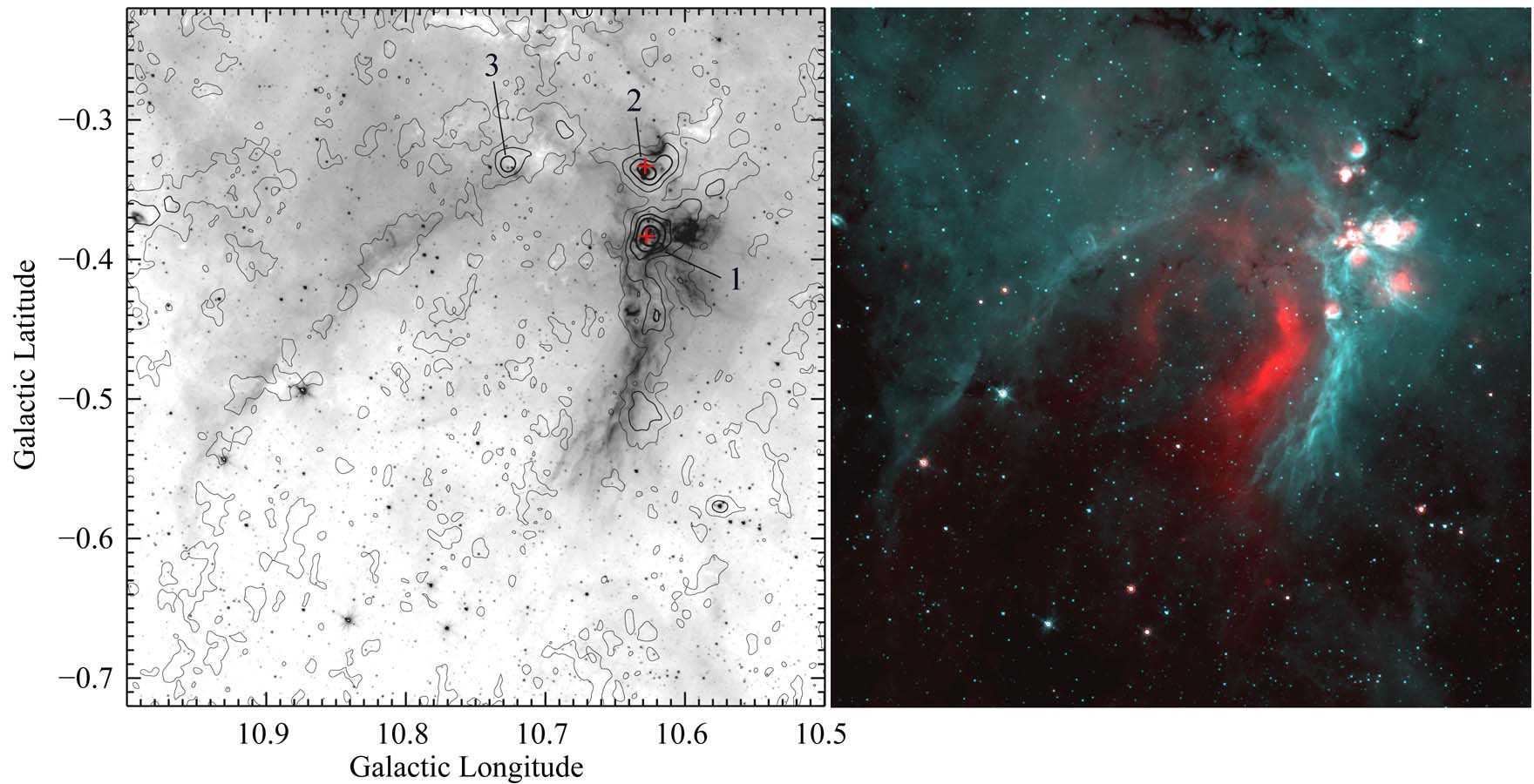}
  \caption{N2: {\it Left:} Contours of the 870~$\mu$m emission
    superimposed on the {\it Spizer}-GLIMPSE 8.0~$\mu$m image. We see
    the faint emission from the cold dust associated with the
    collected neutral material, on the borders of N2; we also see
    bright condensations containing various young stellar objects and
    UC \HII\ regions; (870~$\mu$m emission: $\sigma$=0.06~Jy/beam;
    contour levels 1, 5, 20, 40, 150~$\sigma$). The red crosses 
    give the position of two methanol masers. {\it Right:} {\it
      Spitzer}-GLIMPSE 8.0~$\mu$m emission in turquoise and 
      -MIPSGAL 24~$\mu$m emission in red.}
  \label{N2}
\end{figure*}

This is a large bubble, elongated along the eastwest
direction, that opens on its eastern side (see Fig.~\ref{N2}). It
surrounds an extended \HII\ region, which probably evolved in a medium
presenting a density gradient as all the emission on the western side
(at 8.0~$\mu$m, 24~$\mu$m, and 20-cm) are brighter than on the eastern
side. The distance to this region is uncertain; it has been discussed
by Corbel \& Eikenberry (\cite{cor04}), who, on the basis of CO
emission lines and NH$_{3}$ absorption features from associated
molecular clouds, suggest a near distance of 4.5$\pm$0.6~kpc.  It has
also been discussed by Pandian et al.~(\cite{pan08}), who also favor a
near distance of 3.4~kpc. The N2 bubble has a diameter of 30~pc (E-W)
$\times$ 15~pc (N-S) (if D=4.5~kpc).  The 870~$\mu$m cold dust
emission shows a shell of  
collected material surrounding the ionized region in the west
(Fig.~\ref{N2}). Several bright condensations are present there.  The
velocity of the neutral material has been measured at several
positions (see Table~A.2) and velocities have been found in the range
$-1.5$~km~s$^{-1}$ to $-4.6$~km~s$^{-1}$.  This is similar to the
velocity of the central \HII\ region, $-2.4$~km~s$^{-1}$.  Because of
this velocity agreement, we conclude that the neutral material in the
shell and the bright condensations are associated with the central
\HII\ region.

Several UC and compact \HII\ regions are present in the direction of
the two main condensations, or are adjacent to them. The peak emission
for condensation \#1 at 10.6234$-$00.3819 is 25.2~Jy/beam at
870~$\mu$m.  This corresponds to a column density N(H$_2$)$=6.5 \times
10^{23}$~cm$^{-2}$ ($\sim$ 350~mag of visual extinction).  The UC
\HII\ region G010.624$-$00.384 lies in the same direction. As
discussed by Keto \& Wood (\cite{ket06}), a cluster of massive stars
that contains massive YSOs still in the accretion phase ionizes this
very young \HII\ region. A 6.7~GHz methanol maser is detected nearby
(Menten, \cite{men91}; Caswell, \cite{cas95}; see Sect.~6.6 and
Table~A.2). MAGPIS detects two additional compact \HII\ regions adjacent
to this condensation, located at G010.598$-$00.384 and
G010.621$-$00.386.  Condensation \#2 peaks at 10.6264$-$00.3357 with a
peak emission at 870~$\mu$m of 2.4~Jy/beam.  This corresponds to a
column density of N(H$_2$)$=6.2 \times 10^{22}$~cm$^{-2}$
($\sim$ 33~mag of visual extinction). MAGPIS detects an UC \HII\ region
in this direction, at G010.629$-$00.338, and a 6.7~GHz methanol maser is
detected nearby (Caswell \cite{cas95}; Sect.~6.6 and Table~A.2).
MAGPIS finds a compact \HII\ region adjacent to this condensation, at
G010.618$-$00.320. Another extended 24~$\mu$m source is surrounded by
a 8.0~$\mu$m shell at G010.639$-$00.434. It lies in the direction of the
collected shell.  This is possibly another \HII\ region.

\subsubsection{N18}

This is a faint, not very clearly defined open bubble with 
spiraling filaments at 8.0~$\mu$m. For a distance of 12.4~kpc,  
its diameter is $\sim$ 50~pc (but the distance is uncertain, 
see Table~A.2).  Several very faint
870~$\mu$m cold dust condensations are located along the 8.0~$\mu$m
filaments.  Two compact radio sources are present at the western
border of the bubble.  They are not extragalactic sources because they have
associated 24~$\mu$m emission. The brightest UC \HII\ region lies at
16.6002$-$00.2759; at 5~GHz, its integrated flux is 12.2~mJy and its
diameter is $3\farcs1$ (Becker et al.~\cite{bec94}). Another fainter
UC \HII\ region lies nearby, at 16.589$-$00.283. No cold dust emission
is detected in their direction.  Their association with the N18 bubble
is uncertain.

\subsubsection{N24}

\begin{figure*}[tb]
 \includegraphics[angle=0,width=180mm]{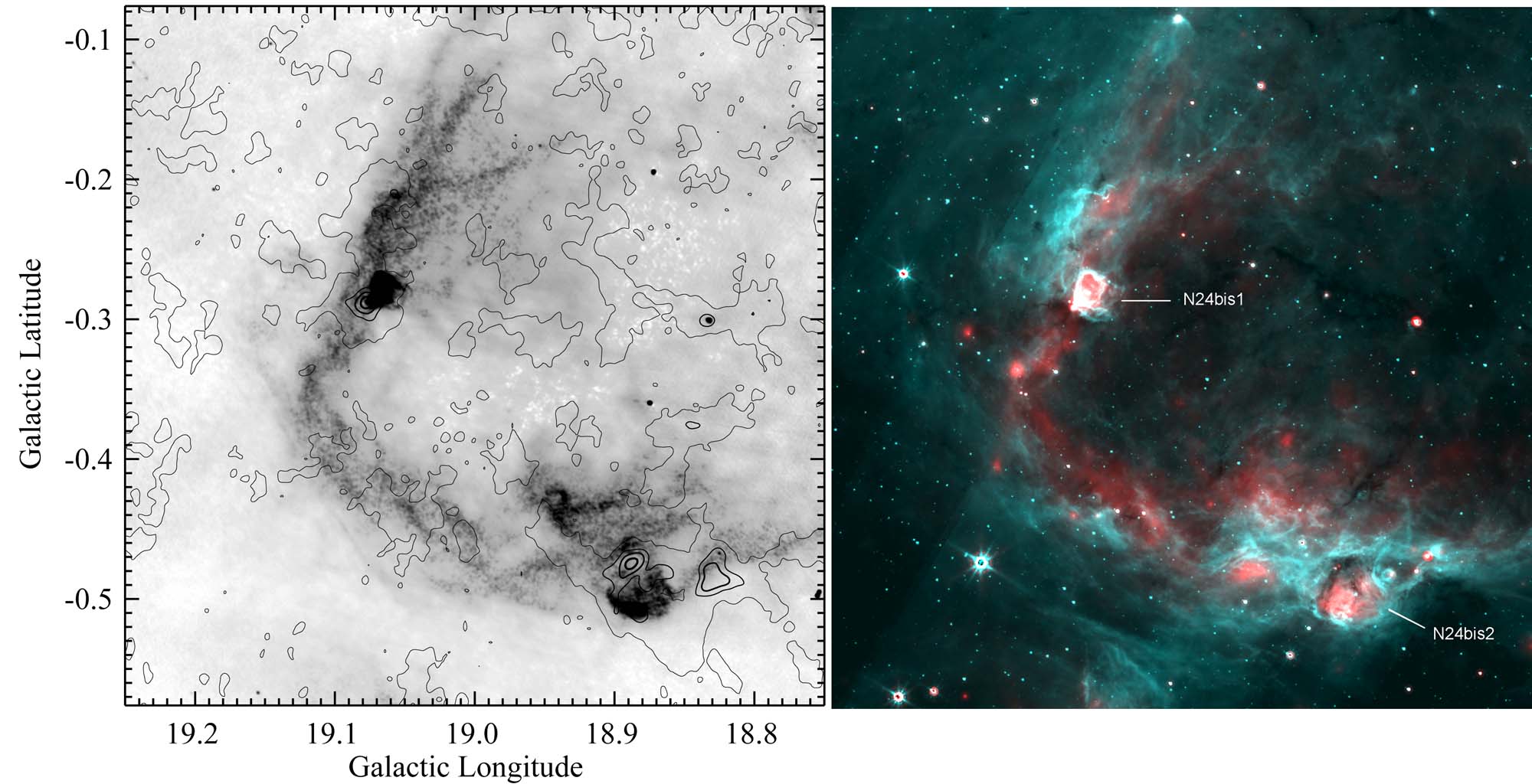}
  \caption{N24: {\it Left:} 870~$\mu$m contours superimposed on the
    MAGPIS radio-continuum image at 20-cm ($sigma$=0.06~Jy/beam;
    contour levels 1, 10, 30, 60, 100~$\sigma$). {\it Right:} colour
    composite image, with red for the {\it Spitzer}-MIPSGAL emission
    at 24~$\mu$m and turquoise for the {\it Spitzer}-GLIMPSE emission
    at 8.0~$\mu$m.  The two second-generation compact \HII\ regions,
    N24bis1 and N24bis2, are identified.}
  \label{N24}
\end{figure*}

This is a large ($\geq$30~pc for a distance of 4.6~kpc; 
Table~A.2), wide-open bubble. The 8.0~$\mu$m emission tracing the
bubble is somewhat confusing, but the radio-continuum emission exhibits a
well-defined \HII\ region (Fig.~\ref{N24}). The radio and 24~$\mu$m
emission follow, but are located interior to, the emission from
filaments at 8.0~$\mu$m. Faint cold dust emission is observed in the
direction of the 8.0~$\mu$m filaments, most likely tracing the
collected neutral material. At least two compact \HII\ regions,
N24bis1 and N24bis2, lie on the borders of N24.  Bright 870~$\mu$m
emission is observed in their direction or vicinity.  All these
objects (the large bubble, the two compact \HII\ regions, and the
870~$\mu$m condensations) are associated because their velocities are
within 3 km~s$^{-1}$ of each other (see Table~A.2). The 870~$\mu$m
emission associated with N24bis2 forms a shell surrounding the compact
central \HII\ region. A class II methanol maser lies in the direction
of this shell (Sect.~6.6 and Table~A.2).

\subsubsection{N35}

This is a large bubble, not well-defined at 8\,\micron,
that is possibly open to the southeast. At a distance of 8.6~kpc,
its size is 23~pc $\times$ 11~pc. Several cold dust condensations are
present, adjacent to the PDR.  Two condensations are bordered by
bright rims, and thus are clearly associated with the bubble.
Furthermore, one of the brightest, at 24.4587$+$00.1968, has a velocity
of 119.08~km~s$^{-1}$ similar to
that of the central \HII\ region (Table~A.2), confirming the
association. A bright radio source with a jet is present at 
24.4539$+$00.2296 (Becker et al.~\cite{bec94}); it is most probably an
extragalactic source as it has no counterpart at 8.0~$\mu$m or
24~$\mu$m. A compact \HII\ region is present at 24.429$+$00.223, in the
direction of an adjacent dust condensation.  Furthermore, a very small
8.0~$\mu$m bubble (diameter~0.5~pc, if at the same distance) enclosing
24~$\mu$m emission lies in the direction of another condensation; we
suggest that it may be another second-generation UC \HII\ region. A
6.7~GHz methanol maser is observed in the direction of the PDR,
associated with faint dust emission (Sect.~6.6 and Table~A.2).

\subsubsection{N36}

\begin{figure*}[tb]
 \includegraphics[angle=0,width=180mm]{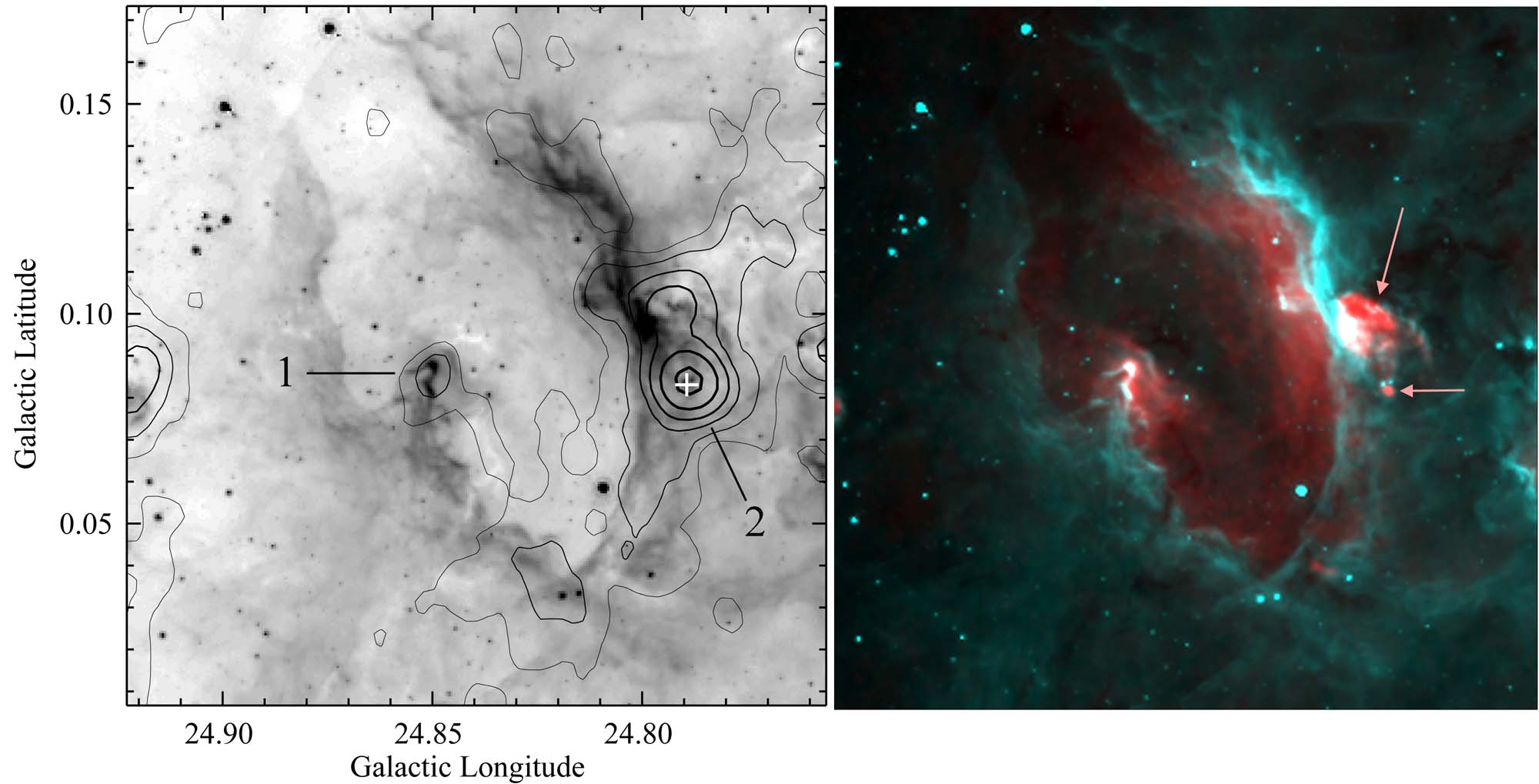}
  \caption{N36: {\it Left:} Contours of the 870~$\mu$m emission 
  superimposed on the 8.0~$\mu$m image ($\sigma$=0.09~Jy/beam; contour 
  levels 1, 3, 5, 10, 20, 60 ~$\sigma$). The white cross shows the 
  position of the methanol maser. {\it Right:} {\it Spitzer}-GLIMPSE 
  8.0~$\mu$m emission in turquoise and 
  MAGPIS 20-cm emission in red. The arrows identify the compact 
  and UC \HII\ regions.}
  \label{N36}
\end{figure*}

This is an open bubble (Fig \ref{N36}), of size
15~pc $\times$ 7.5~pc for a distance of 6.4~kpc (Table~A.2).  It is
elongated southeast -- northwest, and open on the north west
side. The 8.0~$\mu$m bubble surrounds a diffuse \HII\ region G24.83
+00.10 (Kantharia et al.~\cite{kan07}; also called G24.81$+$0.10 by
Lockman \cite{loc89}).  Figure~\ref{N36} shows faint 870~$\mu$m emission
in the direction of the PDR, most likely from cold dust in the neutral
collected shell. On the east side, a small condensation (\#1) lies in
the direction of a 8.0~$\mu$m bright rim.  Examination of emission
from MAGPIS indicates that this radio source at G24.849$+$00.088 likely
arises from the dense ionized layer bordering this condensation, and is not
a different \HII\ region. A very bright condensation is situated on
the opposite side (\#2). High resolution radio observations (Kantharia
et al. \cite{kan07}, MAGPIS) show that it contains a compact cometary
\HII\ region (G24.796$+$00.098) and two UC ones in the 
direction 24.789$+$00.82; Furuya et al.~(\cite{fur02}) discuss the presence of CO
outflows and of various maser emission in this direction (among them a
6.7~GHz methanol maser, Sect.~6.6 and Table~A.2). All these objects
are probably associated as they present similar velocities
(Table~A.2).

\subsubsection{N39}

This is a bipolar nebula (Fig.~\ref{N39} and
Sect.~6.1). Several bright 870~$\mu$m condensations are observed at 
its periphery. The compact \HII\ region G25.30$-$00.14 lies in the
direction of one of them (in this direction, N(H$_2$)$=1.10 \times
10^{23}$~cm$^{-2}$ or A$_V \sim 59$~mag). The velocity of
G25.30$-$00.14 (98.4~km~s$^{-1}$) differs consideraably from that of the
central \HII\ region G25.38$-$00.18 (57.1~km~s$^{-1}$); thus they are
not associated.  A 6.7~GHz methanol maser lies at the waist of the
bipolar nebula; both objects have similar velocities (see Sect.~6.6
and Table~A.2).

\subsubsection{N49}

\begin{figure*}[tb]
 \includegraphics[angle=0,width=180mm]{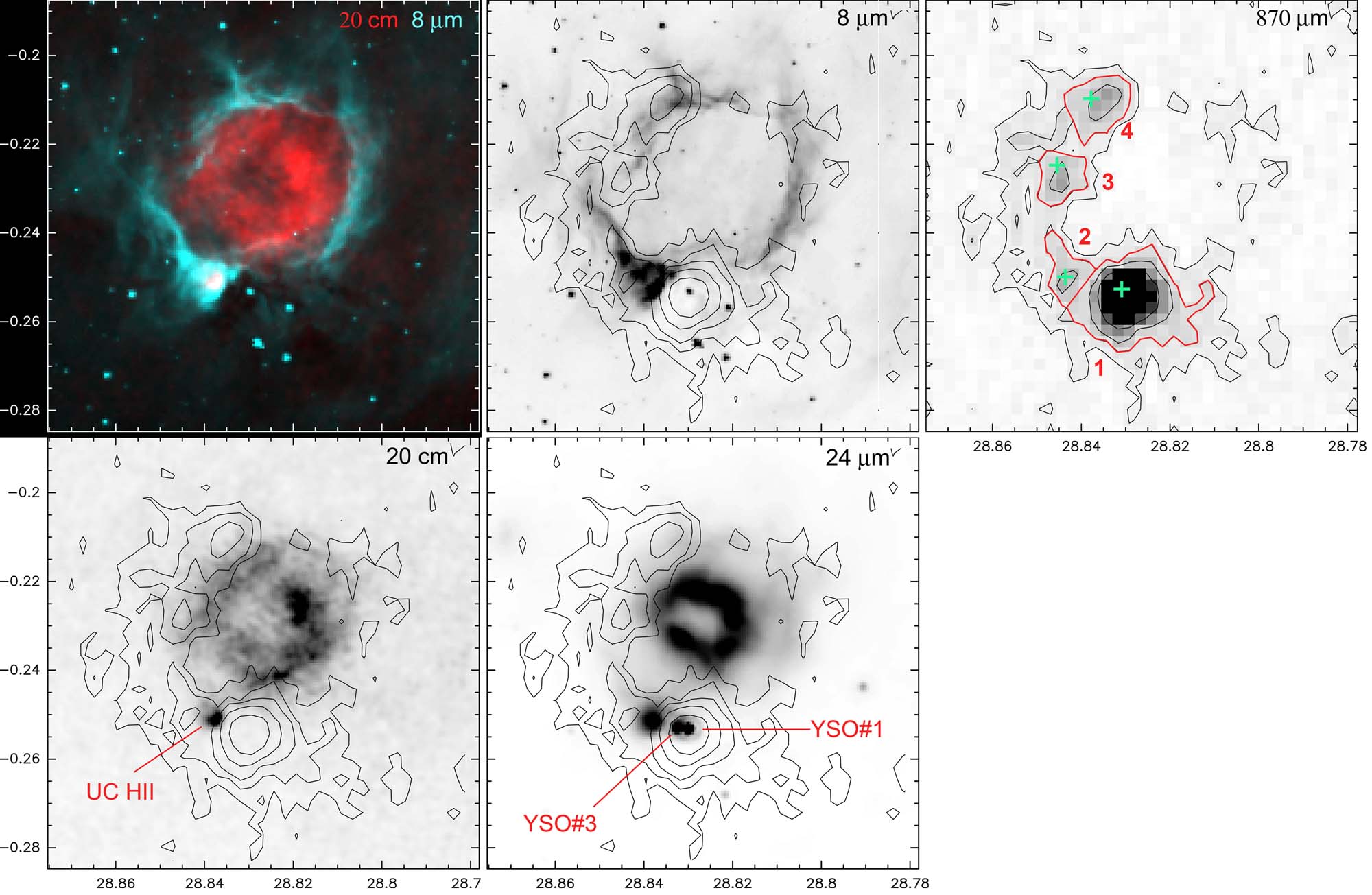}
  \caption{N49 {\it Top left:} {\it Spitzer}-GLIMPSE 8.0~$\mu$m
    emission in turquoise and 24~$\mu$m emission in red. {\it Others:}
    contours of 870~$\mu$m emission superimposed to greyscale images
    at 8.0~$\mu$m, 870~$\mu$m, 20-cm, and 24~$\mu$m; the 870~$\mu$m
    contour levels are 0.08~Jy/beam (used to define the limits of the
    collected shell), 0.25, 0.5, 1.0, and 2.0 Jy/beam. The red
    contours at 0.25~Jy/beam have been used to measure the integrated
    870~$\mu$m flux of condensations 1 to 4.  The green crosses
    indicate the NH$_3$ positions measured by Wyrowski \& Wienen (in
    preparation; Table~A.2). The massive stage I YSOs found by Watson
    et al.~(\cite{wat08}) are indicated.}
  \label{N49}
\end{figure*}

This is one of the most well-defined bubbles of the
sample (Fig.~\ref{N49}). Its internal diameter is 4.0~pc (for a
distance of 5.5~kpc, Table~A.2). The central \HII\ region has a
radio-continuum flux density of 1.14~Jy at 2.695~GHz (Reich et
al.~\cite{rei84}), 1.16~Jy at 4.875~GHz (Wink et al.~\cite{win82}),
and 1.38~Jy at 10.05~GHz (Handa et al.~\cite{han87}). This radio emission 
corresponds to an ionizing photon flux of
$\sim$ 3 $\times$ 10$^{48}$~s$^{-1}$.  This flux corresponds to a main
exciting star of spectral type O7V--O7.5V (Martins et
al.~\cite{mar05}).  MAGPIS shows that a compact \HII\ region lies in
the direction of the PDR. For this region, we measure a radio flux
(20-cm) $\leq$0.1~Jy.  This flux corresponds to a B0V exciting star,
assuming that its radio emission is optically thin at 20-cm.  This
compact \HII\ region is also surrounded by bright 8.0~$\mu$m emission.
N49 has been studied in the mid-IR by Watson et al.~(\cite{wat08}).
Possible exciting stars for the central \HII\ region have been
identified; the main one lies in the direction of the radio and
24~$\mu$m central hole, and, based on its spectral energy distribution
(SED), is either an O5V or O8III star (thus emitting more ionizing 
photons than necessary to account for the radio flux).

ATLASGAL shows that a shell of collected material surrounds the
central \HII\ region.  This shell is brighter on its eastern and
northern sides, probably because a density gradient was present in the
parental cloud. The whole shell, defined by a 870~$\mu$m emission
brighter than 0.08~Jy/beam (the rms noise in this field), has a total
flux of 24.75~Jy, thus a mass of 4200~\msol\ (Appendix B). The
shell is fragmented. The most massive fragment has a peak flux of
4.4~Jy/beam (at 18:44:51.05 $-$03:46:02 (J2000); condensation \#1 on
Fig.~\ref{N49}) corresponding to a column density
1.1 $\times$ 10$^{23}$~cm$^{-2}$ (or a visual extinction of 61~mag). Its
flux density is 13.70~Jy, which corresponds to a mass of
2300~\msol. The other neutral condensations, numbered \#2, \#3, and
\#4, are less bright.  They have peak intensities in the range
0.64--0.75~Jy/beam (and thus column densities in the range
1.65--1.9 $\times$ 10$^{22}$~cm$^{-2}$), integrated flux densities of
1.14~Jy, 1.41~Jy, and 2.07~Jy, thus masses of 190~\msol,
240~\msol, and 350~\msol, respectively. The velocity of the four
condensations (Table~A.2) are similar (in the range 85.89 --
96.34~km~s$^{-1}$) to the mean velocity of the ionized gas
(90.6~km~s$^{-1}$), confirming the association of the entire neutral
shell with the \HII\ region.

The compact \HII\ region lies between condensations \#1 and \#2,
adjacent to, but outside the most massive condensation. Two bright
YSOs identified by Watson et al.~(\cite{wat08}; their Fig.~16) lie in
the direction of the center of condensation 1. According to their
SEDs, both YSOs are massive (more than 10~\msol\ for the central
sources), and in an early evolutionary stage (stage I; the 
luminosity is dominated by the accreting envelope). 
Furthermore, YSO~\#3 in Watson et al. is
associated with an extended 4.5~$\mu$m jet (extended green object or
EGO; according to Cyganowski et al. \cite{cyg08} extended 4.5~$\mu$m
emission traces shocked molecular gas in protostellar outflows).  Two
6.7~GHz methanol masers were detected by Caswell et
al.~(\cite{cas95}) in this region. The brightest one is observed in
the direction of condensation \#1, the other, much fainter, in the
direction of condensation \#3. Their velocities (Table~A.2), which are
similar to those of the associated condensations , indicate that they
are associated with the shell surrounding the N49
\HII\ region. Cyganowski et al.~(\cite{cyg09}) present high angular
resolution methanol maser observations at 6.7~GHz and 44~GHz in the
direction of YSO \#3 (their Figs~1 and 5).  They show that the 6.7~GHz
methanol emission is observed in the direction of the center of the
4.5~$\mu$m jet.  This emission is linearly distributed along an axis
extending over 3000~AU and has a velocity gradient that is indicative 
of a rotating disk.

\subsubsection{N52}

This bubble encloses a giant \HII\ region and is part
of the W43 complex. The distance of this region is uncertain; Anderson
\& Bania (\cite{and09}) place it at the near distance, 5.7~kpc,
whereas Pandian et al.  (\cite{pan08}) favour the far distance,
9.0~kpc. In the following, we adopt, rather arbitrarily, the near
distance. Numerous studies have been performed of the W43 complex. The flux
density of the \HII\ region at 5~GHz, 86.5~Jy (Smith et
al. \cite{smi78}) corresponds to a ionizing photons flux
$\sim$ 2.3 $\times$ 10$^{50}$~s$^{-1}$; this points to an exciting
cluster containing $\sim$ 13 O5V stars or $\sim$ 55 O7V stars (according
to the calibration of Martins et al. \cite{mar05}). The ionized gas
was studied by Lester et al.~(\cite{les85}) and Balser et
al.~(\cite{bal01} and references therein); the radio continuum map
shows several filamentary components or sources; a detailed velocity
field contains velocities ranging from 60~km~s$^{-1}$ to
110~km~s$^{-1}$. The exciting cluster was analysed by Blum et
al. (\cite{blu99}), using deep $JHK$ images; this cluster is highly
reddened (A$_{\rm V}$ in the range 0 to 45~mag), and contains at least
a WN7 and two early (supergiant) O stars; these stars emit strong
winds.  The associated molecular material was studied by Motte et
al. (\cite{mot03} and references therein). The molecular cloud has a
mass $\sim$ 10$^6$~\msol.  The cold dust emission (at 350~$\mu$m and
1.3-mm) and the molecular HCO$^+$ emission have been mapped at high
resolutions; they show a filamentary structure, with massive
condensations (40--4000~\msol), large nonthermal velocities
($\sim$ 5~km~s$^{-1}$), and high densities
($\sim 10^6$~cm$^{-3}$). Stellar sources with a near-IR excess (Blum et
al. \cite{blu99}) and several maser sources (OH, H$_2$O, and methanol;
see their location on Fig.~1 in Motte et al. \cite{mot03}) indicate
that star formation is still taking place in this complex. Several
papers (especially Balser et al. \cite{bal01} and Motte et
al. \cite{mot03}) suggest that the exciting cluster has a large impact
on the surrounding material (both ionized and molecular), blowing it
away from the center of the nebula, and compressing it to filaments.

\begin{figure*}[tb]
 \includegraphics[angle=0,width=180mm]{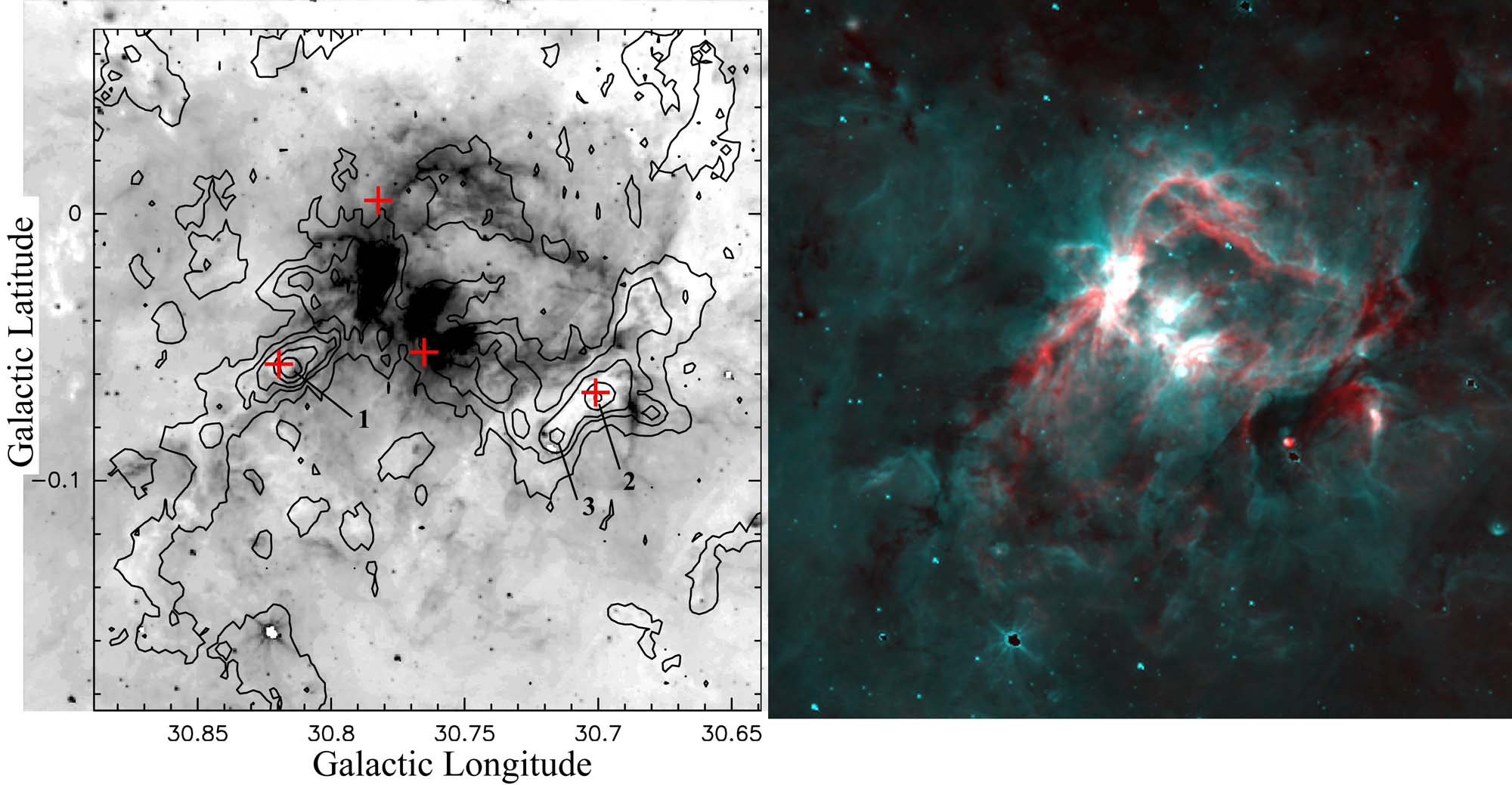}
  \caption{ N52: {\it Left:} contours of the 870~$\mu$m emission 
  superimposed on the {\it Spitzer}-GLIMPSE 8.0~$\mu$m image 
  ( the contour levels are 0.15~Jy/beam, 1, 2, 5, and 10~Jy/beam). 
  The red crosses show the position of the methanol masers. 
   The condensations \#1, \#2, \#3 contain the cores W43-MM1, 
   -MM2, -MM3 of Motte et al. (\cite{mot03}). 
  {\it Right:} colour composite image, with the MAGPIS
    20-cm emission in red and the 8.0~$\mu$m emission in
    turquoise. Note the bipolar and filamentary structure of this
    nebula.}
  \label{N52}
\end{figure*}

Thus, N52 is not a bubble enclosing a classical, almost spherical
\HII\ region.  Since the Motte et al. \cite{mot03} study, 
observations have shed new light on this complex. The {\it
  Spitzer}-8.0~$\mu$m and the MAGPIS-radio observations (the 24~$\mu$m
emission is saturated over the field), both show, at high resolution
and over a wide field, many filamentary structures linked to the
complex. Both observations indicate a bipolar morphology (Fig.~\ref{N52}),
with the exciting cluster at the center. Furthermore, the ionized gas
is distributed in filaments following exactly, on the inside, the
8.0~$\mu$m filaments. We propose the following origin and evolution for this
complex: i) a massive cluster formed in a massive and rather flat
molecular cloud and it ionized an expanding \HII\ region; ii) with time,
the massive stars evolved  and become Wolf-Rayet and
supergiant stars,  emitting very strong winds;
iii) a very low density, very hot ionized component  
developed inside the \HII\ region (according to the simulations of
Freyer \cite{fre03}); the expansion accelerated, and the classical
\HII\ region was reduced to a thin layer surrounding the central hot
ionized gas; iv) presently, the ionized region extends in directions
of lower density (perpendicular to the parental cloud), and neutral
(collected) material is seen all around.  More material is collected
at the waist of the bipolar nebula, because this is where the density
of the neutral gas is the highest.

This explanation of the bipolar structure agrees
with the observed cold dust emission.  This emission comes both from
material at the waist of the bipolar nebula, and from filamentary
structures elongated along the northeast and the southwest PDRs
(Fig.~\ref{N52}). The 870~$\mu$m emission is very similar to the
350~$\mu$m and 1.3-mm emission observed by Motte et
al. (\cite{mot03}).  For the entire region, we measure a flux density
at 870~$\mu$m of 408~Jy, indicating a mass of
7.5 $\times$ 10$^4$~\msol. At least three condensations are very bright
and well defined (\#1, \#2, \#3; Fig.~\ref{N52}); they contain the 
bright cores W43-MM1, -MM2, and -MM3 of Motte et al.(\cite{mot03}). 
We measure for these condensations 870~$\mu$m peak intensities of
respectively 21.4~Jy/beam, 11.6~Jy/beam, and 6.3~Jy/beam,
corresponding to column densities N(H$_2$) of
11.4 $\times$ 10$^{23}$~cm$^{-2}$ (A$_{\rm V}=295$~mag),
6.2 $\times$ 10$^{23}$~cm$^{-2}$ (A$_{\rm V}=160$~mag), and
3.3 $\times$ 10$^{23}$~cm$^{-2}$ (A$_{\rm V}=87$~mag). The 870~$\mu$m 
flux densities integrated over regions limited by the 1~Jy/beam  
contour level (Fig.~\ref{N52}) are, respectively, 76.7~Jy, 59.3~Jy, 
and 25.3~Jy, for masses of 14000~\msol, 4700~\msol, 
and 11000~\msol\ (the W43-MM1, -MM2, -MM3 cores have smaller sizes 
and masses). Methanol masers
at 6.7~GHz are observed in the direction of W43-MM1 and W43-MM2.  A
third methanol maser is observed in the direction of W43-MM4 
(situated at the waist of the bipolar nebula) for which
we measured a peak signal of 4.4~Jy/beam
(N(H$_2$)=2.3 $\times$ 10$^{23}$~cm$^{-2}$).  A fourth maser is given by
Xu et al. (\cite{xu09}) toward the PDR, but in a direction of very low
870~$\mu$m emission (see Table~A.2).

\subsubsection{G31.155 and N53}

\begin{figure*}[tb]
 \includegraphics[angle=0,width=180mm]{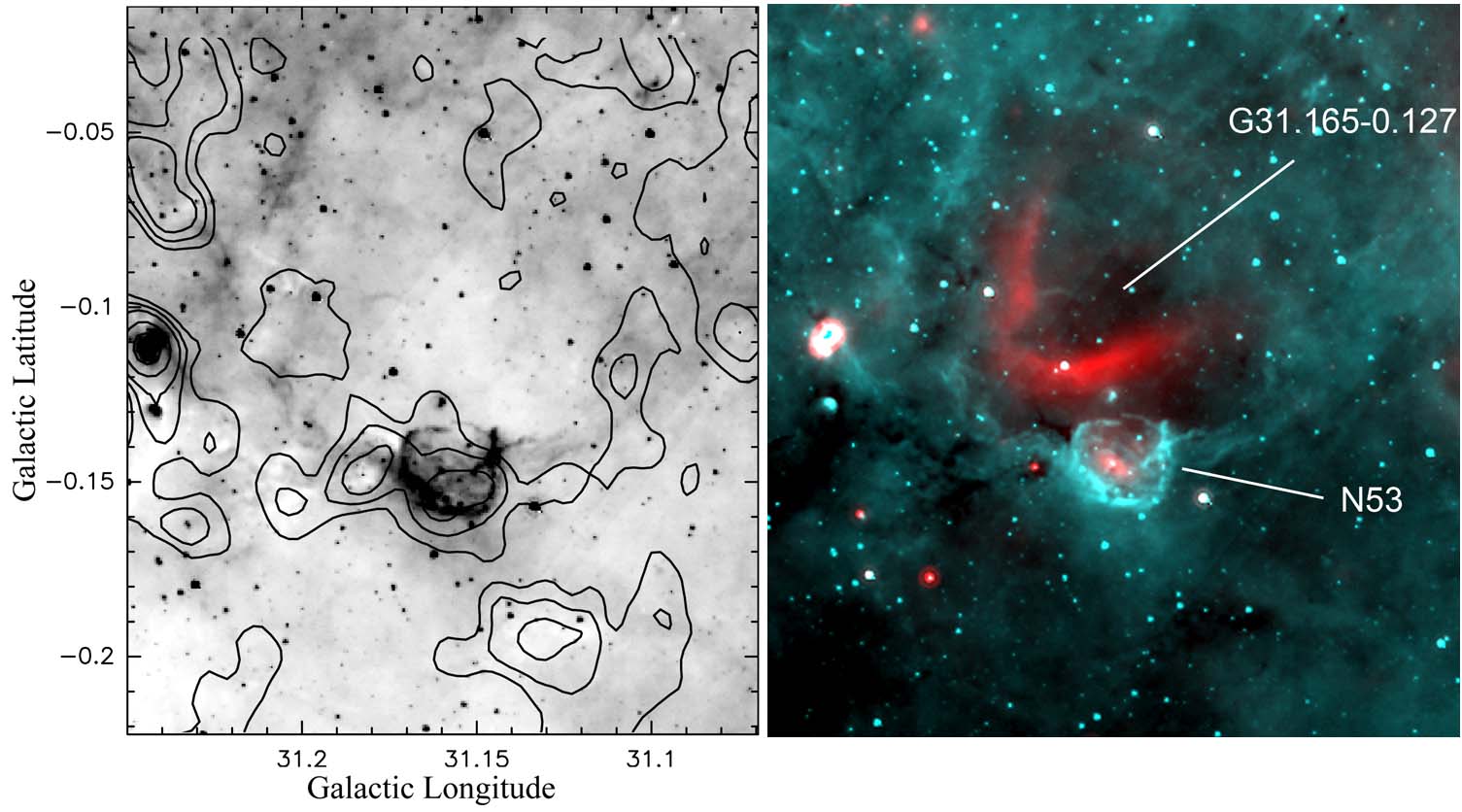}
  \caption{G31.165-00.127 and N53: {\it Left:} The 870~$\mu$m contours
    levels are superimposed on the 8~$\mu$m image; the levels are
    0.03, 0.1, 0.2, 0.5 and 1.0~Jy/beam. There is a faint emission
    structure surrounding the bottom part of G31.165-00.12, with a
    condensation in the direction of N53. {\it Right:} Composite
    colour image with the 24~$\mu$m emission in red and the 8.0~$\mu$m
    emission in turquoise.}
  \label{N53}
\end{figure*}

The bubble N53 encloses a relatively small
\HII\ region (diameter $\sim$ 4.8~pc for a distance of 11.9~kpc; 
Table~A.2).  Figure~\ref{N53} shows that it lies on the
border of a larger \HII\ region, G31.165$-$00.127 (diameter $\sim$ 20~pc
for the same distance).  The two \HII\ regions are probably linked as
they have similar velocities (43.6 and 41.4~km~s$^{-1}$, Table~A.2) and 
N53 is centered
on the PDR surrounding G31.165$-$00.127. In addition N53 lies in the direction of a
faint 870~$\mu$m emission structure following the PDR of
G31.165$-$00.127. Thus, the exciting star of the \HII\ region at the
center of N53 is a good candidate to be a second-generation massive-star,
the formation of which has been triggered by the expansion of
G31.165$-$00.127.

\subsubsection{N59}

This is a rather faint bubble that is not
well-defined. It is open to the south. If at 5.6~kpc (Table~A.2), its
diameter is $\sim$ 20~pc. A bipolar nebula lies on its northern border,
which is possibly associated with N59, but its velocity differs by
some 10~km~s$^{-1}$ from that of N59. Several cold dust condensations
are present adjacent to the PDR. The brightest one contains several
radio sources; three of them are in the Becker et al.~(\cite{bec94})
catalogue. The radio source at 33.1441$-$00.0665 (S(5GHz)=33.9~mJy,
diameter=4.8$\arcsec$) is most probably extragalactic as it has no
24~$\mu$m counterpart. An UC \HII\ region at 33.1330$-$00.0926
(S(5GHz)=385.2~mJy, diameter=$3\farcs7$) has an associated H$_2$O
maser (Kurtz \& Hofner \cite{kur05}), and an associated 6.7~GHz
methanol maser (Sect.~6.6 and Table~A.2).  A compact \HII\ region lies
at 33.1420$-$00.0863 (S(5GHz)=26.5~mJy, diameter=$13\farcs0$). A faint
UC \HII\ region, which is not catalogued by Becker at
al. (\cite{bec94}) lies at 33.111$-$00.095.  The small nearby bubble
N58 is possibly associated with N59, as they have very similar
velocities (but see the discussion about their distances in Appendix A).

\subsubsection{N65}

\begin{figure*}[tb]
 \includegraphics[angle=0,width=180mm]{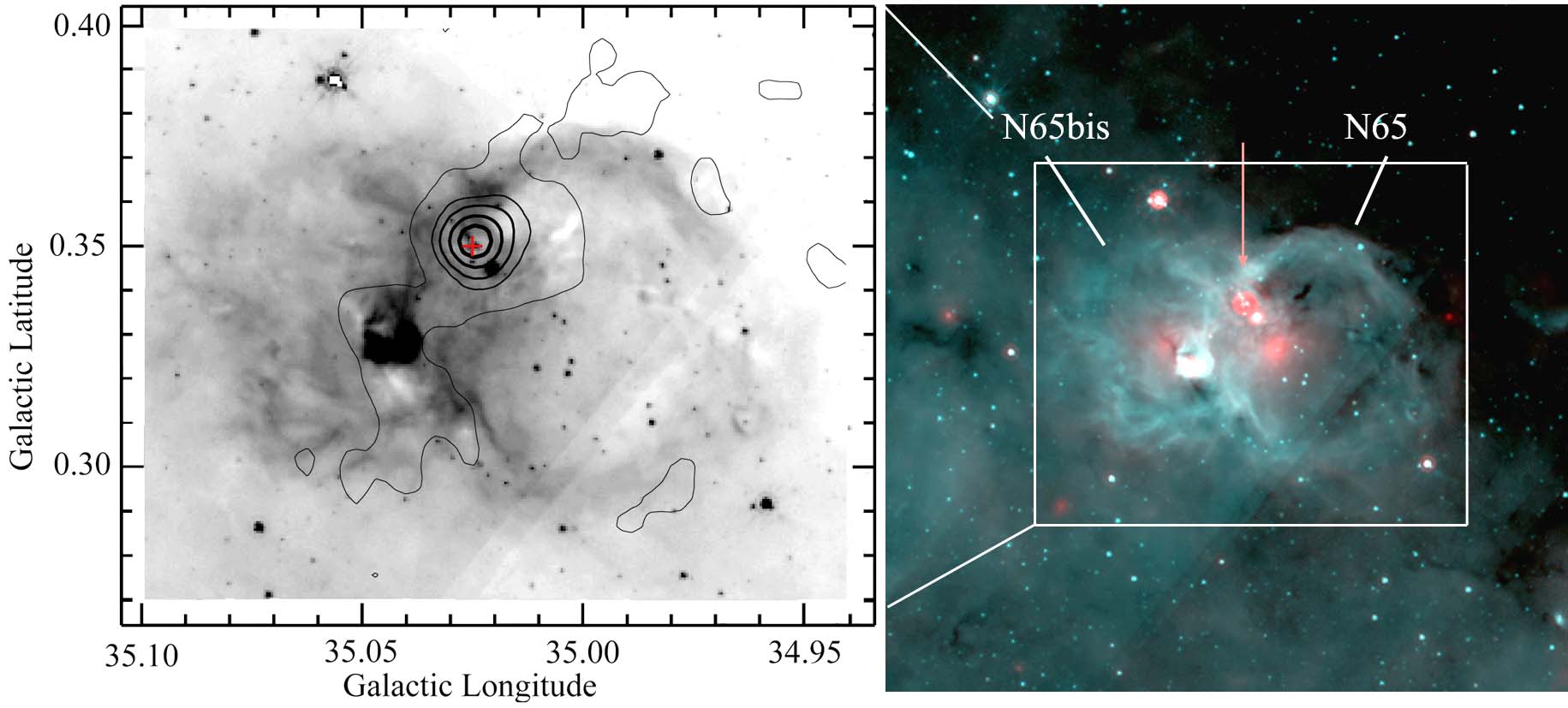}
  \caption{N65 \& N65bis. {\it Left:} Contours of the 870~$\mu$m
    emission superimposed on the 8.0~$\mu$m image
    ($\sigma$=0.04~Jy/beam; contour levels 1, 5, 15, 30, 50~$\sigma$);
    {\it Right:} {\it Spitzer}-GLIMPSE 8~$\mu$m emission in
    turquoise and -MIPSGAL 24~$\mu$m emission in red. We see both the faint
    emission from the cold dust associated with the collected material
    surrounding N65, and a bright condensation between the two
    \HII\ regions, suggesting a zone of compression. The bright
    condensation contains an UC \HII\ region and a class II methanol
    maser. The arrow points to a bright 24~$\mu$m source which is also 
    a UC \HII\ region. The red cross gives the position of the methanol 
    maser.}
  \label{N65}
\end{figure*}

This is a well-defined bubble of medium size (diameter
$\sim$ 5~pc for a distance of 3.6~kpc). It is possibly adjacent to
another bubble that we have named N65bis. A shell of collected
material surrounds N65 (Fig.~\ref{N65}). A bright dust condensation
lies between N65 and N65bis. A very bright 24~$\mu$m source lies in
its center, cospatial with an UC \HII\ region detected via its radio
continuum emission.  The dust condensation has been observed at 1.2-mm
and has a mass of 560~M\sun\ (Hill et al.~\cite{hil05}, for a distance
of 3.6~kpc (Table~A.2). The mass of its molecular counterpart was
estimated from $^{13}$CO (1-0) observations by Petriella et al.
(\cite{pet10}) to be $\sim$ 2000~\msol . From the 870~$\mu$m emission, we
estimate the mass of this condensation to be 1060~\msol\ (see
Sect.~6.4 for more details). The UC \HII\ region lies at
35.024$+$00.350; it has a flux density of 14~mJy at 3.6-cm, and a size
(HPBW) of ~$\sim$ $0\farcs9$ (Kurtz, Churchwell, \&
Wood~\cite{kur05}). The OH , H$_2$O, and 6.7~GHz methanol maser emission 
are oberved in a nearby direction (respectively, Argon et
al.~\cite{arg00}; Kurtz \& Hofner~\cite{kur05}; Pandian et
al.~\cite{pan07}; see also Sect.~6.6).

A small 8.0~$\mu$m bubble around an extended 24~$\mu$m source is also
present in the direction of the collected layer between N65 and
N65bis, at 35.044$+$00.327. No radio emission is detected in its
direction; it is possibly a faint compact \HII\ region.

The N65 - N65bis complex may be similar to the Sh~255 - Sh~257 complex
where several young massive objects have been found in the compressed zone
between the two \HII\ regions (Chavarria et al.~\cite{cha08} and
references therein).

\subsubsection{N68}

This is a large bubble, elongated northwest to
southeast, and possibly open to the southeast. Its size is 34~pc
$\times$ 17~pc for a distance of 10.6~kpc (Table~A.2). It is very
similar in appearance to N36. Faint 870~$\mu$m emission follows the
8.0~$\mu$m PDR, probably corresponding to cold dust emission in the
collected neutral shell. One dust condensation corresponds to a bright
rim observed at 8.0~$\mu$m, bordered by bright 20-cm emission. On
the other side lies a bright 870~$\mu$m condensation.  This
condensation is associated with the central \HII\ region, as it has a
similar velocity (V$_{\rm LSR}$(CS)=54.1~km~s$^{-1}$, Plume
\cite{plu92}; V$_{\rm LSR}$(H110$\alpha$)=51.8~km~s$^{-1}$, Araya et
al.  \cite{ara02}). In the direction of this condensation lies a
compact \HII\ region, G35.590$-$00.025, and an UC \HII\ region
G35.578$-$00.031 (Kurtz et al.~\cite{kur99}).  OH and H$_2$O maser
emission are observed in the direction of the UC \HII\ region
(Forster \& Caswell \cite{for99}).

\subsubsection{N81}

\begin{figure}[tb]
 \includegraphics[angle=0,width=90mm]{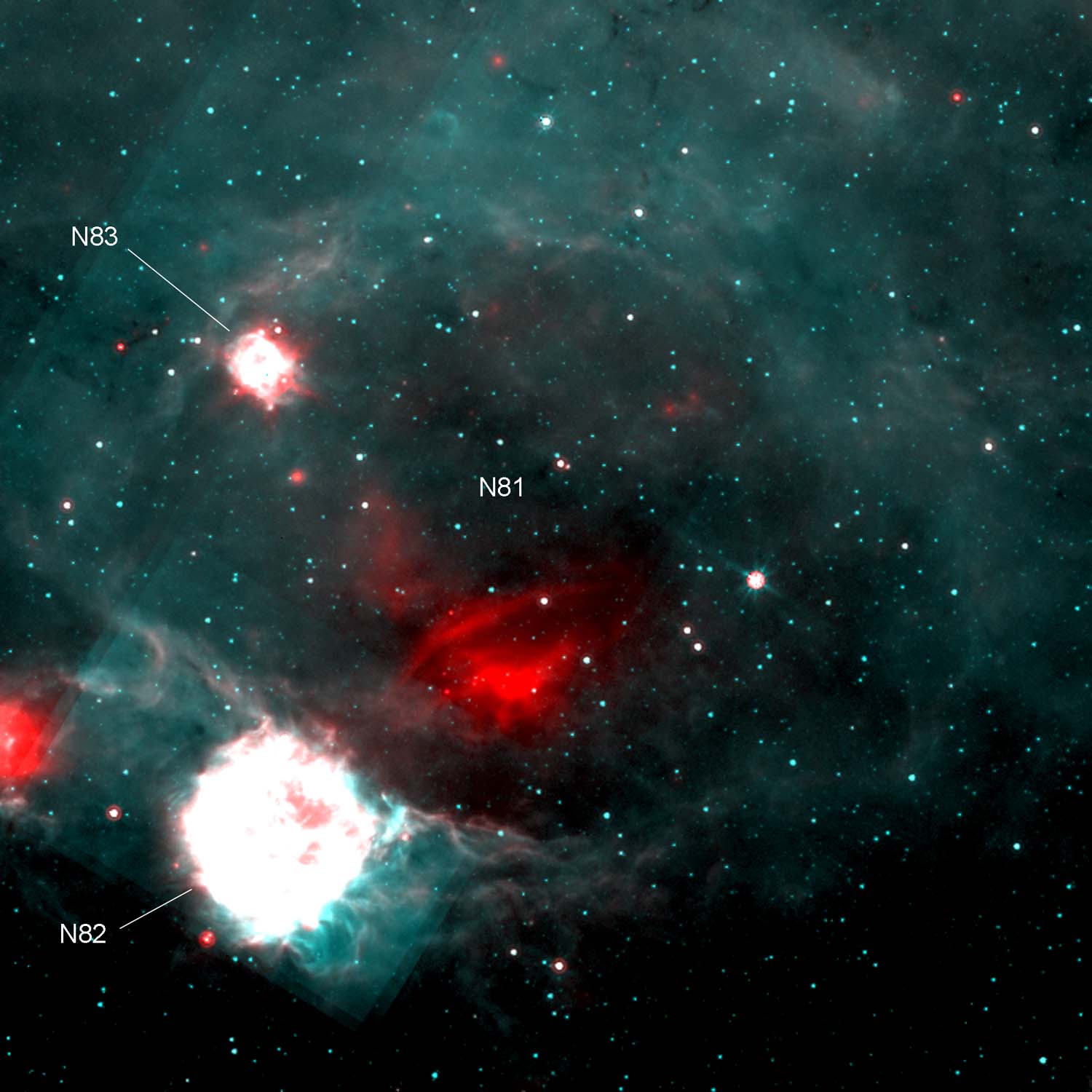}
  \caption{N81 and the possible second generation \HII\ region N82.
    Colour composite image: red represents the {\it Spitzer-}MIPSGAL
    emission at 24~$\mu$m and turquoise represents the {\it
      Spitzer-}GLIMPSE emission at 8.0~$\mu$m.}
  \label{N81}
\end{figure}

This is a very faint, large bubble.  For a distance of
8.1~kpc (Table~A.2), its size is 45~pc $\times$ 32~pc.  Two smaller bubbles, 
N82 and N83 lie on its border. N82 (diameter 7.5~pc) is clearly 
associated with N81 on morphological basis (Fig.~\ref{N81}). The velocity 
of N83 differs from that of N82 by more than 10~km~s$^{-1}$; thus, N83 is 
probably not associated with N81 and N82.

\subsubsection{N91}

\begin{figure*}[tb]
 \includegraphics[angle=0,width=180mm]{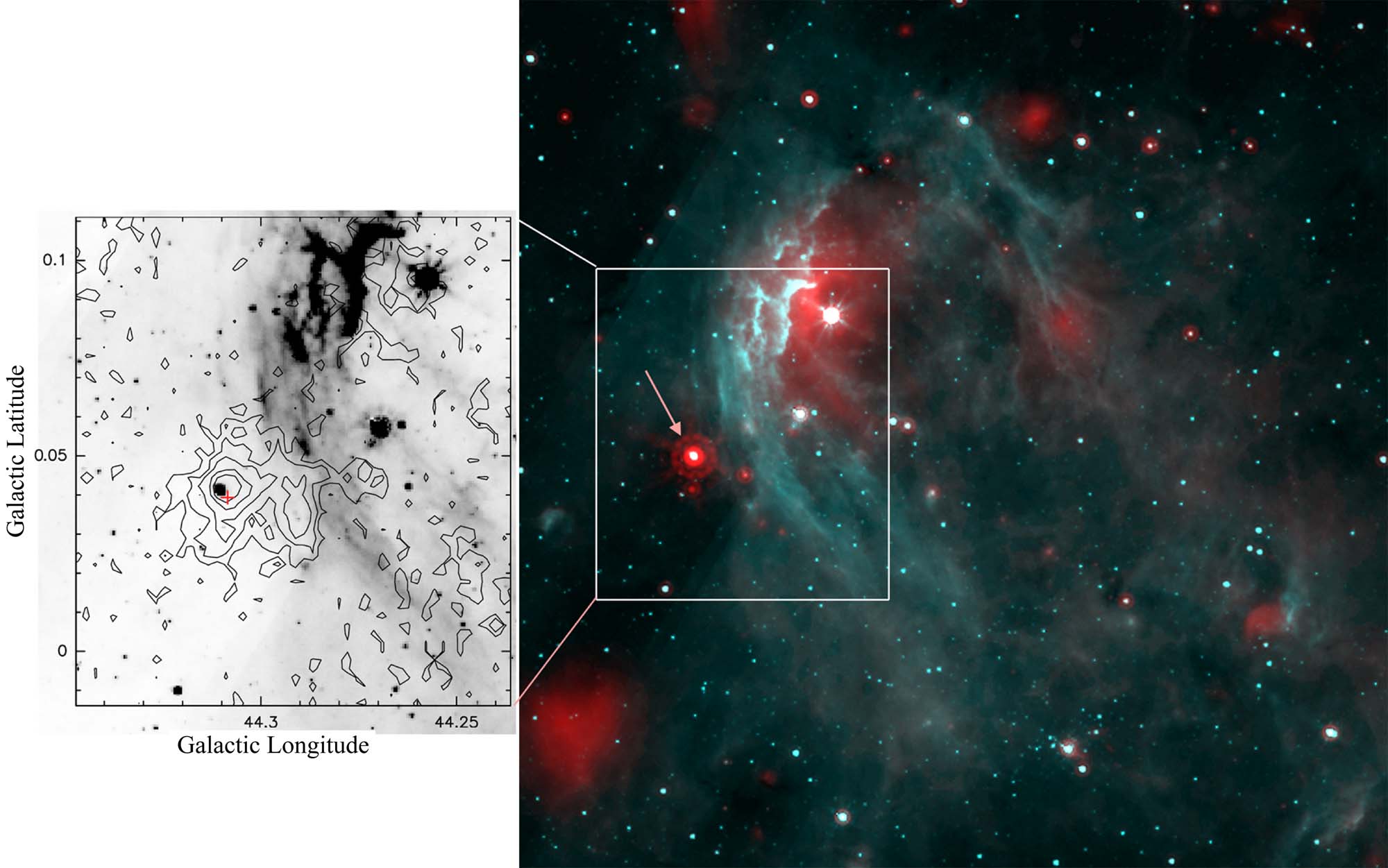}
  \caption{N91. {\it Left:} The 870~$\mu$m contour levels 
  (0.1, 0.25, 0.5, 1.0, 1.5~Jy/beam) are superimposed on the 
  {\it Spitzer}-GLIMPSE 8~$\mu$m image. The red cross 
  gives the position of the methanol maser. {\it Right:} 
  Colour composite image with the 24~$\mu$m emission in red 
  and the 8~$\mu$m emission in turquoise. The arrow points 
  to a bright 24~$\mu$m source, which is also a UC \HII\ region.} 
  \label{N91}
\end{figure*}

This is a large bubble (28~pc $\times$ 14~pc for 
D=8.1~kpc; Table~A.2), open in the south, with a bright PDR on its east side 
(Fig.~\ref{N91}). 
A faint elongated 870~$\mu$m structure follows this part of the 
PDR. Two dust condensations are present in the direction of the PDR, 
probably slightly in the foreground as a faint absorption is 
observed at 8~$\mu$m in their directions. An UC \HII\ region 
(which is also a bright 24~$\mu$m source) lies near the peak of 
the brightest condensation. Several other  24~$\mu$m sources 
and a 6.7~GHz methanol maser are also present there. 
We have no velocity for the UC \HII\ region or 
for this dust condensation, but the methanol maser has a velocity 
similar to that of the central \HII\ region, confirming the 
association (Table~A.2).

\subsubsection{N94}

This is an open bubble. At a distance of 9.3~kpc
(Table~A.2), its size is 20~pc $\times$ 12~pc. The nearby small bubble
N93 (diameter 4.0~pc if at the same distance) lies in the direction 
of a filament, which is adjacent to N94.  This association, however, is 
uncertain and a measurement of the velocity of N93 is required.

\subsubsection{N97}

This is an angularly large (diameter $\sim$ 16$\arcmin$)
and faint bubble that is not well defined at 8\,\micron, except along
its north border. The small N96 bubble (diameter $\sim$ 1$\arcmin$)
lies on this border and is clearly associated with a bright rim
bordering a dust condensation. We have no velocities and no distances
for these regions.\\

\subsubsection{Conclusions}

To conclude, 13 bubbles have associated UC \HII\ regions in the
direction of their PDRs and in neutral condensations adjacent to their
IFs.  These bubbles are N2, N24, N35, N36, N49, N52, G31.16, N59, N65,
N68, N81, N91, and N97. These regions are good candidates for triggered
massive-star formation. The N94 bubble is another possible candidate
for this process, but the association between the large \HII\ region
and the UC \HII\ region needs to be confirmed by velocity
measurements.  The case for N18 is more uncertain as the nearby UC
\HII\ regions are not observed in the direction of dust condensations.

Most of these bubble candidates for triggered massive-star formation
are large bubbles (size $\geq$~15~pc), which could indicate that they
represent an evolved population. However, N49 and N65, perhaps our
best cases of triggered massive-star formation, are exceptions.  Their
diameters are in the range 4--5~pc. A possible explanation is that
they formed and evolved in a medium of higher density, which has
restricted their expansion.

\subsection{Other signposts of massive-star formation}

The 6.7~GHz methanol
masers (class II), discovered by Menten (\cite{men91}), are associated
with massive-star formation.  Some are observed in the vicinity of UC
\HII\ regions (e.g. the W3(OH) source, Menten et al. \cite{men92}).
Other methanol masers originate in the direction of (sub)-millimeter
condensations containing luminous very red mid- or far-IR sources
({\it Spitzer}-GLIMPSE sources: Ellingsen \cite{ell06}; 24~$\mu$m
sources: Xu et al. \cite{xu09} and Minier et al. \cite{min05}, and
references therein).  Breen et al. (\cite{bre10}) present an
evolutionary sequence for masers in high-mass star formation regions:
6.7~GHz masers have a lifetime in the range $2.5 \times 10^4$~yr to
$4.5 \times 10^4$~yr; they are coeval with H$_2$O masers, and appear
before the formation of a detectable UC \HII\ region.\\

We searched for 6.7~GHz methanol masers in the vicinity of the
bubbles, using the compilations by Pestalozzi et al. (\cite{pes05})
and Xu et al. (\cite{xu09}). Twenty-nine 6.7~GHz methanol masers have
been detected in the vicinity of 23 bubbles.  In the following, we
comment briefly on the location of these masers with respect to nearby
870~$\mu$m condensations, UC \HII\ regions, and mid-IR sources. The
positions and velocities of the masers are given in Appendix A,
Table~A.2. The main difficulty when attempting to associate masers
with other signposts of star formation is their positional accuracy,
which varies widely from one study to another.  The most accurate
positions come from interferometric observations.

Walsh et al. (\cite{wal98}) used the Australia Telescope Compact
Array (ATCA), which gives absolute positions with an uncertainty $\sim
1\arcsec$.  These masers lie near:

- N2: Two bright 870~$\mu$m condensations lie at one extremity of this
large open bubble (Fig.~\ref{N2}, Sect.~6.5); both contain a maser. 
The first maser lies $\sim 10\arcsec$ away from the
center of condensation \#1 (peak emission 25.2~Jy/beam).  This
condensation contains several UC \HII\ regions, which are also
24~$\mu$m sources; the maser lies $\sim 9\arcsec$ away from the
closest 24~$\mu$m source. The second maser lies $\sim 11\arcsec$ away
from the center of condensation \#2 (peak emission 2.4~Jy/beam), and
$\sim 17\arcsec$ away from the associated UC \HII\ region.

- N36: A maser lies in the direction of the bright adjacent 870~$\mu$m
condensation (peak emission 15.8~Jy/beam, see Fig.~\ref{N36}), $\sim
6\arcsec$ from the sub-mm peak. This is also the direction of a
cluster of four YSOs (Furuya et al. \cite{fur02}). As discussed by
these authors, the methanol maser spots correspond to an UC
\HII\ region that is also associated with a CO outflow.

- N49: A maser lies in the direction of the bright 870~$\mu$m
condensation \#1 (peak emission 4.4~Jy/beam, Fig.~\ref{N49} and
Sect.~6.5), $\sim 12\arcsec$ from its center. This condensation
contains two bright mid-IR sources. The maser lies in the direction of
one of them, which is also an extended green object (EGO, Cyganowski 
et al. \cite{cyg09}) with jets. A nearby UC \HII\ region lies outside
the condensation, within $\sim 20\arcsec$ of the maser's position. A
second fainter maser was detected by Caswell et
al. (\cite{cas95}; positional accuracy $\sim 10\arcsec$), at the border
of condensation \#3 (peak emission 0.73~Jy/beam); it has no radio or
mid-IR counterpart.

- N52 (W43): A maser lies in the direction of the brightest 870~$\mu$m
condensation (\#1, peak emission 23.2~Jy/beam), $\sim 12\arcsec$ from its
center, and $\sim 6\arcsec$ from two faint 24~$\mu$m sources. This
condensation is adjacent to the bright PDR of W43 (Sect.~6.5 and
Fig.~\ref{N52}).  A second maser also lies in the direction of the
PDR, but at the waist of the bipolar nebula, $\sim 24\arcsec$ from a
870~$\mu$m condensation (emission peak 4.7~Jy/beam).  In this
direction, the 24~$\mu$m emission is saturated. A third maser lies
$\sim 4\arcsec$ away from the center of another bright 870~$\mu$m
condensation (\#2, emission peak 12.6~Jy/beam), and in the direction of the
PDR of a compact associated \HII\ region; it has no 24~$\mu$m
counterpart.

\begin{figure*}[tb]
 \includegraphics[angle=0,width=180mm]{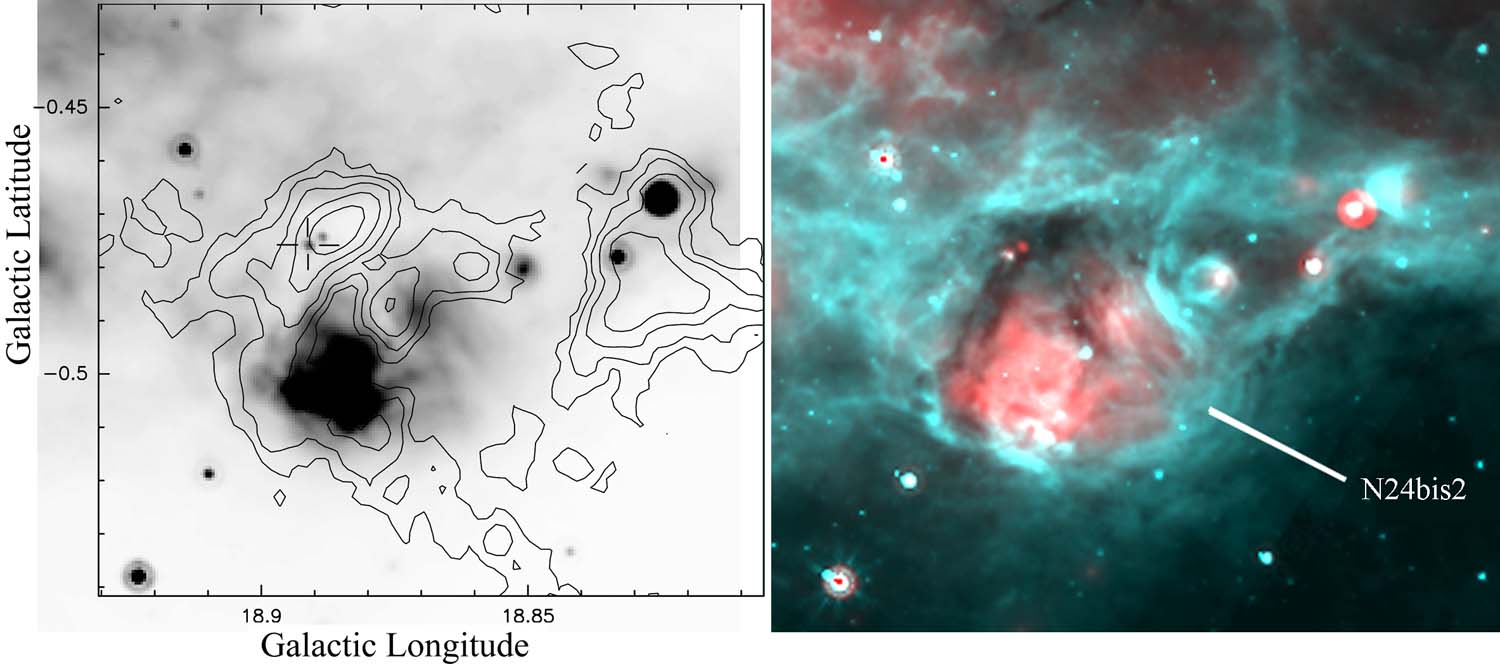}
  \caption{N24bis2. {\it Left:} The 870~$\mu$m contour levels 
  (0.1, 0.25, 0.5, 1.0, and 2.0~Jy/beam) are superimposed on 
  the 24 $\mu$m image. The 6.7~GHz methanol maser, observed in 
  the direction of an EGO, is at the center of the cross. 
  {\it Right:} Composite colour image with the 24~$\mu$m  
  emission in red and the 8~$\mu$m emission in turquoise.} 
  \label{N24bis2}
\end{figure*}

Two more regions were observed at high resolution with the VLA 
by Cyganowsky et al. (\cite{cyg09}). They are:

- N24bis2: The maser lies in the direction of a 870~$\mu$m condensation, part 
of the dust shell surrounding the central compact \HII\ region (peak 
emission 3.7~Jy/beam; Fig.~\ref{N24bis2}). Two mid-IR sources lie nearby. The maser is observed 
in the direction of one of them, which is also an EGO 
(Cyganowski et al. \cite{cyg09}). No radio-continuum 
source is detected nearby.

- N65: The maser is observed in the direction of the center of a
870~$\mu$m condensation (emission peak 5.15~Jy/beam) located between
the two bubbles N65 and N65bis (Fig.~\ref{N65} and Sect.~6.5).  An UC
\HII\ region, which is also a bright 24~$\mu$m and 70~$\mu$m source,
also lies at the center of this condensation.  It was found by
Kurtz et al. (\cite{kur94}) to be composed of two unresolved 
(FWHM of the beam $\sim 0\farcs9$) radio-continuum
sources separated by $2\arcsec$. The high resolution observations 
of Cyganowsky et al. (\cite{cyg09}) 
show that the maser is observed in the direction of
an EGO with jets, possibly centered on one of the radio sources. As
discussed by Cyganowsky et al. (\cite{cyg09}), the nature of this radio
source is uncertain; it may be a hypercompact \HII\ region or a dust core.

The masers reported by Caswell et al. (\cite{cas95}) were 
observed with the Parkes 64-m telescope. Their positions have errors
of smaller than $10\arcsec$ in each coordinate. They are:

-N59: A maser lies in the direction of the brightest 870~$\mu$m condensation 
(peak emission 4.0~Jy/beam), $\sim 17\arcsec$ away from its center. This 
condensation contains an UC \HII\ region in its very center, which is 
also a bright 24~$\mu$m source. A second maser is associated with a 
secondary fainter 870~$\mu$m condensation (peak emission 0.50~Jy/beam), 
which is surrounded by a bright rim at 8.0~$\mu$m; the maser lies 
$\sim 14\arcsec$ away from the condensation's center and $\sim 8\arcsec$ 
away from a 24~$\mu$m source situated at the tip of the bright rim.

- N61: A very bright 870~$\mu$m condensation (peak emission 51.4~Jy/beam) is
adjacent to this faint bubble. It contains a group of UC \HII\ regions 
(G34.3+0.2 A, B, C; Campbell et al. \cite{cam04}) and 
two more classical (extended) \HII\ regions. The maser lies $\sim 9\arcsec$ 
from the condensation's peak, on the border of the group of 
UC  \HII\ regions.  (The UC \HII\ regions 
lie near the center of the 870~$\mu$m condensation.)  The 24~$\mu$m
emission is saturated. This condensation is a very active massive-star
formation region.

Six more masers associated with N67bis, N70, N71, N75, N79, and N91
were detected by Pandian et al. (\cite{pan07}). This unbiased
survey, carried out using the Arecibo radio telescope, is complete at
the level of 0.27~Jy over the region $l = 35.2\degr$ to $53.7\degr$,
$|b| \le 0.41\degr$. The Arecibo beam was $40\arcsec$ (FWHM) and the
pointing accuracy was $7\arcsec$.

\begin{figure*}[tb]
 \includegraphics[angle=0,width=180mm]{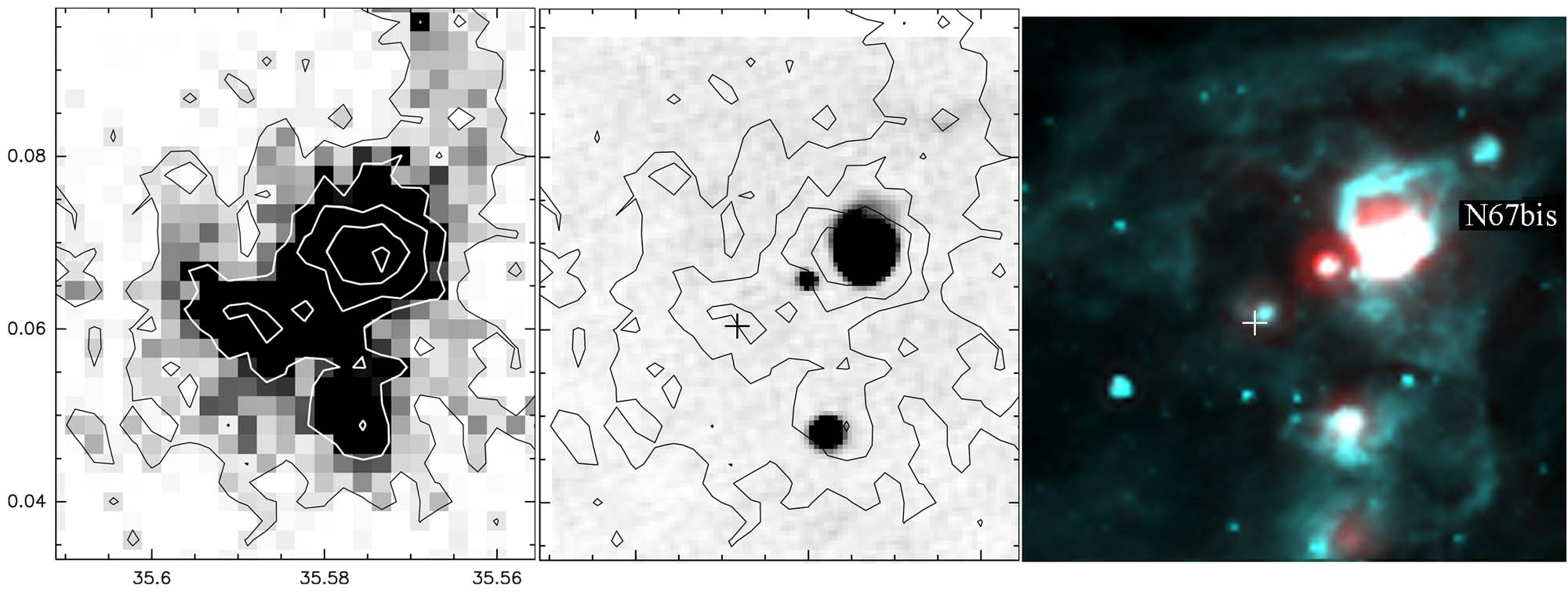}
  \caption{N67bis, a region of active massive-star formation.  
  {\it Left:} 870~$\mu$m emission; the contour levels are 0.1, 0.5, 
  1.0, 1.5, and 2.0 Jy/beam. {\it Middle:} the 870~$\mu$m contour levels 
  are superimposed on the MAGPIS 20-cm image. {\it Right:} Colour 
  composite image with the 24~$\mu$m emission in red and the 
  8~$\mu$m emission in turquoise. The cross gives the position of 
  the 6.7~GHz methanol maser (Pandian et al., \cite{pan07}).}
  \label{N67bis}
\end{figure*}

- The bubble N67bis is observed in the direction of a bright
870~$\mu$m condensation (peak intensity 2.1~Jy/beam) that is composed
of three substructures. The whole condensation contains three radio
sources (maximum separation $\sim 80\arcsec$, $\sim$ 4~pc for a
distance of 10.6~kpc, Table~A.2). The methanol maser is observed in
the direction of one of the secondary peaks; it is not associated with
an UC \HII\ region, but lies $\sim 5\arcsec$ away from a 24~$\mu$m
source (Fig.~\ref{N67bis}).  This region is an active massive-star
formation, with three UC \HII\ regions and a class~II methanol maser
inside a massive cloud of about 15000~\msol\ (the 870~$\mu$m emission,
integrated over a region enclosed by the 0.1~Jy/beam level, gives a
flux density of 23.5~Jy when we assume a distance of 10.6~kpc,
Table~A.2).

- N70: An 870~$\mu$m condensation  is adjacent to N70 (emission peak 
2.55~Jy/beam). An UC \HII\ region lies at the condensation's center. 
The maser lies $\sim 17\arcsec$ from the condensation, in the direction of 
a faint 24~$\mu$m and 70~$\mu$m source.

- N71: Two methanol masers are detected on the border of the large and 
faint N71 bubble. They have very different velocities, and thus cannot 
both be associated with N71. The velocity of the \HII\ region enclosed by 
N71 is unknown. We suggest that maser \#1 is associated with the bubble 
as it is observed in the direction of a 870~$\mu$m condensation possibly 
linked to structures of the ionization front.  
The 870~$\mu$m peak emission is 1.6~Jy/beam. The condensation does not 
contain an UC \HII\ region, but only a 24~$\mu$m source at the very 
center of the condensation. The maser lies $17\arcsec$ away from the center. 

- N75 is adjacent to a 870~$\mu$m condensation (peak emission
2.8~Jy/beam) that contains a compact \HII\ region and bright 24~$\mu$m
sources.  A methanol maser lies on the border of the condensation,
$\sim 26\arcsec$ away from its center, on the border of the compact
\HII\ region.  The maser lies $\sim 10\arcsec$ away from the closest
24~$\mu$m source.

- N79: This is a region of very faint 870~$\mu$m emission. The methanol 
maser lies in the direction of the PDR; it is not associated with a  
870~$\mu$m condensation, or an UC \HII\ region, and lies  
$\sim 8\arcsec$ away from a faint 24~$\mu$m source.

- N91: A bright 870~$\mu$m condensation containing an UC \HII\ region 
and a methanol maser lies adjacent to the PDR of N91 
(maximum emission 2.6~Jy/beam; Fig.\ref{N91}) 
The methanol maser lies $\sim 17\arcsec$ away from the center of 
the condensation, and  $\sim 13\arcsec$ away from the UC \HII\ region.  

Three masers associated with N35, N58, and N69 were observed by
Szymczak et al. (\cite{szy02}), using the Torun 32-m telescope.
According to the authors, the positional accuracy is better than
$30\arcsec$.

- N35: A maser lies in the direction of a faint filament (870~$\mu$m 
emission 0.24~Jy/beam), adjacent to the PDR. There is no nearby UC \HII\ 
region or 24~$\mu$m source.

- N58: This small bubble lies near the center of a 870~$\mu$m condensation 
(peak emission 2.4~Jy/beam). The maser is on the border of this condensation, 
$\sim 34\arcsec$ away from its center, outside the ionized region. No 
24~$\mu$m source is present nearby.

- N69: The maser lies on the border of an 870~$\mu$m condensation
(peak emission 1.2~Jy/beam), situated on the open side of N69.  The
condensation contains no UC \HII\ region, but does have a bright 24~$\mu$m
source. The maser lies $\sim 30\arcsec$ away from the center of the
condensation; it has no 24~$\mu$m counterpart.

Two more masers were detected by Xu et al. (\cite{xu08})
associated with N32 and N39. They have been observed with the
Effelsberg 100~m-telescope. According to Xu et al. (\cite{xu08}), the
masers' positions may have an error as high as $1\arcmin$.

- N32: The maser is situated on the border of N32. No 24~$\mu$m source
is observed nearby. There is an 870~$\mu$m condensation (peak emission
1.6~Jy/beam) adjacent to N32, containing an UC \HII\ region.  The
maser lies $\sim 50\arcsec$ away from the condensation's peak emission
and $\sim 38\arcsec$ away from the UC \HII\ region's center.

- N39: A maser lies on the border of an 870~$\mu$m condensation
located at the waist of the bipolar nebula (peak emission
3.0~Jy/beam). The maser lies $\sim 23\arcsec$ away from the
condensation's peak, in a region of high radio emission and 
saturated 24~$\mu$m emission.

Four more masers were observed in the direction of IRAS sources
that heve characteristics similsr to UC \HII\ regions, by Szymczak et
al. (\cite{szy00}) and Slysh et al. (\cite{sly99}). Because of the
relatively poor positional accuracy of IRAS, the positions of these
masers are rather uncertain. They are:

- N10: A maser is observed on the border of one of the two bright
870~$\mu$m condensations adjacent to the bubble. It lies $\sim
31\arcsec$ away from the peak emission (signal 4.7~Jy/beam).  No UC
\HII\ region and no 24~$\mu$m sources are observed nearby (the nearest
24~$\mu$m source lies $\sim 18\arcsec$ away from the maser).

- N12: A maser lies not far from the PDR, but inside the bubble, in a 
region where no 870~$\mu$m emission is detected, and far from any 
24~$\mu$m source.

- N22: The maser is not linked to the bubble, as it has a very 
different velocity (see Table~A.2).

- N55: A maser lies inside the adjacent 870~$\mu$m condensation (peak
emission 2.3~Jy/beam), $\sim 9\arcsec$ from the condensation's
center. This condensation is an active massive-star formation region
as it contains two UC \HII\ regions and several bright 24~$\mu$m
sources.  The maser lies between the two UC \HII\ regions, within
$10\arcsec$ of the closest one.\\

\begin{figure*}[tb]
\includegraphics[angle=0,width=180mm]{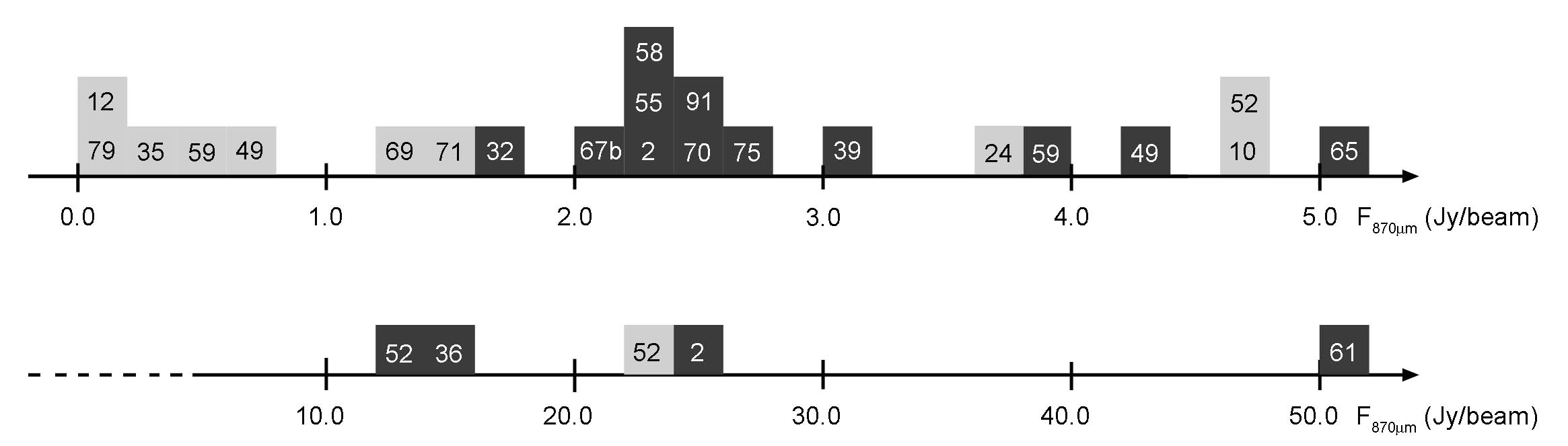}
  \caption{Methanol masers: 870~$\mu$m flux density at the peak of the
    condensations in the directions of which 6.7~GHz methanol 
    masers are detected. The
    nearby bubbles are identified by their number (e.g. ``12'' for N12).  The
    black condensations also contain one or several UC \HII\ regions;
    no UC \HII\ regions have been detected in the direction of the
    light grey condensations. Note that the scale along the two axes
    differ by a factor ten.}
  \label{masers}
\end{figure*}

These class II methanol masers are most often observed in the
direction of 870~$\mu$m condensations. Figure~\ref{masers} gives the
870~$\mu$m flux density at the peak of these condensations. The
condensations in the directions of UC \HII\ regions (detected by
MAGPIS) appear in black. Condensations without any UC radio sources
appear in light grey. Figure \ref{masers} shows that most of the
condensations containing UC \HII\ regions and class II methanol masers
have high column densities, in the range $5 \times 10^{22}$ to $1.3
\times 10^{24}$~cm$^{-2}$. These column densities are higher 
by a factor $\sim$10 than 
the column densities estimated in the direction of the shells of collected 
material. (The 870~$\mu$m flux densities in the collected shells, 
outside the condensations, are in the range 0.1--0.5~Jy/beam 
indicating column densities in the range 
2.6 $\times 10^{21}$--1.3 $\times 10^{22}$~cm$^{-2}$).

Many of the bright 870~$\mu$m condensations containing a class II methanol 
maser are very active massive-star forming regions. They often contain  
several young massive objects in different evolutionary stages: various 
masers, and UC and compact \HII\ regions. The condensation 
associated with N67bis is a good illustration of such a region 
(Fig. \ref{N67bis}). Similar condensations are found near N2, N36, N55, N59,  
and N61.

A few class II methanol masers are found in quite different regions,
associated with a bright rim (N59, N71), in a region of high
compression between two bubbles (N65). Some are even seen in
directions of dim 870~$\mu$m emission such as those associated with 
N12, N35, and N79 (their positions, however, are possibly not very 
accurate).\\

The presence of 6.7~GHz methanol masers in dust condensations 
adjacent to the bubbles is a strong indicator of massive-star 
formation presently at work. As discussed by Breen et al. (\cite{bre10}),  
6.7~GHz methanol masers appear before the formation of a detectable 
UC \HII\ region, and are still present at the beginning of their evolution;  
their lifetime is short, $\leq5 \times 10^4$~yr. Thus, they are much 
younger than the classical \HII\ regions enclosed by the bubbles 
(with ages of a few Myrs). 

At least 12 bubbles show class II methanol masers in the 
direction of condensations interacting with the enclosed \HII\ regions. 
They  are good candidates for triggered massive-star formation. They are 
N2, N10, N24bis2, N36, N39, N49, N52, N59, N65, N69, N71, and N91.

\subsection{Infrared dark clouds (IRDCs)}

IRDCs were discovered by the ISO and MSX surveys as structures seen in
absorption against the bright Galactic mid-IR background (Perault et
al.  \cite{per96}; Carey et al. \cite{car98}). IRDCs may contain dense,
cold, and massive cores which are often presented as the locations
where clusters and high-mass stars form (Chambers et al. \cite{cha09},
and references therein; Peretto \&Fuller \cite{per09}).\\

\begin{figure}[tb]
\includegraphics[angle=0,width=90mm]{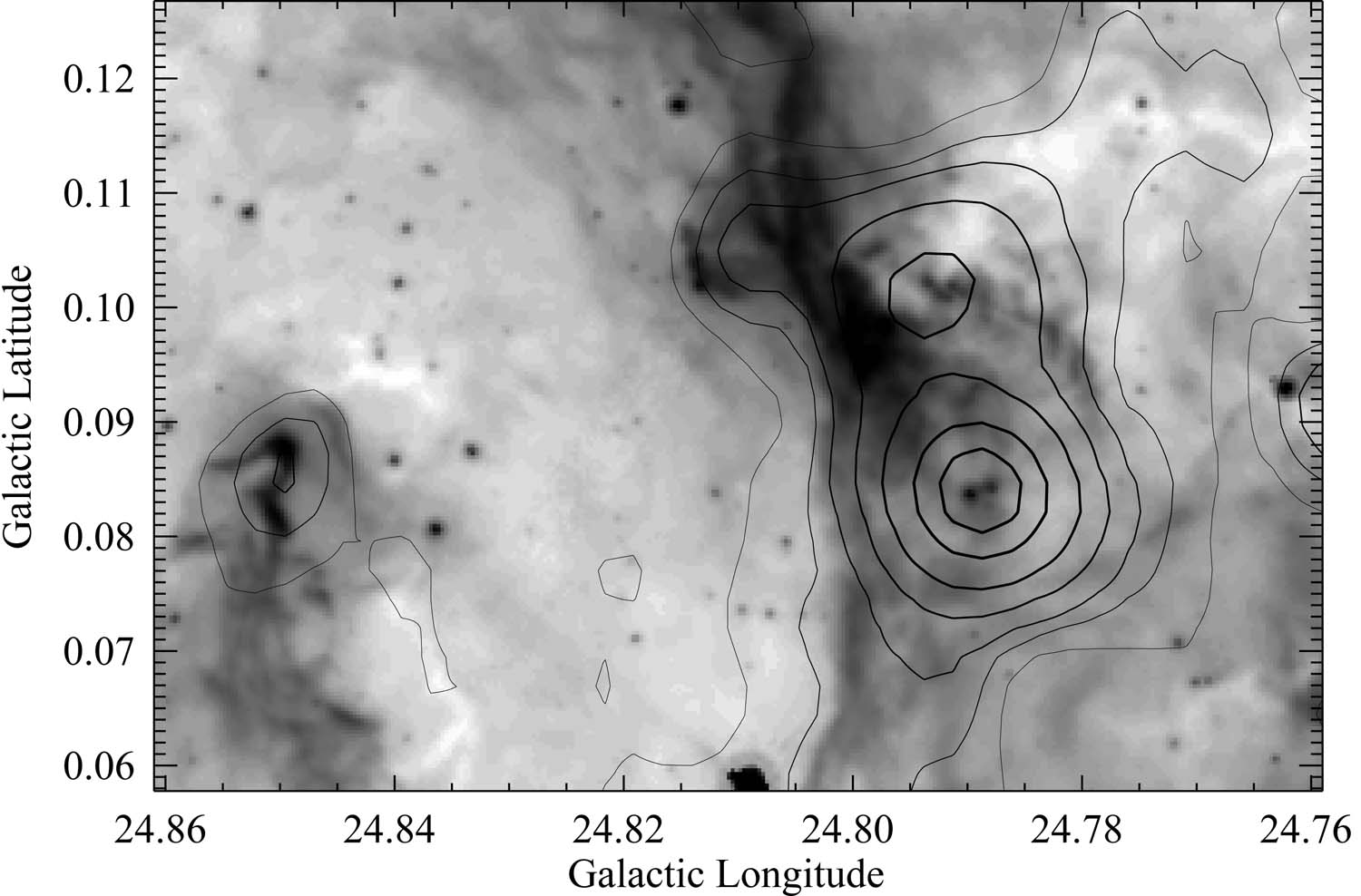}
  \caption{Dense condensations and IRDCs at the border of N36.  
    The 870~$\mu$m contour levels are superimposed on the 
    8~$\mu$m image ($sigma$=0.09~Jy/beam; contour levels 
    are 1, 3, 5, 10, 20, 40, and 60~$\sigma$). The
    dense condensation in the direction of the possible
    second-generation UC \HII\ region G24.789$+$00.082 does not appear
    as an IRDC. It lies in the direction of the bright PDR surrounding
    N36, and probably slightly behind it. The foreground bright
    8.0~$\mu$m emission of the PDR hides the background absorption
    caused by the condensation.}
  \label{IRDC}
\end{figure}

Several of the possible second-generation UC \HII\ regions described
above are seen in the direction of bright molecular condensations,
with high column densities at the peak. This is the case for the UC
\HII\ region G24.789$+$00.082 on the border of N36.  The column density
in this direction is N(H$_2$)$=4.13 \times 10^{23}$~cm$^{-2}$
(corresponding to a visual extinction A$_{\rm V} \sim
221$~mag). However, as shown in Fig.~\ref{IRDC}, this condensation
does not appear to be an IRDC at 8.0~$\mu$m (or 24~$\mu$m) because it
lies in the direction of the bright PDR of N36, and slightly at the
back of the \HII\ region.  The foreground PDR emission hides any
absorption. An IRDC is observed nearby (also detected in emission at
870~$\mu$m) that corresponds to the outer part of the condensation,
far from the PDR. The same situation is observed for the condensation
on the opposite side of N36.

The situation is the same for the bright condensations at the border 
of N2 (N(H$_2$)$=6.51 \times 10^{23}$~cm$^{-2}$), at the border of N59 
(N(H$_2$)$=1.02 \times 10^{23}$~cm$^{-2}$), at the border 
of N65 (N(H$_2$)$=1.33 \times 10^{23}$~cm$^{-2}$), or at the 
border of N68 (N(H$_2$)$=1.07 \times 10^{23}$~cm$^{-2}$). All these 
bright dust condensations do not appear as IRDCs.
  
This situation was already observed in the RCW~120 bubble (Deharveng
et al.~\cite{deh09}; their Fig.~3) where the brightest condensation
adjacent to the ionization front (condensation 1, with a peak column
density of $4.2 \times 10^{23}$~cm$^{-2}$) is not observed as an IRDC,
whereas the outskirts of this condensation, far from the IF, are
observed in absorption at 8.0~$\mu$m.

To conclude, not all massive condensations (where high-mass star
formation may occur) appear as IRDCs.  Many massive condensations,
adjacent to \HII\ regions, and slightly behind the ionized gas are not
seen in absorption at 8.0~$\mu$m or 24~$\mu$m; this absorption is
hidden by the emission of the PDR at these wavelengths. 


\section{Conclusions}

We have studied 102 regions identified as ``bubbles'' on the 
{\it Spitzer}-GLIMPSE images at 8.0~$\mu$m. Most of these regions were 
catalogued by Churchwell et al.~(2006). This large sample allows
us to draw the following statistical conclusions:

$\bullet$ At least 86\% of the bubbles enclose \HII\ regions, 
detected in the MAGPIS and VGPS surveys at 20-cm as radio-continuum 
sources. Therefore, bubbles are clearly associated with \HII\ regions and 
their massive OB2 exciting stars.

$\bullet$ Ninety-eight percent of the bubbles enclose hot dust
radiating at 24~$\mu$m.  This emission is more centrally peaked than
the radio-continuum emission of the ionized gas. Therefore, dust
grains are present inside the ionized region, where they are heated by
the absorption of Lyman continuum photons emitted by the central
exciting stars.  However, this study alone cannot say wether this hot dust
emission comes from big grains in thermal equilibrium or from very
small grains out of equilibrium.

$\bullet$ The cold dust emission at 870~$\mu$m, as traced by ATLASGAL,
allows us to detect the dense (and often massive) condensations that 
are potential sites of star formation. In 65 regions, we have high
enough angular resolution to study the distribution of neutral
material and dust in the vicinity of the bubbles. Forty percent
of these 65 regions are surrounded by collected material forming a
(partial) shell or a ``necklace'' of condensations around the ionized
gas. Another 28\% show interacting condensations, bordered by bright
rims.  These are possibly pre-existing condensations that are being
compressed by the adjacent expanding \HII\ regions.  The association
is uncertain for 24\% of the bubbles as we lack velocity information,
and 8\% of the bubbles present no detectable cold dust emission.

$\bullet$ At least 13 bubbles (possibly 14) have been found to 
contain an UC \HII\ region (or several of them) 
in the direction of their PDRs.
These additional \HII\ regions are located in dense condensations
adjacent to the IFs. At least 12 bubbles show class II
methanol masers in dense adjacent condensations (7 bubbles with 
associated UC \HII\ regions and methanol masers, plus 5 more bubbles 
with only methanol masers). These regions are good candidates for triggered
massive-star formation. Thus, among the 65 \HII\ regions  enclosed 
by bubbles, and for which the angular resolution of ATLASGAL is high  
enough to study the distribution of cold dust, more than
a quarter may have triggered the formation of massive-stars.   
This indicates that massive-star formation triggered by \HII\ regions 
may be a rather
efficient process to form massive stars (either via collect and
collapse or by radiative compression of pre-existing clumps).  A
detailed study of several of these regions will be given in a
forthcoming paper. \\

The very simple morphology of these regions also allows us to draw 
some morphological conclusions:

$\bullet$ We confirm that PAHs are destroyed in the ionized gas. PAHs
are strong emitters at 8.0~$\mu$m in the PDRs surrounding the
\HII\ regions for two reasons: i) they are excited by the far UV
radiation leaking from the ionized regions; ii) there is a strong
overdensity of neutral material just beyond the IF (by a factor 10 to
100, Hosokawa \& Inutsuka, \cite{hos05}, \cite{hos06a}).

$\bullet$ Hot dust is present inside the ionized zones. In many bubbles,  
more than half of 
the 24~$\mu$m continuum emission comes from dust inside 
the \HII\ region. Models of \HII\ regions must take into account the 
absorption of Lyman continuum photons there.

\begin{acknowledgements}

We thank D. Russeil for useful discussions about distance
determinations, C. Beaumont and J. Williams for allowing us to use
their CO observations, and J. Arthur for providing beautiful   
images of simulated \HII\ regions. We thank Paola Caselli, our referee, 
for suggesting a number of clarifications. We thank the APEX staff 
for their help during the observations with APEX-LABOCA. 
This research has made use of the SIMBAD
data base, operated at CDS, Strasbourg, France, and of the interactive
sky atlas Aladin (Bonnarel et al.\ \cite{bon00}). This work is based
in part on observations made with the {\it Spitzer Space Telescope},
which is operated by the Jet Propulsion Laboratory, California
Institute of Technology, under contract with NASA. We have made use of
the NASA/IPAC Infrared Science Archive to obtain data products from
the 2MASS, {\it Spitzer}-GLIMPSE, and {\it Spitzer}-MIPSGAL
surveys. We have used images extracted from the MAGPIS survey. These
images were generated with data from telescopes of the National Radio
Astronomy Observatory, a National Science Foundation Facility, managed
by Associated Universities Inc.  The research presented in this paper
has also used data from the Canadian Galactic Plane Survey, a Canadian
project with international partners supported by the Natural Sciences
and Engineering Research Council. L.A. acknowledges support by the
ANR {\emph Agence Nationale pour la recherche} project ``PROBeS'',
number ANR-08-BLAN-0241. L.B. acknowledge support from the
Center of Excellence in Astrophysics and Associated Technologies
(PFB06) and by FONDAP Center for Astrophysics 15010003.

\end{acknowledgements}



\begin{appendices}
\appendix
\section{The distance to the bubbles}

The distances to the \HII\ regions in the center of the bubbles (or
associated with them, see the footnotes) are given in Table~A.2. These
distances are kinematic distances (based on the velocity of the
ionized gas measured mostly using radio recombination lines, and
assuming circular rotation around the Galactic center). All the
regions belong to the first quadrant. Therefore, each positive
velocity has two possible distance solutions.  This is known as the
kinematic distance ambiguity (KDA).

Seventy-four \HII\ regions have a measured velocity and fifty-seven
of them have the KDA resolved.  Most of the distances given in
Table~A.2 are from Anderson \& Bania (\cite{and09}), who
resolved the KDA using two methods, one comparing the velocity of the
ionized gas with the maximum velocity of \HI\ absorption
(\HI\ emission/absorption), and one looking for \HI\ absorption at the
velocity of molecular emission (\HI\ self-absorption).  We also
used distances from Sewilo et al.~(\cite{sew04}), who resolved the
distance ambiguity by comparing the velocity of the ionized gas with
that of the H$_2$CO absorbing material along the same line of sight.
Additional velocity measurements are from the Green Bank Telescope
\HII\ Region Discovery Survey (Anderson et al., 2010a, in prep.).  For
these new measurements, the distance was derived using the
\HI\ emission/absorption method (Anderson et al., 2010b, in prep).
When the KDA was not resolved, we indicate in Table~A.2 the two
possible kinematic distances obtained using the Galactic rotation
curve of Brand \& Blitz (\cite{bra93}). In a few cases, two velocity
components have been measured; they are both given in Table~A.2.

Class II methanol masers (emitters at 6.7~GHz) are considered as 
signposts of recent massive-star formation. Twenty-nine methanol 
masers have been detected in the vicinity of the bubbles studied 
in this paper. Their coordinates and velocities are also given 
in Table~A.2. We also report in this Table the distances 
estimated by Pandian et al. (\cite{pan08}) for these masers. The 
KDA is resolved by comparing the maser velocity with 
the maximum velocity of the \HI\ absorption in the same direction. 

The velocities of several bright 870~$\mu$m condensations have been 
measured by F. Wyrowsky and M. Wienen, using the NH$_3$ (1,1) 
inversion line (Wienen, Wyrowski, Menten et al., in preparation). 
This information allows to confirm 
the association of the dust condensations with the nearby 
bubbles and \HII\ regions. These velocities are also given in 
Table~A.2.\\

The distances to two bubbles are especially problematic.  The bipolar
bubble N52 (W43) has a recombination line velocity of 91.6~km\,s$^{-1}$ 
(Lockman \cite{loc89}), which is near the tangent point velocity
in this direction (105\,\kms\ according to Brand \& Blitz \cite{bra93}).  
Differentiating between the near and the far kinematic
distances using \HI\ absorption, as A\&B have done, relies on the
detection of \HI\ absorption between the recombination line velocity
and the tangent point velocity.  If these two velocities cannot be 
clearly separated, this method gives less reliable results than when 
they can.  A\&B also used
the \HI\ self-absorption method, this analysis favoring the near
distance for N52, in agreement with   
Downes et al. (\cite{dow80}) and Kuchar \& Bania (\cite{kuc94}).

The \HII\ region enclosed by N59 has a recombination line velocity 
of 87.4~km\,s$^{-1}$ 
(Araya et al. \cite{ara02}), which is near the tangent point 
velocity of 97.9~km\,s$^{-1}$ according to the Brand \& Blitz (\cite{bra93}) 
rotation curve.  There is, however, \HI\ absorption from the \HII\
region's continuum beyond the tangent point velocity, which favors the
far distance.  A\&B assigned the near distance based on \HI\
self-absorption alone since the \HI\ absorption method is less
reliable for sources with velocities near the tangent point velocity 
than distinct from it. 
Pandian et al. (\cite{pan08}) favor the far distance based 
on \HI\ absorption analysis.  It
seems likely, given the strong absorption at more than 10~km\,s$^{-1}$ 
from the recombination line velocity, that N59 is at the far distance.

\section{Column density and mass estimated from the 870~$\mu$m dust emission}

In the following, we assume that the 870~$\mu$m emission arises from
thermal dust grains. According to Hildebrand (\cite{hil83}), the total
(gas+dust) mass of a condensation is related to its flux density
$S_\nu$ by
\begin{eqnarray*}
M_{\mathrm{(gas+dust)}} = 
100\,\,\frac{S_{\mathrm{870\,\mu m}}\,\,
D^2}{\kappa_{\mathrm{870\,\mu m}}\,\,
B_{\mathrm{870\,\mu m}}(T_{\mathrm{dust}})},
\end{eqnarray*}
where $D$ is the distance of the source, $\kappa_{\mathrm{870\,\mu m}}$ is 
the dust opacity per unit mass at 870\,$\mu$m, 
and $B_{\mathrm{870\,\mu m}}(T_{\mathrm{dust}})$
is the Planck function for a temperature $T_{\mathrm{dust}}$. 
For all the estimates in this paper, we assumed a gas-to-dust 
ratio of 100, a dust temperature of 20~K, and an opacity 
$\kappa_{\mathrm{870\,\mu m}}$~=~1.8 cm$^{2}$~g$^{-1}$. All these values
are uncertain and the subject of discussion.

With the same assumptions, we can estimate the H$_2$ column density, 
$N(\mathrm H_2)$, from the surface 
brightness $F_{\mathrm{870\,\mu m}}$, using the formula 
\begin{eqnarray*}
N(\mathrm H_2) =
\frac{100\,\, F_{\mathrm{870\,\mu m}}}
{\kappa_{\mathrm{870\,\mu m}}\,\,
B_{\mathrm{870\,\mu m}}(T_{\mathrm{dust}})\,\,
2.8\,\,m_{\mathrm H}\,\,\Omega_\mathrm{beam}}\mathrm,
\end{eqnarray*}
where $F_{\mathrm{870\,\mu m}}$ is expressed in 
Jy~beam$^{-1}$, $B_{\mathrm{870\,\mu m}}$ in Jy, 
$N(\mathrm H_2)$ is per square centimetre, the hydrogen atom mass 
$m_{\mathrm H}$ is in grams, and the beam solid angle 
$\Omega_\mathrm{beam}$ is in steradians. 

Adopting $\Omega_\mathrm{beam} = 9.817 \times 10^{-9}$~sr  
corresponding to a beam of $19\farcs2$ (FWHM), and assuming 
$T_{\mathrm{dust}} = 20$~K, this gives
\begin{eqnarray*}
N(\mathrm H_2) = 2.581 \times 10^{22}\,\, F_{\mathrm{870\,\mu m}}.
\end{eqnarray*}

From the classical relations,
$N(\mathrm H+H_2)/E(B-V)=5.8 \times 10^{21}$~ particles~cm$^{-2}$~mag$^{-1}$
(Bohlin et al.~\cite{boh78}) and $A_V=3.1\,\,E(B-V)$, we obtain  
$A_V=5.34 \times 10^{-22}~N(\mathrm H_2)=13.78\,\,F_{\mathrm 870\,\mu m}$.

\end{appendices}


\clearpage
\onecolumn
\begin{longtable}{lllllll}
\caption{Nature of the bubbles}\\ 
\hline\hline\\
Name & l ($\degr$) & b ($\degr$) & size ($\arcmin$)  
& radio emission & 24~$\mu$m emission & nature \\
\hline 
\endfirsthead
\caption{continued.}\\
\hline\hline\\
Name & l ($\degr$) & b ($\degr$) & size ($\arcmin$) & radio 
emission & 24~$\mu$m emission  & nature \\
\hline
\endhead
\hline
\endfoot
N1\footnotemark[2] & 10.231 & $-$0.305 & 10 & Yes & Yes & \HII\ region \\
N2 & 10.747 & $-$0.468 & 30 & Yes & Yes & \HII\ region, open bubble \\
N3 & 11.570 & +0.395 & 10 & No & Yes & ? \\
N4 & 11.892 & +0.748 & 10 & Yes & Yes & \HII\ region \\
N6\footnotemark[3] & 12.512 & $-$0.609 & 25 & ? & Yes (faint) & ?, two sources? \\
N7 & 12.787 & $-$0.167 & 5 & No & No & ? \\
N8 & 12.805 & $-$0.312 & 5 & Yes & Yes & \HII\ region \\
N8bis & 12.795 & $-$0.326 & idem & No & Yes & ? \\
N9 & 12.890 & $-$0.046 & 5 & No & Yes & ? \\
N10\footnotemark[1] & 13.188 & +0.039 & 15 & Yes & Yes (saturated) & \HII\ region \\
N11 & 13.218 & +0.082 & idem & Yes & Yes & \HII\ region \\
N12 & 13.727 & $-$0.015 & 20 & Yes & Yes &  \HII\ region \\
N13 & 13.900 & $-$0.014 & 5 & Yes & Yes & \HII\ region \\
N14\footnotemark[1] & 14.002 & $-$0.135 & 10 & Yes & Yes (saturated) & \HII\ region = Sh~44 \\
N15\footnotemark[1] & 15.010 & $-$0.611 & 10 & No & ? (faint) & ? \\
N16\footnotemark[1] & 15.017 & +0.056 & 10 & Yes & Yes & \HII\ region, open bubble \\
G15.68 & 15.680 & $-$0.280 & 30 & Yes & Yes & \HII\ region \\
N18 & 16.679 & $-$0.366 & 20 & Yes & Yes & \HII\ region \\
N20\footnotemark[1] & 17.917 & $-$0.687 & 10 & Yes & Yes & \HII\ region \\
N21\footnotemark[1] \footnotemark[4] & 18.190 & $-$0.396 & 10 & Yes & Yes & \HII\ region \\
N22\footnotemark[1] & 18.254 & $-$0.305 & 10 & Yes & Yes (saturated) & \HII\ region \\
N23 & 18.679 & $-$0.237 & 5 & Yes & Yes & \HII\ region \\
N24 \footnotemark[5] & 19.000 & $-$0.326 & 30 & Yes & Yes & \HII\ region, open bubble \\
N24bis1 & 19.066 & $-$0.278 & idem & Yes & Yes (saturated) & UC \HII\ region \\
N24bis2 & 18.885 & $-$0.492 & idem & Yes & Yes & \HII\ region \\
N25 & 19.507 & $-$0.191 & 5 & Yes & Yes & \HII\ region \\
N25bis & 19.494 & $-$0.206 & idem & Yes & Yes & UC \HII\ region\\
N26 & 19.587 & $-$0.051 & 5 & Yes & Yes & \HII\ region \\
G19.82 & 19.821 & $-$0.322 & 5 & ? & Yes & ? \\
N27\footnotemark[1] & 19.814 & +0.017 & 10 & Yes & Yes & \HII\ region \\
N28 & 21.351 & $-$0.137 & 10 & Yes & Yes & \HII\ region \\
N29\footnotemark[1] & 23.055 & +0.559 & 25 & Yes & Yes & \HII\ region \\
N31 & 23.843 & +0.097 & 10 & Yes & Yes & \HII\ region \\
N32 & 23.905 & +0.070 & 5 & Yes & Yes & \HII\ region \\
N33 & 24.215 & $-$0.044 & 5 & Yes & Yes & \HII\ region \\
N34\footnotemark[1] & 24.295 & $-$0.170 & 10 & Yes & Yes & \HII\ region, two sources? \\
N35\footnotemark[1] & 24.513 & +0.241 & 15 & Yes & Yes & \HII\ region \\
N36\footnotemark[1] & 24.837 & +0.090 & 10 & Yes & Yes (saturated) & \HII\ region, open bubble \\
N37\footnotemark[1] & 25.292 & +0.293 & 10 & Yes & Yes & \HII\ region, open bubble \\
N39\footnotemark[1] & 25.364 & $-$0.160 & 10 & Yes & Yes (saturated) & bipolar \HII\ region \\
N40\footnotemark[1] \footnotemark[6] & 25.369 & $-$0.365 & 10 & Yes & Yes & \HII\ region, two sources? \\
N41 & 26.266 & +0.282 & 5 & Yes & Yes & \HII\ region \\
N42 & 26.329 & $-$0.071 & 5 & Yes & Yes & \HII\ region \\
N43 & 26.597 & +0.098 & 5 & Yes & Yes & \HII\ region \\
N44\footnotemark[1] & 26.821 & +0.382 & 10 & Yes (faint) & Yes & \HII\ region \\
N45\footnotemark[1] & 26.991 & $-$0.053 & 10 & Yes & Yes & \HII\ region \\
N46\footnotemark[1] & 27.310 & $-$0.110 & 10 & Yes & Yes & \HII\ region, open bubble \\
N47\footnotemark[1] & 28.025 & $-$0.031 & 20 & Yes & Yes & \HII\ region \\
N48 & 28.332 & +0.154 & 10 & ? & Yes & ? \\
N49\footnotemark[1] & 28.827 & $-$0.229 &  10 & Yes & Yes & \HII\ region \\
N50\footnotemark[1] & 29.001 & +0.097 & 10 & Yes & Yes & \HII\ region, open bubble \\
N51\footnotemark[1] & 29.158 & $-$0.262 & 10 & No & Yes & ? \\
G30.250 & 30.250 & +0.240 &  5 & Yes & Yes & \HII\ region \\
N52\footnotemark[1] \footnotemark[7] & 30.749 & $-$0.019 & 20 & Yes  & Yes (saturated)& \HII\ region \\
N53\footnotemark[1] & 31.157 & $-$0.145 & 5 & Yes & Yes & \HII\ region \\
G31.16 & 31.165 & $-$0.127 & 15 & Yes & Yes & \HII\ region \\
N54\footnotemark[1] & 31.164 & +0.292 & 10 & Yes & Yes & \HII\ region \\
N55 & 32.098 & +0.091 & 5 & Yes & Yes & \HII\ region \\
N56\footnotemark[1] & 32.583 & +0.002 & 10 & Yes & Yes & \HII\ region \\
N57 & 32.763 & $-$0.150 & 5 & Yes & Yes & \HII\ region \\
N58 & 32.990 & +0.040 & 5 & Yes & Yes &  \HII\ region \\
N59 & 33.071 & $-$0.075 & 20 & Yes & Yes & \HII\ region \\
N60 & 33.814 & $-$0.150 & 5 & Yes & Yes & \HII\ region \\
N61\footnotemark[1] & 34.157 & +0.140 & 20 & Yes & Yes & \HII\ region \\
N62\footnotemark[1] & 34.334 & +0.216 & 10 & Yes & Yes & \HII\ region \\
N64 & 34.760 & $-$0.669 & 20 & Yes & Yes & \HII\ region  \\
N64bis & 34.695 & $-$0.653 & idem & ? & Yes & ? \\
N65\footnotemark[1] & 35.000 & +0.332 & 15 & Yes & Yes & \HII\ region \\
N65bis & 35.062 & +0.336 & idem & Yes & Yes & \HII\ region  \\
N66 & 35.259 & +0.119 & 5 & Yes & Yes & \HII\ region \\
N67 & 35.543 & +0.011 & 10 & Yes & Yes & \HII\ region \\
N67bis & 35.573 & +0.068 & idem & Yes & Yes & \HII\ region \\
N68 & 35.654 & $-$0.062 & 15 & Yes & Yes & \HII\ region \\
N69 & 36.187 & +0.648 &  30 & Yes & Yes & \HII\ region, open bubble \\
N70 & 37.751 & $-$0.112 & 5 & Yes & Yes & \HII\ region \\
N71 & 38.290 & +0.007 & 20 & ? & Yes (faint)& ? \\
N72 & 38.352 & $-$0.133 & 5 & Yes & Yes & \HII\ region \\
N73 & 38.739 & $-$0.137 & 10 & Yes & Yes & \HII\ region \\
N74\footnotemark[1] & 38.909 & $-$0.437 & 20 & Yes & Yes & \HII\ region \\
N75 & 38.928 & $-$0.387 & & Yes & Yes & \HII\ region \\
N76 & 38.955 & $-$0.731 & 20 & Yes & Yes & \HII\ region \\
N77\footnotemark[1] & 40.421 & $-$0.056 & 10 & Yes & Yes & \HII\ region \\
N78 & 41.229 & +0.170 & 5 & Yes & Yes & \HII\ region \\
N79\footnotemark[1] & 41.514 & +0.030 & 10 & Yes & Yes & \HII\ region \\
N80\footnotemark[1] & 41.930 & +0.031 & 10 & Yes & Yes & \HII\ region \\
N81 & 42.003 & $-$0.512 & 25 & Yes & Yes (faint) & \HII\ region \\
N82\footnotemark[1] & 42.102 & $-$0.623 & idem & Yes & Yes & \HII\ region \\
N83 & 42.113 & $-$0.442 & idem & Yes & Yes & \HII\ region \\
N84\footnotemark[1] & 42.831 & $-$0.161 & 10 & ? & Yes & ? \\
N85 & 43.074 & $-$0.017 & 10 & ? & Yes & ? \\
N86 & 43.097 & $-$0.040 & idem & Yes & Yes & \HII\ region \\
N87 & 43.218 & +0.107 & 5 & ? & Yes & ? \\
N89 & 43.739 & +0.114 & 25 & Yes & Yes & \HII\ region \\
N90\footnotemark[1] & 43.774 & +0.059 & idem & Yes & Yes & \HII\ region \\
N91 & 44.211 & +0.050 & 20 & Yes & Yes & \HII\ region, open bubble \\
N92\footnotemark[1] & 44.333 & $-$0.839 & 10 & Yes & Yes & \HII\ region, open bubble \\
N93 & 44.777 & $-$0.550 & 10 & Yes & Yes & \HII\ region \\
N94\footnotemark[1] & 44.818 & $-$0.500 & 15 & Yes & Yes & \HII\ region, open bubble  \\
N95 & 45.393 & $-$0.717 & 10 & Yes & Yes & \HII\ region \\
N96 & 46.947 & +0.368 & 5 & Yes & Yes & \HII\ region \\
N97 & 46.951 & +0.312 & 20 & Yes & Yes & \HII\ region, open bubble \\
N98 & 47.027 & +0.219 & 10 & Yes & Yes & \HII\ region \\
\footnotetext[1]{The CO environment of these bubbles was studied 
by Beaumont \& Williams (\cite{bea10}).}
\footnotetext[2]{N1 belongs to the W31 complex, which contains the nearby 
bright \HII\ region G10.16$-$0.35.}
\footnotetext[3]{N6 possibly consists of two different structures, an open 
bubble in the north and a filament in the south; the radio emission 
seems to be associated with the filament.}
\footnotetext[4]{N21 \& N22  belong to a large group of \HII\ regions, among them SH2-53 
(Sharpless~\cite{sha59}).}
\footnotetext[5]{N24 is in the test field of the general paper describing 
the ATLASGAL survey (Schuller et al.~\cite{sch09}).}
\footnotetext[6]{N40 probably encloses two distinct \HII\ regions (centered at 
25.384$-$0.328 and 25.377$-$0.357), as two regions of extended 24~$\mu$m 
emission are observed in its direction.} 
\footnotetext[7]{N52 is part of the W43 complex.}
\end{longtable}

\clearpage
\onecolumn
\begin{longtable}{lllllll}
\caption{Velocities and distances}\\
\hline\hline\\
Name & CH06 coordinates  & coordinates & refs\footnotemark[1] & velocity & distance & comments \\
     &                   &  (2000)           &      &(km s$^{-1}$)& kpc)  & \\
\hline
\endfirsthead
\caption{continued.}\\
\hline\hline\\
Name & CH06 coordinates  & coordinates & refs\footnotemark[1]  & velocity      & distance & comments \\
     &                   & (2000)      &       & (km s$^{-1}$) & (kpc)    &          \\
\hline
\endhead
\hline
\endfoot
N1 \footnotemark[2] & 10.231$-$0.305 & 10.232$-$0.301           & A+  & 11.7  & 2.0/14.8 & \HII\ region \\
                    &               & 10.16$-$0.35            & S+  & 12.5  & 14.6 & nearby \HII\ region \\
                    &               & 18:09:20.63 -20:15:04.5 & WW  & 12.01 & & dust condensation \\
                    &               & 18:09:24.52 -20:15:41.5 & WW  & 11.23 & & dust condensation \\
N2  & 10.737$-$0.468 & 10.664$-$0.467 & L89 & $-2.4$ &  & \HII\ region \\
    &               & 18:10:25.71 -19:56:04.5 & WW & $-3.21$ & & dust condensation \#1 \\
    &               & 18:10:35.00 -19:55:51.6 & WW & $-2.91$ & & dust condensation \#1 \\
    &                & 18:10:29.24 -19:55:41.3 & W98 & 2.2--5.9  & & methanol maser \\
    &                & 18:10:18.58 -19:54:24.6 & WW & $-4.19$ & & dust condensation \#2 \\
    &                & 18:10:15.63 -19:54:46.7 & WW & $-4.62$ & & dust condensation \#2 \\
    &                & 18:10:17.99 -19:54:04.2 & W98 & $-6.2$, $-8.0$ & & methanol maser \\
    &                & 18:10:29.80 -19:49:03.6 & WW  & $-1.49$ & & dust condensation \#3 \\
    &                & 18:10:18.60 -19:50:16.1 & WW & $-1.53$ & & dust shell \\
    &                & 18:10:36.53 -19:57:05.9 & WW & $-1.60$ & & dust shell \\
    &                & 18:10:55.17 -19:58:35.1 & WW & $-2.89$ & & dust shell \\
     &               & 18:10:56.40 -19:59:19.3 & WW & $-2.82$ & & dust shell \\
     &               & 18:10:58.19 -19:58:58.3 & WW & $-3.76$ & & dust shell \\
N4 \footnotemark[3] & 11.892 +0.748 & 11.898 +0.747 & L89 & 25.1 & 3.15/13.48 & \HII\ region \\
                    &               & 18:08:55.05 -18:14:32.6 & WW & 24.84 & & dust condensation \\
N6  & 12.512$-$0.609 & 12.446$-$0.619 & L89 & 40.0 & 4.12/12.49 & \HII\ region \\
    &               & 18:14:24.04 -18:24:43.9 & WW & 45.99 & & dust condensation \\
N8  & 12.805$-$0.312 & 18:14:40.10 -17:59:17.7 & WW & 13.81 & & dust condensation \\
N10 & 13.188+0.039 & 13.186+0.045 & L89 & 54.1 & & \HII\ region \\
    &               & 18:14:00.92 -17:28:41.3 & WW & 49.16 & & dust condensation \#1 \\
    &               &               & P08+  & 48--49 & 4.6 & methanol maser \\
    &               & 18:14:00.8 -17:28:05 & SZY00 & 48.5 & & methanol maser \\
    &               & 18:14:10.16 -17:27:20.6 & WW & 51.84 & & dust condensation \#2 \\
N11 & 13.218+0.082 & 13.231+0.082        & A+ & 54.0 & 4.7/11.8 & \HII\ region \\  
N12 & 13.727$-$0.015 & 18:15:30.2 -17:03:58 & SZY00 & 51.9 & 4.4/11.9 & methanol maser \\    
    &               & 18:15:37.63 -17:04:24.7 & WW & 47.13 & & dust condensation \\
N13 & 13.900$-$0.014 & 13.886 -0.017 & L89 & 32.6 & 3.43/13.07 & \HII\ region \\
N14 \footnotemark[4] & 14.002 -0.135 & 13.998 -0.128 & L89  & 36.0 & 3.65/12.85 & \HII\ region \\
                     &               & 18:16:22.91 -16:51:56.4 & WW & 39.97 & & dust condensation \\
                     &               & 18:16:24.23 -16:49:49.1 & WW & 41.06 & & dust condensation \\
                     &               & 18:16:32.89 -16:51:22.1 & WW & 41.32 & & dust concensation \\
N16 & 15.017+0.056 & 15.00+0.05 & A\&B & 26.5 & 13.7 & \HII\ region \\
    &               & 18:17:50.52 -15:53:33.0 & WW & 24.58 & & dust condensation \\
G15.68 & 15.680$-$0.28 & 15.64$-$0.24 & A\&B & 61.8 & 11.5 & \HII\ region \\
N18 \footnotemark[5] & 16.679$-$0.366 & 16.61$-$0.32 & A\&B & 44.9 & 12.4 & \HII\ region \\
N20 & 17.917$-$0.687 & 17.928$-$0.677 & A+ & 39.1  & 12.8 & \HII\ region \\
N21 & 18.190$-$0.396 & 18.19$-$0.40 & A\&B & 43.2 & 3.6 & \HII\ region \\
    &               & 18:25:19.37 -13:12:46.9 & WW & 49.1 & & dust condensation \\
    &               & 18:25:21.56 -13:13:39.6 & WW & 46.95 & & dust condensation \\
N22 & 18.254$-$0.305 & 18.26$-$0.30 & A\&B & 50.9 & 4.0 & \HII\ region \\
    &               & 18:25:01.48 -13:09:06.4 & WW & 66.7 & & dust condensation \\
    &               & 18:25:05.54 -13:08:19.8 & WW & 68.31 & & dust condensation \\
    &               & 18:25:06.45 -13:08:51.3 & WW & 67.75 & & dust condensation \\
    &               & 18:25:11.25 -13:08:04.4 & WW & 68.55 & & dust condensation \\
    &               & 18:25:13.3 -13:09:16 & SZY00 & 75.2 &  & methanol maser \\ 
Sh2-53 &  & 18.143$-$0.289 & K & 53.9 & 4.3 & \HII\ region near N21, N22 \\
N23 & 18.679$-$0.237 & 18.677$-$0.236 & A+ & 42.6, 61.9 & & \HII\ region \\
N24 \footnotemark[6] & 19.000$-$0.326 & 19.044$-$0.431 & L89 & 65.8 & & large \HII\ region \\
        &               & 19.12 -0.34 & S+ & 63.5 & 4.6 & large \HII\ region \\
        &               & 18:26:15.14 -12:41:36.4 & WW & 65.49 & & dust condensation \\
        &               & 18:26:23.47 -12:39:42.2 & WW & 63.45 & & dust condensations \\
N24bis1 & 19.070$-$0.279 & 19.07$-$0.27 & A\&B & 64.2 & 4.6 & UC \HII\ region \\
        &               & 19.07$-$0.28 & S+   & 64.2 & 4.7 & UC \HII\ region \\
        &               & 18:26:46.31 -12:26:27.3 & WW & 66.0 & & dust condensation \\
N24bis2 & 19.070$-$0.279 & 18.881$-$0.493 & L89 & 65.5 & & compact \HII\ region \\
        &               & 18:26:58.87 -12:44:49.3 & WW & 63.27 & & dust condensation \\
        &               & 18:27:02.74 -12:45:12.2 & WW & 65.2 & & dust condensation \\
        &               & 18:27:05.69 -12:43:10.8 & WW & 65.43 & & dust condensation \\
        &               & 18:27:07.49 -12:41:41.7 & WW & 66.08 & & dust condensation \\
        &               & 18:27:07.83 -12:41:35.9 & CY & 56.46 & & methanol maser \\
        &               & 18:27:09.27 -12:42:36.6 & WW & 65.5 & & dust condensation \\
        &               & 18:27:14.60 -12:43:10.8 & WW & 66.21 & & dust condensation \\
N25 & 19.507$-$0.191 & 19.504$-$0.193 & A+ & 37.8 & 12.8 & \HII\ region \\
G19.82 & 19.821$-$0.322 & 18:28:23.79 -11:47:36.1 & WW & 44.75 & & dust condensation \\
N27 & 19.814+0.017 & 19.813 +0.010 & A+ & 60.4, 118.0 & & \HII\ region \\
N29 & 23.055+0.559 & 23.12 +0.56 & A\&B & 29.5 & 2.2 & \HII\ region \\
N31 & 23.843+0.097 & 23.836 +0.104 & A+ & 41.9, 114.3 & & \HII\ region \\
N32 & 23.905+0.070 & 23.91 +0.07 & A\&B & 32.8 & 13.1 & \HII\ region \\
    &               & 23.91 +0.07 & A\&B & 103.4 & 9.4 & \HII\ region \\
    &               & 18:34:36.18 -08:01:02.4 & WW & 41.03 & & dust condensation \\
    &               & 18:34:37.06 -08:00:21.4 & WW & 39.25 & & dust condensation \\
    &               & 18:34:38.2 -07:59:35 & X+ & 35.7 & & methanol maser \\
N33 & 24.215$-$0.044 & 24.22$-$0.05 & A\&B & 82.0 & 10.4 & \HII\ region  \\
N34 & 24.295$-$0.170 & 24.30$-$0.15 & A\&B & 55.5 & 11.7 & \HII\ region \\
    &               & 18:36:07.97 -07:45:01.6 & WW & 53.64 & & dust condensation \\
N35 & 24.513+0.241 & 24.48+0.21 & A\&B & 115.7 & 8.6 & \HII\ region \\
    &               & 18:35:11.28 -07:26:41.1 & WW & 119.08 & & dust condensation \\
    &               & 18:34:53.0 -07:19:11 & SZY02 & 106.2 &  & methanol maser \\ 
N36 & 24.837+0.090 & 24.81+0.10 & A\&B & 108.6 & 6.4 & \HII\ region \\
    &               & 24.78 +0.08 & P08+ & 106--115 & 6.1 & methanol maser \\
    &               & 18:36:12.57 -07:12:11.5 & W98 & 106.7--114.6 &  & methanol maser \\
N37 & 25.292+0.293 & 25.29+0.31 & A\&B & 39.6 & 12.6 & \HII\ region \\
N39 \footnotemark[7] & 25.364$-$0.160 & 25.38$-$0.18 & A\&B & 57.1 & 3.8 & large \HII\ region \\    
    &               & 25.30$-$0.14 & A\&B & 98.4 & 5.9 & compact \HII\ region\\
    &               & 18:38:14.3 -06:47:47 & X+ & 58.2 & & methanol maser \\
N40 & 25.369$-$0.365 & 25.39$-$0.35 & A\&B & 62.1 & 11.3 & \HII\ region \\
N41 & 26.266+0.282 & 26.261+0.280 & A+ & 88.6, 117.0 & & \HII\ region \\
N42 & 26.329$-$0.071 & 26.330$-$0.071 & A+ & 100.9 & 5.9/9.3 & \HII\ region \\
N43 & 26.597+0.098 & 26.60+0.09 & A\&B & 102.5 & 9.0 & \HII\ region \\
N44 & 26.821+0.382 & 26.824+0.380 & A+ & 82.0 & 5.0/10.1 & \HII\ region \\
N45 & 26.991$-$0.053 & 26.98$-$0.07 & A\&B & 79.9 & 10.2 & \HII\ region \\
    &               & 18:40:46.55 -05:21:05.4 & WW & 81.85 & & dust condensation \\
N46 & 27.310$-$0.110 & 27.31$-$0.14 & A\&B & 92.3 & 5.6 & \HII\ region \\
    &               & 18:41:51.15 -05:01:42.5 & WW & 91.21 & & dust condensation \\
N47 & 28.025$-$0.031 & 28.00$-$0.03 & A\&B & 99.9 & 8.9 & \HII\ region \\ 
N49 & 28.827$-$0.229 & 28.82$-$0.23 & A\&B & 90.6 & 5.5 & \HII\ region \\
    &               & 18:44:51.0 -03:45:53.5 & WW & 86.71 & &  dust condensation \#1 \\
    &               & 18:44:51.08 -03:45:48.5& CY & 83.48 & & methanol maser \\    
    &               & 18:44:51.8 -03:45:10 & WW & 87.99 & & dust condensation \#2 \\
    &               & 18:44:46.6 -03:44:20 & WW & 96.34 & & dust condensation \#3 \\
    &               & 18:44:47.6 -03:44:49 & C+ & 100   & & methanol maser \\
    &               & 18:44:42.5 -03:44:21 & WW & 85.89 & & dust condensation \#4 \\
N50 & 29.001+0.097 & 29.007+0.076 & A+ & 67.7 & 10.6 & \HII\ region \\
G30.250 & 30.250+0.240 & 30.249+0.243 & A+ & 8.9, (90.8) & 14.0 & \HII\ region \\
N52 \footnotemark[8] & 30.749$-$0.019 & 30.78$-$0.03 & A\&B & 91.6 & 5.7 & large \HII\ region \\
    &                & 18:47:28.42 -02:01:00 & WW & 89.48 & & dust condensation \\
    &                & 18:47:36.15 -02:00:58 & WW & 91.27 & & dust condensation \\                    
    &                & 18:47:38.93 -01:58:32 & WW & 94.53 & & dust condensation \\
    &                & 18:47:39.73 -01:57:21.9 & W98 & 91.8 & & methanol maser \\
     &               & 18:47:41.42 -02:00:42 & WW & 93.89 & & dust condensation \\
    &                & 18:47:36.80 -02:00:49.0 & W98 & 88.0 & & methanol maser \\
    &                & 18:47:46.98 -01:54:19.6 & W98 & 101.2 & & metanol maser \\
    &                & 30.82$-$0.05             & P08+ & 105.7 & 9.0 & methanol maser \\
    &                & 18:47:29.9 -01:54:39    & X+   & 95.1 & & methanol maser \\
N53 \footnotemark[9] & 31.157$-$0.145 & 31.157$-$0.148 & A+ & 43.6, (101.6) & 11.6 & \HII\ region \\
G31.16 & 31.165$-$0.127 & 31.17$-$0.13 & A\&B & 41.4 & 11.9 & \HII\ region \\
N54 & 31.164+0.292 & 31.13+0.28 & A\&B & 104.7 & 7.3 & \HII\ region \\
N55 \footnotemark[10] & 32.098+0.091 & 32.11+0.09 & A\&B & 93.0 & 8.4 & nearby UC \HII\ region \\
    &               & 18:49:36.7 -00:41:05 & SLY99 & 93.2 &     & methanol maser \\
N56 & 32.583+0.002 & 32.587+0.006 & A+ & 78.1 & 9.4 & \HII\ region \\
N57 & 32.763$-$0.150 & 32.761$-$0.151 & A+ & 30.1 & 2.1/12.2 & \HII\ region \\
N58 & 32.990+0.040 & 32.99+0.04        & A\&B & 89.0 & 8.6 & \HII\ region \\
    &               & 18:51:23 +00:03:46 & SZY02 & 92.0 &     & methanol maser \\
    &               &                    & P08+ & 89--93 & 8.7 & methanol maser \\
N59 \footnotemark[11] & 33.071$-$0.075 & 33.13$-$0.09          & A\&B & 87.4 & 5.6 & \HII\ region + UC \HII\ \\
    &               & 33.129$-$0.094        & L89  & 93.8 &     & \HII\ region + UC \HII\ \\
    &               & 18:52:07.3 +00:08:05 & C+ & 73 &     & methanol maser \\
    &               &                      & P08+ & 71--81 & 8.7 & methanol maser \\
    &               & 33.20$-$0.01          & A\&B & 105.8 & 7.1   & nearby compact \HII\ region \\
    &               & 18:51:58.8 +00:06:31 & C+ & 96 & & methanol maser \\
N60 & 33.814$-$0.150 & 33.810$-$0.154 & A+ & 50.0 & 10.8 & \HII\ region \\
N61 \footnotemark[12] & 34.157+0.140 & 34.26+0.15          & A\&B & 54.0 & 3.4 & nearby UC  \HII\ region \\
    &               & 18:53:19.0 +01:14:52 & C+ & 58 &     & methanol maser \\
N62 \footnotemark[13] & 34.334+0.216 & 34.325+0.211 & A+   & 62.9 & 4.1/10.0 & \HII\ region \\
    &               & 34.40+0.23   & A\&B & 60.1 & 10.3 & nearby UC \HII\ region \\
    &               & 34.4+0.2     & R+ &      & 3.7 & nearby IRDC \\
N64 & 34.760$-$0.669 & 34.76$-$0.68 & A\&B & 52.1 & 3.2 & \HII\ region \\
N65 & 35.000+0.332 & 35.02+0.35          & A\&B & 57.2 & 3.6 & \HII\ region + UC \HII\ \\
    &               & 18:54:00.66 +02:01:19.3 & CY & 44.36   &     & methanol maser \\
    &               & 18:54:01.3 +02:01:28 & P+ & 44.4 & & methanol maser \\
N66 & 35.259+0.119 & 35.260+0.122 & A+ & 36.2 & 11.4 & \HII\ region \\
N67 & 35.543+0.011 & 35.541+0.005 & A+ & 57.6 & 10.1 & \HII\ region \\
N67bis & 35.573+0.068 & 35.57+0.07       & A\&B & 51.8 & 10.6 & \HII\ region \\
       &               & 18:56:04.3 +02:23:28 & P+ & 45.9 &  & methanol maser \\
N68 & 35.654$-$0.062 & 35.67$-$0.04          & A\&B & 51.9 & 10.6 & \HII\ region \\
N69 & 36.187+0.648 & 36.29+0.73          & A\&B & 76.5 & 4.9 &  \HII\ region \\
    &               & 18:55:15.6 +03:04:42 & SZY02 & 73.0 &     & methanol maser \\
N70 & 37.751$-$0.112 & 37.75$-$0.10          & A\&B & 49.7 & 10.4 & \HII\ region \\
    &               & 19:00:38.0 +04:13:18 & P+ & 50.3 &      & methanol maser \\
N71 & 38.290+0.007 & 19:01:20.1 +04:39:37 & P+ & 79.6 &      & methanol maser 1 \\
    &               & 19:01:27.2 +04:42:09 & P+ & 15.4 &      & methanol maser 2 \\
N73 & 38.739$-$0.137 & 38.738$-$0.140 & A+ & 60.9 & 9.2 & \HII\ region \\
N74 & 38.909$-$0.437 & 19:03:39.7 +05:09:36 & P+ & 31.9 &      & methanol maser \\
N75 & 38.928$-$0.387 & 38.930$-$0.386 & A+ & 42.1 & 2.8/10.4 & \HII\ region \\
N78 & 41.229+0.170 & 41.229+0.170 & A+ & 22.9 & 1.5/11.2 & \HII\ region \\
N79 & 41.514+0.030 & 41.52+0.03          & A\&B & 17.7 & 12.1 & \HII\ region \\
    &               & 19:07:09.4 +07:42:19 & P+ & 11.9 &      & methanol maser \\
N80 & 41.930+0.031 & 41.928+0.029 & A+ & 20.7 & 11.2 & \HII\ region \\
N81 \footnotemark[14] & 42.003$-$0.512 &                      &    &      & 8.1 & \\
N82 & 42.102$-$0.623 & 42.1$-$-0.62          & A\&B & 66.0 & 8.1 & \HII\ region \\
N83 & 42.113$-$0.442 & 42.11$-$0.44          & A\&B & 53.4 & 9.2 & \HII\ region \\
N90 & 43.774+0.059 & 43.770+0.070 & A+ & 70.5 & 6.1 & \HII\ region \\
N91 & 44.211+0.050 & 44.26+0.10          & A\&B & 59.6 & 8.1 & \HII\ region \\
    &               & 19:12:16.4 +10:07:44 & P+ & 55.7 &     & methanol maser \\
N92 & 44.333$-$0.839 & 44.339$-$0.827 & A+ & 62.5 & 6.1 & \HII\ region \\
N94 & 44.818$-$0.500 & 44.79$-$0.49          & A\&B & 44.8 & 9.3 & \HII\ region \\
N95 & 45.393$-$0.717 & 45.386$-$0.726 & A+ & 52.5 & 8.0 & \HII\ region \\
N96 & 46.947+0.368 & 46.948+0.374 & A+ & $-45.2$ & 16.2 & \HII\ region \\
N98 & 47.027+0.219 & 47.028+0.232 & A+ & 56.9 & 5.8 & \HII\ region \\
\footnotetext[1]{References: A\&B: Anderson and Bania (\cite{and09}); 
C+: Caswell et al. (\cite{cas95}); CY: Cyganowsky et al. (\cite{cyg09}); 
K: Kolpak et al. (\cite{kol03}; A+: Anderson et al. (in preparation); 
L89: Lockman (\cite{loc89}); P+: Pandian et al. (\cite{pan07}); P08+: 
Pandian et al. (\cite{pan08}); R+: Rathborne et al. (\cite{rat05}); 
S+: Sewilo et al. (\cite{sew04}); 
SLY: Slysh et al. (\cite{sly99}); SZY00: Szymczak et al. (\cite{szy00}); 
SZY02: Szymczak et al. (\cite{szy02}); WW: Wienen \& Wyrowski, 
in preparation; W98: Walsh et al. (\cite{wal98}); X+: Xu et al. (\cite{xu08}).}
\footnotetext[2]{N1: the first velocity measurement concerns the nearby 
\HII\ region G10.16$-$0.35, and not directly N1. The distance of 3.4~kpc is 
that of the exciting cluster of G10.16$-$0.35 (Blum et al. \cite{blu01}).}
\footnotetext[3]{N4: we favor the near distance as H$\alpha$ emission is 
observed.}
\footnotetext[4]{N14: we favor the near distance as H$\alpha$ emission is 
observed.}
\footnotetext[5]{N18: the distance is uncertain as the confidence parameters 
given by A\&B are not good.}
\footnotetext[6]{N24 is a large open bubble. The velocity has been measured 
in several directions at its border. Two of these are the directions of the 
UC \HII\ regions (N24bis1 and N24bis2). These UC \HII\ regions are most 
probably physically associated with N24, as they present 
very similar velocities.}
\footnotetext[7]{N39 is a large bipolar nebula, on the border of which 
lies an UC \HII\ region at 25.397$-$0.140. This \HII\ region is 
probably not associated with N39, as it presents a different velocity.} 
\footnotetext[8]{N52: the detailed velocity field of the ionized gas is 
given by Balser et al.~(\cite{bal01}), and the velocity field of the 
molecular material by Motte et al. (\cite{mot03}).} 
\footnotetext[9]{N53 lies on the border of the large 
G31.165$-$0.127 \HII\ region; they are probably linked as they present 
similar velocities.}
\footnotetext[10]{N55 lies adjacent to a dense core containing several 
UC \HII\ regions. The measured velocity is that of one of these 
UC \HII\ region.} 
\footnotetext[11]{N59 is a large bubble. The second observed direction is 
that of a compact \HII\ region, on the border of the large bubble. They 
are possibly associated.} 
\footnotetext[12]{N61 is a large and faint bubble. The measured velocity is 
that of a nearby UC \HII\ region. They are most probably associated as they 
lie on each side of an interacting filament.} 
\footnotetext[13]{The distance of N62 is uncertain. Rathborne 
et al. (\cite{rat05}) assumed that the nearby IRDC, adjacent to N62 and 
containing the G34.4$+$0.2 UC \HII\ region, is at the near kinematical 
distance; A\&B put the UC \HII\ region at the far distance, but their 
confidence parameter is not good.}
\footnotetext[14]{N81 is a large and faint bubble. N82 and N83 are observed 
respectively on the border and inside the bubble. N82 is clearly linked 
to N81, on morphological basis. The association is doubful for N83. We 
adopt for N81 the distance of N82.} 
\end{longtable}


\end{document}